MICROSPHERICAL PHOTONICS: GIANT RESONANT LIGHT FORCES,
SPECTRALLY RESOLVED OPTICAL MANIPULATION,
AND COUPLED MODES OF MICROCAVITY ARRAYS

by

Yangcheng Li

A dissertation submitted to the faculty of
The University of North Carolina at Charlotte
in partial fulfillment of the requirements
for the degree of Doctor of Philosophy in
Optical Science and Engineering

Charlotte

2015


Approved by:

______________________________
Dr. Vasily N. Astratov

______________________________
Dr. Michael A. Fiddy

______________________________
Dr. Angela D. Davies

______________________________
Dr. Yong Zhang

______________________________
Dr. Stuart T. Smith










ABSTRACT


YANGCHENG LI. Microspherical photonics: Giant resonant light forces, spectrally resolved optical manipulation, and coupled modes of microcavity arrays. (Under the direction of DR. VASILY N. ASTRATOV)

Microspherical photonics emerged in recent years in the context of fundamentally new optical properties of structures formed by coupled dielectric microspheres. These include coupling between whispering gallery modes (WGMs), photonic nanojets, nanojet-induced modes, resonant light pressure and optical super-resolution effects. The bottleneck problem in this area is connected with size disorder of individual microspheres which leads to optical losses and degraded performance of coupled devices.

In this dissertation novel resonant propulsion of dielectric microspheres is studied with the goal of sorting spheres with identical resonances, which are critical for developing microspherical photonics. First, evanescent field couplers were developed by fixing tapered microfibers in mechanically robust platforms. The tapers with ~1 μm waist diameter were obtained by chemical etching techniques. Using these platforms, WGMs modal numbers, coupling regimes and quality factors were determined for various spheres and compared with theory. Second, the spectroscopic properties of photonic molecules formed by spheres sorted by individual characterization with better than 0.05% uniformity of WGM resonances were studied. It was shown that various spatial configurations of coupled-cavities present relatively stable mode splitting patterns in the fiber transmission spectra which can be used as spectral signatures to distinguish such photonic molecules. The third part of the dissertation is devoted to the observation and study of giant resonant propulsion forces exerted on microspheres in the evanescent




microfiber couplers. This effect was observed in suspensions of polystyrene spheres with sufficiently large diameters ($D > 10$ µm). By integrating optical tweezers for individual sphere manipulation, the wavelength detuning between a tunable laser and WGMs in each of the spheres was precisely controlled. Resonant enhancement of optical forces was directly demonstrated in experiments. The spectral shape, position and magnitude of the observed propulsion force peaks were explained by efficient transfer of light momentum to microspheres under resonant conditions. The peak magnitude of the resonant force is shown to approach total absorption limit imposed by the conservation of momentum. The transverse movement of the spheres during the propulsion process was studied and the existence of a stable radial trap was demonstrated. Giant resonant propulsion forces can be used for large-scale sorting of microspheres with ultrahigh uniform resonant properties.



## ACKNOWLEDGMENTS

I am deeply grateful to my advisor, Prof. Vasily N. Astratov. During my entire period of study and research in the graduate program he continuously advised and supported my research efforts. Our research was guided by his sharp intuition and insight. His knowledgeable and enthusiastic approach to research, which often went beyond science itself, was an inspiration for me and other group members. He has always been diligent in preparing me for all of my publications and presentations. It has been my privilege to have numerous scientific discussions with him which make this dissertation work possible.

I would like to thank my committee members, Profs. Michael A. Fiddy, Angela D. Davies, Yong Zhang, and Stuart T. Smith for serving on my dissertation committee. I am grateful for their time and appreciate their comments and supports for the completion of this dissertation.

I would also like to thank Prof. Angela D. Davies as the graduate program director for her help and guidance of my study at UNC Charlotte, and thank Prof. Glenn D. Boreman as the department chair for his care and support.

I am grateful to Dr. David Carnegie and Prof. Edik U. Rafailov from University of Dundee in the UK for providing us the tunable laser for optical propulsion experiments, and to Dr. Alexey V. Maslov from University of Nizhny Novgorod in Russia for his insight in microphotonics modeling and for the discussions and help in interpretation of the results of resonantly enhanced optical propulsion.

During my graduate studies I have enjoyed productive interactions and discussions with my fellow Mesophotonics Laboratory members, Dr. Oleksiy V. Svitelskiy, Arash



Darafsheh, Kenneth W. Allen, S. Adam Burand, Farzaneh Abolmaali, and Navid Farahi, over a broad range of research projects, which have led to great collaboration works.

I am thankful to Scott Williams, Dr. Lou Deguzman, and Dr. Robert Hudgins for their help of operating equipments in the Optoelectronics Center at UNCC. Also, I would like to thank Mark Clayton and Wendy Ramirez for assisting me throughout many administrative issues.

I was supported by the GASP tuition award from UNCC Graduate School throughout my graduate studies. I am thankful for the teaching assistantship provided by the Department of Physics and Optical Sciences. My research assistantship in Prof. Vasily N. Astratov's Mesophotonics Laboratory was supported through grants from Army Research Office and National Science Foundation.

At last, I wish to express my deep gratitude to my parents for their endless love and support and to my wife for her unconditional love and encouragement which have always been the comfort and the motivation for me.



# TABLE OF CONTENTS













LIST OF FIGURES

































# LIST OF ABBREVIATIONS

| | |
|---|---|
| ~ | approximate |
| 1-D | one-dimensional |
| 2-D | two-dimensional |
| 3-D | three-dimensional |
| BTG | barium titanate glass, $BaTiO_3$ |
| CROW | coupled resonator optical waveguide |
| D | diameter |
| dB | decibel |
| FDTD | finite-difference time-domain |
| FWHM | full width at half-maximum |
| HF | hydrofluoric acid |
| IR | infrared |
| NA | numerical aperture |
| Q | quality factor |
| TE | transverse electric |
| TM | transverse magnetic |
| WGM | whispering gallery mode |

# CHAPTER 1: INTRODUCTION

## 1.1. Outline and Overview of the Dissertation

In this dissertation, we develop and study new technological, theoretical and experimental approach to building coupled cavity structures and devices. We address the main technological bottleneck problem of contemporary physics and technology of coupled microstructures – the fact that the building blocks of such structures are inevitably different and they have different resonant properties. The dissertation deals with basic principles of coupling light in such structures and basic phenomena responsible for the efficiency of light hopping or tunneling between such building blocks. As a solution of this bottleneck problem, we develop in this work new approach to sorting microspheres with almost identical resonant positions of whispering gallery modes. This is based on developing pioneering work of fathers of optical tweezers, Ashkin and Dzedzic [1] who observed weak resonances in radiation pressure effects induced by the coupling of light to WGMs in spheres almost four decades ago. Since that time, this work was not actively developing until very recently when the studies of this effect became a cornerstone of cavity optomechanics [2-4]. In the present work, we studied radiation pressure effects in the context of optical propulsion of microspheres with a tapered microfiber coupler and we observed totally new properties of these forces. First of all, these include observation of giant magnitude of resonant peaks of optical forces [5, 6] which was largely missed in the previous studies of radiation pressure. We



showed that the magnitude of these peaks corresponds to theoretical limits imposed for light-pressure effects by the conservation of light momentum [5, 6]. We showed that due to giant magnitude and large peak-to-background ratios, these effects can be used for developing novel technology of sorting microspheres with uniquely resonant properties [5-8] .These sorted spheres can be treated as photonic atoms due to their nearly indistinguishable positions of WGMs. Photonic atoms in photonic applications can behave similar to classical quantum mechanical atoms in solid state physics in the sense that they can be used as building blocks for photonic structures where the tight-binding approximation can be implemented to engineer photonic dispersions.

Chapter 1 is devoted to be an introduction of the dissertation. The scope of original research carried out in each following chapters will be briefly discussed. A literature review will be given on relevant materials and important articles existing in the field of study. The sub-sections in the introductory chapter can serve as introductions corresponding to the following original chapters in the dissertation.

In Chapter 2, we characterized WGMs in microspheres with various materials and different sizes by using evanescent coupling to a tapered fiber coupler. A platform integrated with tapered fiber was first designed and fabricated to enable the characterization experiments both in air and in water environments [9]. Spheres can be picked up and attached to the tapered region by a micromanipulator equipped with a fiber tip under microscope. Fiber transmission spectrum was obtained with broadband white light source and optical spectral analyzer. Coupling parameters were analyzed and WGMs quantum mode numbers were determined by fitting analytical formulas [10]. Dependence on sphere's refractive index and diameter was studied for the coupling



strength and the quality factor ($Q$-factor). The dependence on the diameter of the tapered fiber was also explored. An exponential increase of $Q$-factor with the sphere diameter was found for relatively small spheres ($D$<20 μm). Transmission spectra were compared with air and water as the medium. And substantially lower $Q$-factor was measured in water environment due to smaller sphere-medium refractive index contrast.

In Chapter 3, we studied various photonic molecules built by identical microspheres with overlapping WGM resonances both theoretically and experimentally [11, 12]. Finite-difference time-domain (FDTD) method was used for numerical simulation of optical coupling and transport properties. We simulated the fiber transmission spectra when it side coupled to a variety of molecule configurations [11, 12]. We summarized the commonly seen mode splitting patterns for each configuration using three different constituent spheres, and presented them as spectral signatures associated with each coupled photonic molecules. The spectral signature is represented by the number of split supermodes and their spectral positions which was found to be a unique and relatively stable property linked to each photonic molecule. The electric field maps of several molecules were presented that shows interesting light coupling and transport phenomena. In experiments, we selected a series of size-matched polystyrene microspheres with mean diameter of 25 μm by spectroscopic characterization and comparison. These pre-selected spheres were used to assemble various photonic molecules in aqueous environment. Experiments of side coupling to these molecules were performed and transmission spectra showing supermode positions were obtained with the modified experimental platform. An excellent agreement was found between transmission spectra measured in the experiment and spectral signatures predicted by the simulation. These experiments



provide a demonstration of potential photonic devices that can be built by microspheres with overlapping resonances. An alternative method of tuning coupling constant between coupled microspheres was also demonstrated in experiments by controlling the spatial position of excitation, which is especially useful when coupling needs to be tuned simultaneously among all spheres in a complicated structure.

In Chapter 4, we observed the optical trapping and propulsion of polystyrene microspheres by the evanescent field of the laser light propagating inside a tapered fiber in experiments [5]. A microfluidic platform integrated with tapered fiber and micropump was prepared to ensure interactions between evanescent fields and spheres in the aqueous environment. The dependence of measured propelling velocities on the sphere diameter was analyzed statistically. Uniform Gaussian distribution of velocities was obtained for small spheres. Some extraordinary high velocities were recorded and large scattering of velocity data was found for large spheres ($D$>15 μm) [5]. Hypothesis of resonant enhancement of optical forces due to pronounced WGMs in large polystyrene microspheres was made based on these experimental observations. Analytical calculations of optical forces were performed to explain the mechanism and also to compare with experimental results.

In Chapter 5, we proved our hypothesis of resonant enhancement of optical forces directly in experiments by individual manipulation of microspheres and spectroscopic control of the relative position of laser wavelength and WGM resonances. We built an optical tweezers set-up and altered the abovementioned platform to enable the precise trap and movement of each individual spheres in the aqueous environment [6]. In contrast to the work in Chapter 3, where the detuning between the fixed laser wavelength and



WGMs in spheres was realized randomly by the 1-2% size deviation of the microspheres, here we can characterize the WGMs while the sphere was held at the vicinity of tapered fiber and tune the wavelength of tunable laser to any desired positions. Therefore we realized a precise control of the wavelength detuning and investigated the spectral properties of optical forces in the optical propulsion experiments with polystyrene spheres of 10 and 20 μm mean diameters [6]. The distribution of propelling velocities was found to be inversely corresponding to the fiber transmission spectrum. A comparison of optical propulsion forces between strong coupled 20 μm spheres and weak coupled 10 μm spheres provides a clear demonstration of the effect of resonant enhancement of optical forces due to WGMs in spheres. Such effect can be applied to develop sorting techniques to obtain microspheres with overlapping resonances with a precision that is not achievable in traditional fabrication techniques. The dynamics during sphere propulsion and the temporal properties were also studied by connecting the fiber output to a photodiode and the oscilloscope, and synchronizing the real time transmission signal with the instantaneous propelling velocities. The temporal behaviors of optical trapping and propulsion were recorded and analyzed.

In Chapter 6, we drew conclusions and proposed future work that could be developed based on the contents of this dissertation.

## 1.2. Resonant Optical Properties of Dielectric Spherical Microcavities

In this Section, we discuss mechanisms of light interactions with dielectric spherical microcavities. Our focus is resonant optical properties based on whispering gallery modes (WGMs) [13] supported in the spherical microcavities.



### 1.2.1. Whispering Gallery Modes in Spherical Microcavities

Electromagnetic waves can be strongly confined in microcavities with circular symmetry due to total internal reflection of light and interference phenomena [13-19]. The resonant modes are called whispering gallery modes (WGMs) in analogy to acoustic whispering gallery waves discovered in architecture. Such confinement of light in cavities takes place in a variety of circular structures such as microrings [20-24], micropillar/microcylinders [25-27], microtoroids [28-31], microdisks [32-35], and microspheres [36-49]. Micropillar resonator, shown in Fig. 1.1(a) is usually a cylinder of high index materials sandwiched between layers of distributed Bragg-grating resonators (DBR). Light is confined in the axial direction by the reflection from DBR and confined in the lateral direction by the total internal reflection from walls of the cylinder. The microdisk resonator shown in Fig. 1.1(b) is a circular resonator in which light travels at near grazing angle and is confined by total internal reflection from the boundary. The microtoroid resonator, as illustrated in Fig. 1.1(c), acts like a fiber-ring resonator, in which the resonant modes are circulating guided waves. The size of the microresonators usually ranges from a few microns to tens of microns.

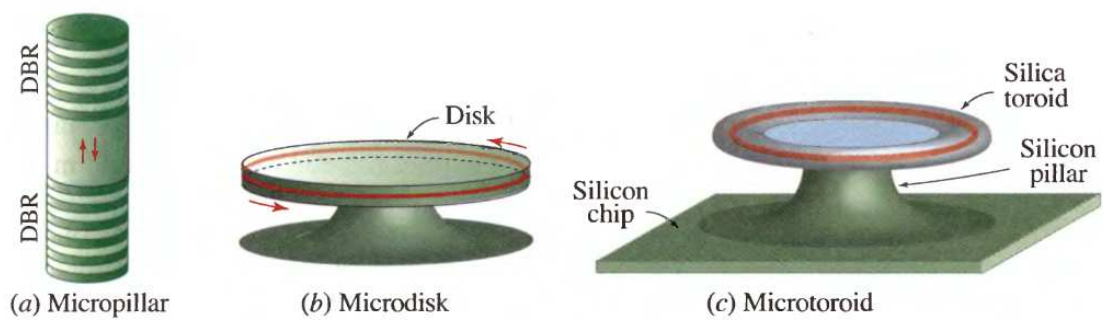

Figure 1.1: Micropillar/microcylinder, microdisk, and microtoroid resonators. [13]

WGMs in microresonators can reach high quality factor ($Q$-factor) and small mode volume due to very small internal loss, which leads to various applications, such as microlasers [32, 46, 50-53], detection of biomedical molecules and viruses [30, 54-57], detection of nanoparticles [58, 59], environmental medium sensors [60-64], temperature sensors [65-67], and force sensors [68, 69].

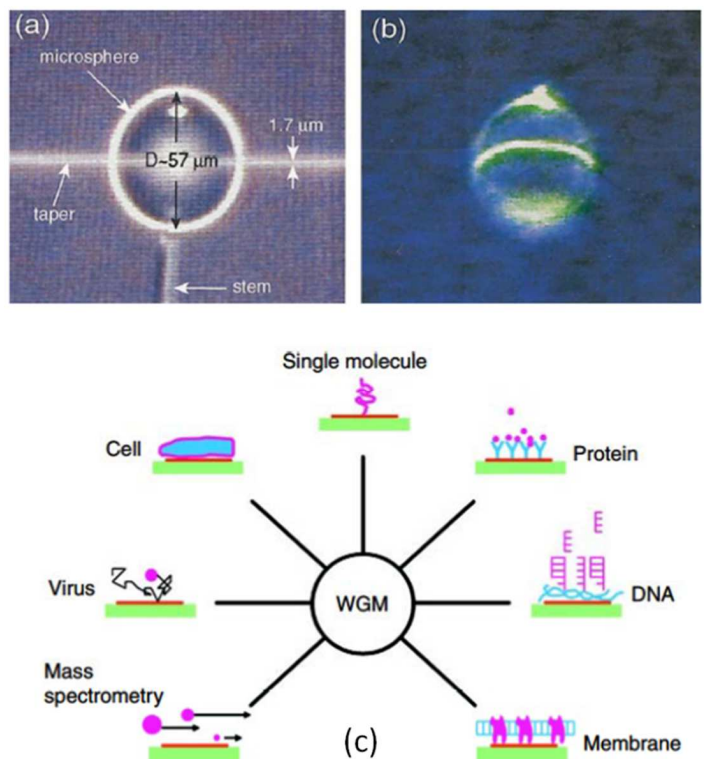

Figure 1.2: (a-b) Images of microsphere laser pumped from a taper [53]. (c) Applications of WGM biosensors for various biological materials [30].



In this dissertation, we focus on WGMs in microspheres due to their higher $Q$-factor [39-41], commercial availability [70, 71], and easier manipulation. Optical waves trapped inside a microsphere with radius $R$ propagate in a circumference near the surface of the sphere, thus they travel a distance of $\sim 2\pi R$ in a round trip. Under the condition of constructive interference, standing waves of WGMs are formed in a circular cavity if $2\pi R \approx l(\lambda_0/n)$ where $\lambda_0/n$ is the wavelength in the medium with refractive index $n$ [72].

Whispering gallery modes (WGMs) in microspheres can be characterized by three quantum mode numbers as radial $q$, angular $l$, and azimuthal $m$ numbers, due to the spherical symmetry [16, 17]. The radial number $q$ indicates the number of WGM intensity maxima along the radial direction. Fig. 1.3 illustrates electric field patterns of WGMs standing waves with radial mode numbers $q$=1, 2, 3 respectively. For WGMs with $q$=1, which are the first-order radial modes, light is concentrated in the circumference close to the sphere surface. The stronger light confinement and smaller leakage lead to the highest $Q$-factor for the first-order radial modes.

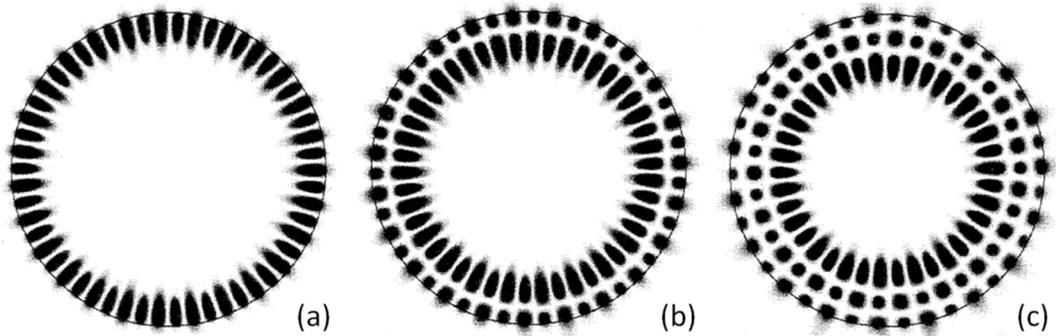

Figure 1.3: Simulated WGMs electric field patterns for different radial mode number $q$: (a) $q$=1, first-order mode; (b) $q$=2, second-order mode; (c) $q$=3, third-order mode. [74]



The angular number *l* represents the number of modal wavelengths that fit into the circumference of the equatorial plan, represented by the number of field maxima along the circle in Fig. 1.3. As we can see in the figure, *l* is much larger than 1. That is the case for high-*Q* microresonators which require strong light confinement. WGMs with the same radial number *q* and different angular number *l* appear as periodic peaks or dips in the spectrum. Fig. 1.4 shows an example of transmission spectrum for 5 $\mu$m diameter microring resonator coupled to straight waveguide, which displays a set of WGMs with radial number *q*=1 and different angular numbers *l* ranging from 25 to 29. Larger *l* number means higher frequency while shorter wavelength. These angular modes are separated by the free spectral range (FSR) as in typical resonators, $\Delta v_{FSR} \approx (c/2\pi R)n_{eff}$, where *R* is the radius and $n_{eff}$ is the effective index of the guided mode.

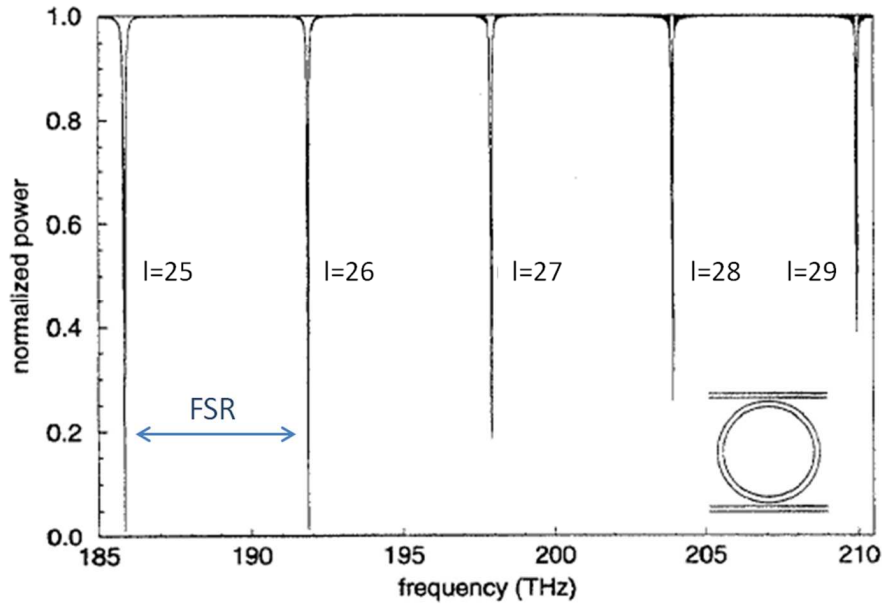

Figure 1.4: Simulated transmission spectrum for 5 $\mu$m diameter microring resonator coupled to straight 0.3 $\mu$m wide waveguide, showing first-order (*q*=1) WGMs with different angular number *l* separated by FSR. [74]



The azimuthal mode number $m$ can take from $-l$ to $+l$ with a total of $2l+1$ values and represents the modes effectively spreading away from the equatorial plane towards the poles of the microsphere. As illustrated in Fig. 1.5(a), a single sphere on the substrate is rotationally invariant around the z axis which is the polar axis. In this geometry, the fundamental WGMs is defined as the first-order radial modes $q=1$ with the azimuthal mode number equal to the angular number $m=l$ in the sphere's equatorial plane parallel to the substrate [75]. Such fundamental modes have the highest $Q$-factors, due to the fact that they are separated from the surface of the substrate by the radius of the sphere. In contrast, the modes with $m \ll l$ are damped, since they have a wider spatial distribution with the maximum intensity approaching to the high index substrate which leads to the leaking of light. Figs. 1.5(b) and 1.5(c) illustrate these two cases of $m=l$ and $m<l$ respectively for the second-order WGMs with $q=2$.

In an ideal free-standing sphere with a perfect spherical shape, the azimuthal modes represented by different $m$ numbers with the same angular number $l$ are degenerate. However, in real experimental situations the degeneracy of azimuthal modes is often lifted by small elliptical deformation of microspheres. In experiments what we measured is usually the spectral overlap of several simultaneously excited azimuthal modes with slight shift, which results in the broadening of the measured resonance width and the decrease of the observed $Q$-factor compared to their intrinsic values calculated from single mode models.



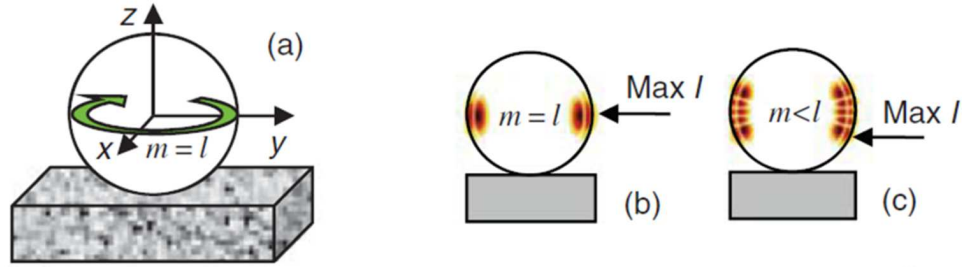

Figure 1.5: (a) Schematic of a single sphere on a substrate. (b-c) Illustration of intensity maxima distributions for azimuthal modes with different *m* numbers. [76]

A typical fluorescence (FL) spectrum of a single dye-doped 5 $\mu$m polystyrene sphere on a glass substrate is presented in Fig. 1.6, where both first-order and second-order modes are shown [76]. The WGMs peaks with orthogonal polarizations are labeled as TE$^q{}_l$ and TM$^q{}_l$ respectively. The *Q*-factors of the first-order modes in Fig. 1.6 (*q*=1) are measured to be $4\times10^3$, which is well below the theoretical limit (*Q*>$10^5$) calculated for perfect 5 $\mu$m spheres.

The spectral positions of fundamental WGMs peaks can be derived analytically from Maxwell's equations solved in the spherical coordinates by using the Mie scattering formalism [10, 17]:

$$P \ \frac{J'_{l+1/2}(nk_0R)}{J_{l+1/2}(nk_0R)} = \frac{N'_{l+1/2}(k_0R)}{N_{l+1/2}(k_0R)} \ , \tag{1.1}$$

where *n* is refractive index, *P*=*n* for TE polarization, and *P*=*n*$^{-1}$ for TM polarization. *J* and *N* are Bessel and Neumann functions, respectively, and the *l*+1/2 term appears from translating the spherical Bessel and Neumann functions to their cylindrical counterparts.



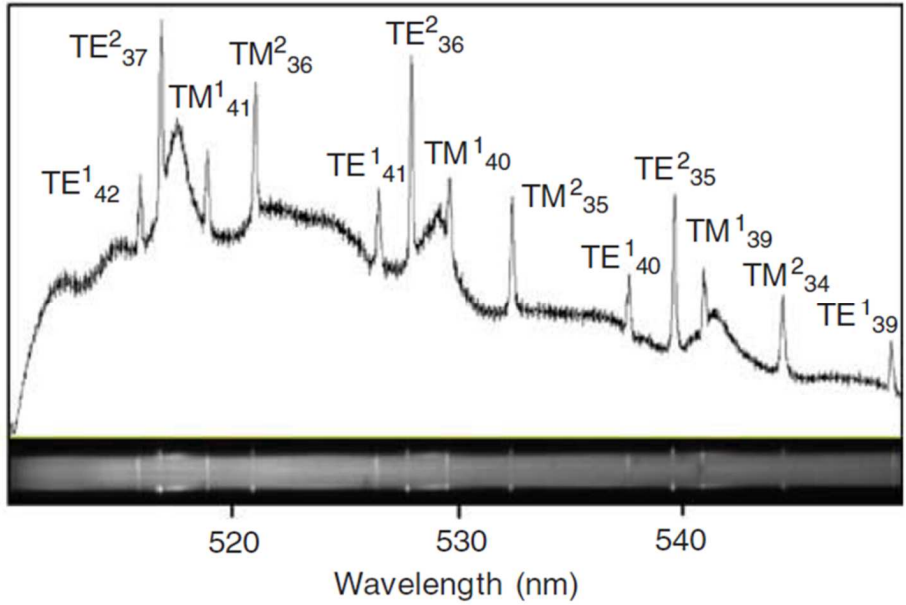

Figure 1.6: Fluorescence spectral image and emission spectrum of a single 5 μm fluorescent sphere with $TE^q_l$ and $TM^q_l$ WGMs peaks. [76]

Solutions of this equation determine an infinite set of wave vectors (eigenfrequencies) for a given radius $R$. Another index $q$ (radial number) is introduced here to indicate the correspondence between roots of Eq. (1.1) and the eigenfrequencies. The characteristic equation is independent of the index $m$, which shows the degeneracy of these modes. Expanding Eq. (1.1) as an asymptotic series in powers of $(l+1/2)^{-1/3}$ and Airy functions [10] gives the first terms of WGM resonances:

$$(nk_0R)_{q,l} = l + \frac{1}{2} - \left(\frac{l+1/2}{2}\right)^{\frac{1}{3}} \alpha_q - \frac{P}{\sqrt{n^2-1}} + \cdots \ , \tag{1.2}$$

where $q$ is the radial number, $l$ is the angular number, and $\alpha_q$ are the roots of the Airy function Ai(-z) associated with the radial number which is equal to 2.338, 4.088, and 5.521 for $q$=1, 2, and 3, respectively.



The solutions lead to a quasi-periodic spectrum of WGM resonances with various angular numbers *l*, whose frequencies are separated by the free spectral range (FSR) for the same polarization [10, 77]:

$$\Delta\nu \approx \frac{c}{2\pi R}\frac{\tan^{-1}\left(\sqrt{n^2-1}\right)}{\sqrt{n^2-1}} \quad,$$
(1.3)

### 1.2.2. Excitation of Whispering Gallery Modes in Microresonators

A crucial aspect in the study of WGMs is that one needs to be able to couple light into and out of the microcavity efficiently. WGMs in microresonators can be excited by illumination of free-space beams. However, due to the nature of light trapping by total internal reflections the excitation would not be efficient. External beams will be partially reflected at the surface of the resonator. The fraction of incident light that transported inside the resonator would escape after multiple reflections at the inner surface. With the principle of optical reciprocity external incident beams cannot be efficiently coupled to WGMs, which leads to light leakage and large loss.

As illustrated in Fig. 1.7, WGMs in microspheres excited by plane wave illumination (a) was compared with coupling via side-polished fiber (b) [78, 79]. The fiber coupler is made from a single-mode optical fiber with a core radius of 1.9 mm (refractive index 1.462) and a cladding radius of 62.5 mm (refractive index 1.457). The cladding below the microsphere is polished down to 0.7 mm to allow the exposure of evanescent fields. The guided mode inside the fiber has an approximately Gaussian intensity profile. The calculated scattering spectra are shown in Figs. 1.7(c) and (d) for corresponding cases. Compared with plane wave scattering results, there is little off-resonance scattering (above the dashed line indicating zero level) for the fiber



coupling. First-order and second-order WGMs were clearly observed in both configurations but the evanescent coupling yields a much higher peak-to-average ratio than the plane wave excitation. A beam focused at the edge of the sphere will result in a better coupling than the plane wave illumination. However it is not as efficient as the evanescent coupling, lacks the modes uniformity, and experimentally not easy to realize [78, 79].

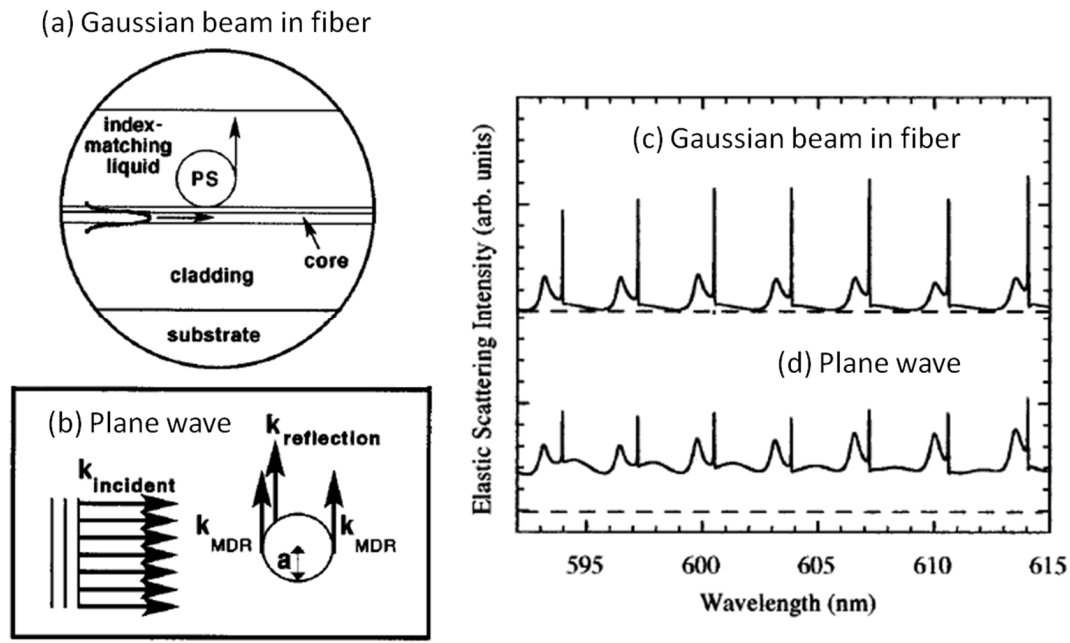

Figure 1.7: Illustration of experiments for (a) polished fiber evanescent coupling and (b) plane wave excitation of a microsphere, respectively. (c) and (d) Calculated scattering spectra corresponding to (a) and (b) [78, 79].

Other evanescent couplers that are widely adopted in experiments include prism coupler [46, 80], surface waveguide [34, 81-83], and tapered fiber [28, 30, 45, 49, 84-86]. The geometries of these couplers are illustrated in Fig. 1.8.



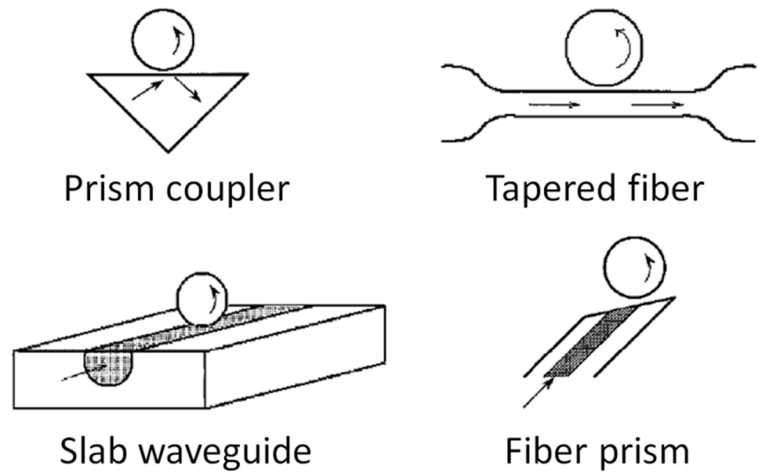

Figure 1.8: Geometries of various evanescent couplers commonly used for excitation of whispering gallery modes [81].

When using the prism coupler, one produces an evanescent field at the surface of a high-index prism under the condition of total internal reflection. The angle of incident light on the internal surface is chosen to match the propagation constant of that required for the excitation of WGMs. A large number of free-space modes can be coupled into the resonant modes. Hence the energy transfer can be efficient. However the set-up is usually bulky with all the beam focusing elements and the manipulation of spheres on prism surface to achieve the optimal coupling is not easy. The tip angle-polished fiber can be considered as an alternative prism [87]. When the light is incident upon the angled surface, the propagating wave inside the core undergoes total internal reflection and escapes the fiber. With the sphere positioned in the range of the evanescent field from the core area, an efficient coupling would occur between fiber modes and WGMs in spheres. For the tapered fiber and slab waveguide couplers, the optical wave propagating inside the waveguide will have an evanescent tail extending outside at the waveguide-medium



interface. These evanescent fields extending to the free space can have efficient coupling to WGMs in spheres if they are placed in the near field.

Loading $Q$-factors were calculated by Gorodetsky and Ilchenko [37] for different types of evanescent couplers with optimized parameters when the silica sphere placed in contact with the couplers and the comparison is presented in Fig. 1.9.

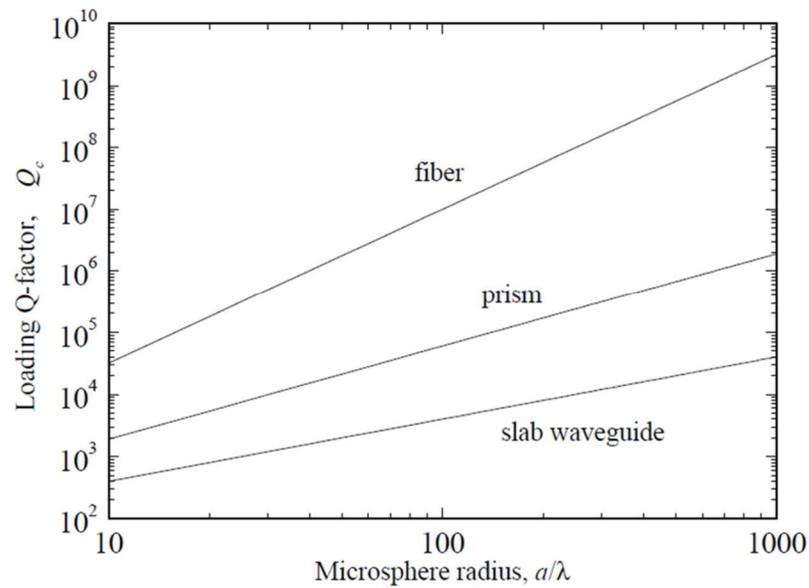

Figure 1.9: Comparison of calculated loading $Q$-factor for different types of evanescent couplers in contact with silica spheres with optimized parameters. [37]

The results in Fig. 1.9 indicate that the loading $Q$-factor increases very quickly with the increase of the sphere size. It can also be seen that the tapered fiber demonstrates the highest coupling efficiency. The tapered fiber is convenient to use as well. Since the taper is usually heat pulled or chemical etched from single-mode telecommunication fibers it is easy to be connected to common fiber integrated devices, eliminating the need of all the



focusing and waveguide coupling elements. Furthermore, the propagation constant in the core of a silica-based optical fiber will always be closest to that of WGMs with the lowest radial mode number which provides desirable coupling properties [84]. The waist region of the taper is usually only a few micrometers in diameter, and can even be sub-micrometer. The light travels in the fundamental fiber mode in the core surrounded by air or other medium. If the waist approaches one micrometer, the fundamental mode will have an evanescent tail significantly extending out. Thus the propagation constant and the coupling efficiency should depend on the taper waist diameter. Generally, a smaller sphere is likely to be phase matched with a thinner tapered fiber. However, it was shown [84] that a relatively small range of taper sizes can match the propagation constants for a wide range of sphere sizes. And because the interaction length is usually small limited by the curvature of the microsphere surface, the overall coupling is not expected to be sensitive to the small variation in the taper diameter or sphere size.

For the above discussed advantages of tapered fiber couplers, they are widely adopted as evanescent couplers for excitation of WGMs in experiments. It can be used as a single waveguide for input as well as output of the microresonator, as shown in Figs. 1.10(a) and (b). This configuration is usually used for excitation of high-$Q$ WGMs for the applications as sensors and lasers. Or two tapered fibers can couple to a microsphere at both sides, as illustrated in Figs. 1.10(c-d), forming a four-port add-drop line device. The signals can be selectively exchanged between two tapered fibers through coupling to the microsphere according to its resonant wavelengths of WGMs. It can also be used as optical switch or efficient energy transfer devices [88].



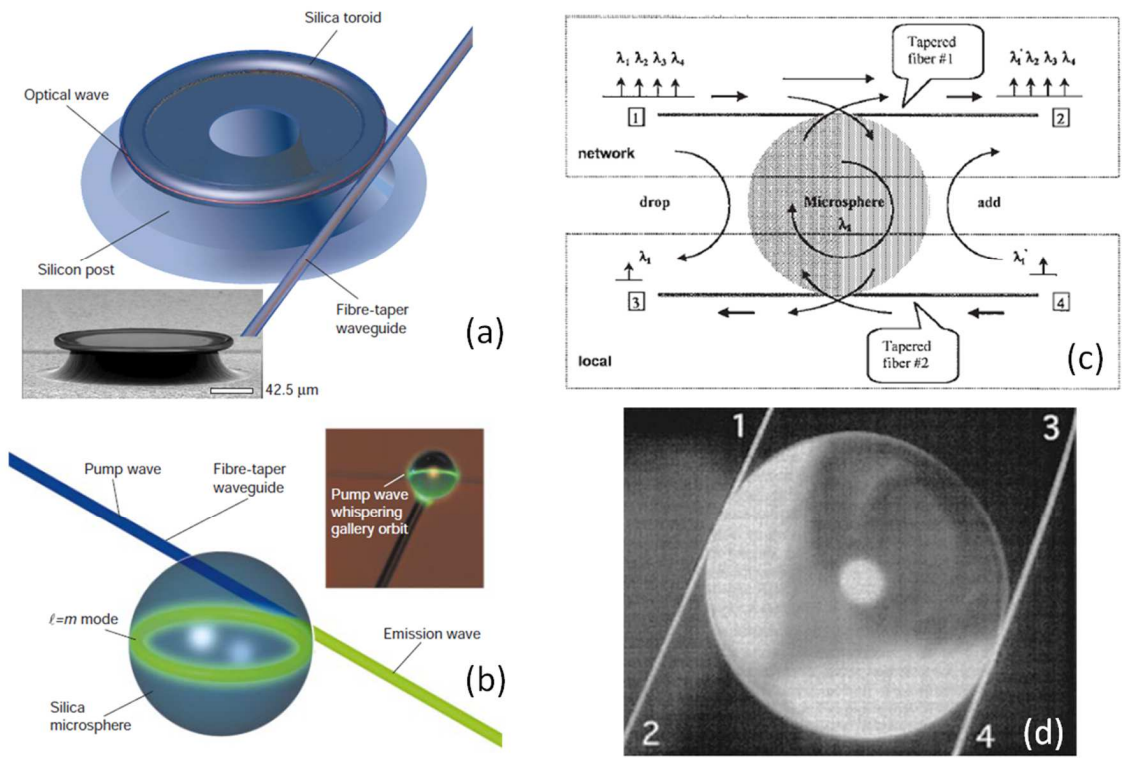

Figure 1.10: Tapered fiber couplers. Single tapered fiber coupled to a microtoroid (a) and a microsphere (b) resonators [15]. (c-d) Two tapered fiber coupled to a microsphere, forming a four-port add-drop line device [49].

## 1.2.3. Critical Coupling Conditions

The manipulation of optical coupling or energy transfer between a waveguide coupler and a resonator is of great interests in research. The ability to achieve complete energy transfer efficiency is fundamental to many quantum-optical studies as well as to practical applications of such coupled structures. To analyze the exchange of optical power, the most basic and generic configuration as shown in Fig. 1.11 was considered by Yariv [89], which consists of a unidirectional coupling between a ring resonator and a waveguide. Under the conditions that a single unidirectional mode of the resonator is



excited and that the coupling is lossless, we can describe the interaction by the matrix relation:

$$\begin{bmatrix} b_1 \\ b_2 \end{bmatrix} = \begin{bmatrix} t & \kappa \\ -\kappa^* & t^* \end{bmatrix} \begin{bmatrix} a_1 \\ a_2 \end{bmatrix}$$
(1.4)

where the complex mode amplitudes $a_i$ and $b_i$ are normalized such that their squared magnitude corresponds to the modal power. The coupling matrix is unitary so that

$$|\kappa|^2 + |t|^2 = 1$$
(1.5)

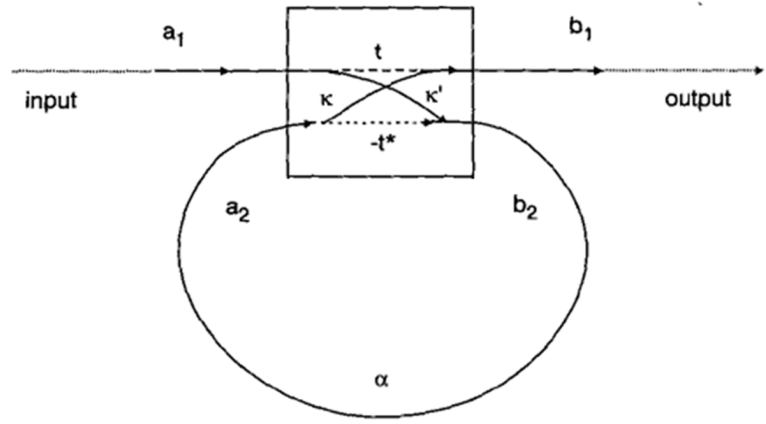

Figure 1.11: Generic description of dielectric waveguide coupled to ring resonator. [89]

At resonances, we can derive the transmission of the waveguide past the resonator and the power circulating inside the resonator as [89]:

$$|b_1|^2 = \frac{(\alpha - |t|)^2}{(1 - \alpha |t|)^2} \quad , \quad |a_2|^2 = \frac{\alpha^2 (1 - |t|)^2}{(1 - \alpha |t|)^2}$$
(1.6)

The first part of Eq. 1.6 shows that when $\alpha = |t|$, i.e. when the internal loss of the resonator (represented by $\alpha$) is equal to the coupling loss of the waveguide (represented by $|t|$) the



transmitted power will vanish, i.e. $|b_1|^2 = 0$. This condition is called critical coupling where all the power is coupled into the resonator thus no transmission at the waveguide output. This result is due to perfect destructive interference between the transmitted field $ta_1$ and the internal field coupled into the output waveguide $\kappa a_2$, similar to the relations described in the Fabry–Perot etalon.

Therefore, two methods can be used to control the waveguide transmission and to achieve critical coupling, either by adjusting internal cavity loss [45, 90] or by controlling the condition at the coupling region [45, 91, 92].

Examples of near critical coupling achieved experimentally by Vahala *et al*. [45] are presented in Fig. 1.12. A silica microsphere of 150 μm diameter was side-coupled to a tapered fiber. To change the resonator's internal loss, a plastic probe tip having comparable size to the microsphere was fabricated and placed near it. The plastic material is lossy at the experiment wavelength of 1.5 μm so that power coupled from the resonator would be absorbed quickly within the probe. The position of the probe was controlled by a piezoelectric motor. By approaching the probe to the microsphere, the total cavity loss of the resonant modes would increase due to the probe absorption and scattering. Fig. 1.12(a) shows the waveguide transmission versus the probe position. A clear transmission minimum approaching zero is apparent in the figure, corresponding to the critical coupling point. For the experiment of adjusting waveguide-resonator coupling, the probe was set at a fixed distance from the sphere, and the gap between the tapered fiber and the sphere was varied incrementally. In this case the taper was very slightly detached from the sphere. A plot of the on-resonance taper transmission versus width of the resonance is shown in Fig. 1.12(b), along with the fit where the resonator round-trip amplitude α was



chosen to match the resonance width at the critical point. Close to zero on-resonance transmission was recorded thus demonstrating critical coupling condition can be realized by matching waveguide-resonator coupling parameter |t| and cavity internal loss α.

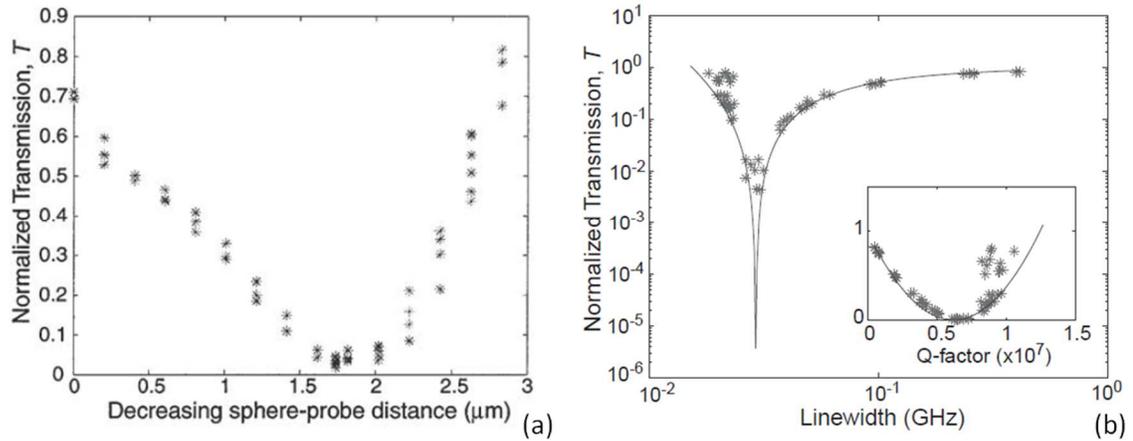

Figure 1.12: (a) Normalized transmission as a function of the probe position for a fundamental WGM of a 150 μm microsphere. (b) Transmission as a function of the linewidth (FWHM) of the mode for varying taper-sphere coupling gap. [45]

## 1.3. Optical Coupling in Photonic Molecules

The concept of photonic atom was brought up by Arnold *et al.* [93] based on the observation that the Maxwell equations describing the modes of spherical resonator can be reduced to a form closely resembling the Schrodinger equation describing real physical atoms in quantum mechanics. This terminology was then used in a broader sense that the individual atom can be used as a basic unit to build coupled cavity structures referred to as photonic molecules. If placed at the vicinity of each other, WGMs in the microresonators can have efficient coupling provided that they are in coherent resonances.



Thus, identical microresonators with the same resonant wavelengths are ideal photonic atoms that can be used as building blocks for resonant coupled structures which are usually referred to as photonic molecules [94, 95] in analogy to the coupled electronic states in chemical molecules described by the quantum mechanics [96].

### 1.3.1. Coupling of WGMs in Resonant Bi-spheres

Dielectric microsphere is known as unique optical microcavity for its high $Q$-factor, long photon lifetime, and small mode volume [36-49]. Light can be confined tightly in a single sphere acting as a photonic atom with the wavelength scale. If multiple coherently resonant spheres are placed close to each other, they form a photonic molecule. Light can propagate through such molecule due to the coupling between the nearest neighbors. This approach is referred to as the tight-binding photon approach [97, 98]. Within the tight-binding photon approach one can guide the optical waves by constructing these atoms into arbitrarily shaped structures.

The simplest structure of such photonic molecules is the resonant bi-sphere, and it was studied by Kuwata-Gonokami *et al*. [99]. Monodispersive polystyrene microspheres (refractive index 1.59) were dye doped in Nile Red solution. The dye fluorescence was collected with a tapered fiber probe and sent to the spectrometer, showing spectrum of WGMs resonances. Spheres of 2 μm in diameter were found to have a $Q$-factor on the order of $10^2$, while spheres of diameters more than 4 μm presented $Q$-factors on the order of $10^3$. WGMs spectra collected from different spheres can be compared and the relative size of the sphere can be judged within an error of $1/Q$. Two spheres of identical sizes were able to be chosen by comparing and matching the wavelengths of their WGMs resonances, and were assembled as bi-sphere in contact position.



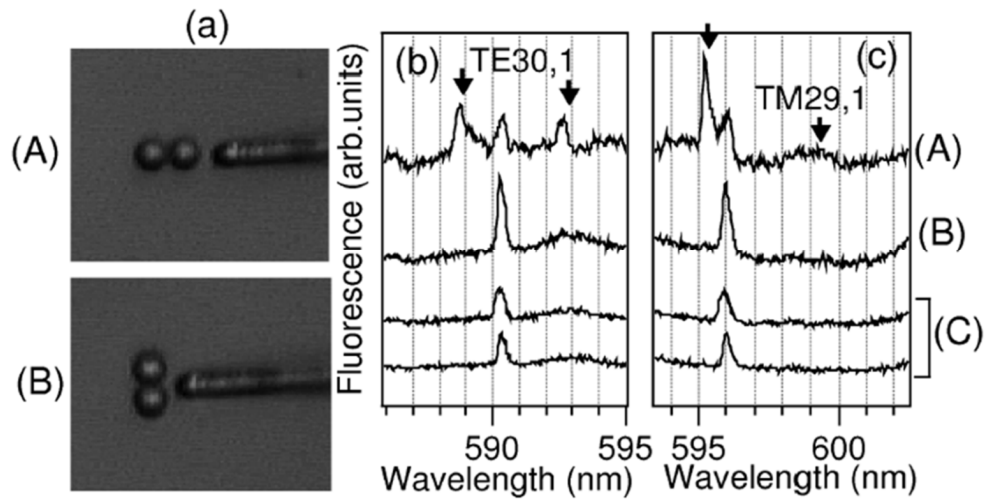

Figure 1.13: (a) Microscope images of the bi-sphere and the fiber probe. The fluorescent emission was measured for two spheres with the same diameter placed in contact in parallel configuration (A) and perpendicular configuration (B). (b), (c) Spectra of bi-sphere resonances in both configurations along with that of individual spheres before contact (C). The arrows indicate the coupled modes. [99]

The microscope images of experimental set-up are shown in Fig. 1.13(a). The fluorescence spectra of bi-sphere were measured in both parallel configuration (as seen in Fig. 1.13(A), the fiber probe is set parallel to the axis of the bi-sphere) and perpendicular configuration (as shown in Fig. 1.13(B), the fiber probe is set perpendicular to the axis of the bi-sphere). Figs. 1.13(b) and (c) show the fluorescence spectra of both configurations of bi-spheres (A, B) along with that of individual spheres (C) near $TE_1^{30}$ and $TM_1^{29}$ WGMs resonances, respectively. For both polarizations, new resonance peaks can be observed in the parallel configuration but not in the perpendicular one [99]. The appearance of new modes on both sides of the original WGM is due to the inter-sphere coupling. Thus the two new modes are referred to as coupled modes among bi-sphere, with the short-wavelength mode called antibonding mode and long-wavelength one



called bonding mode.

The dependence of the mode coupling on the orientation of the orbital plane breaks the $(2l+1)$-fold degeneracy of azimuthal modes. Coupled modes in bi-sphere originating from the coupling of WGMs with different azimuthal numbers contribute differently to the fluorescent spectrum, which lead to the asymmetric line shape of the coupled modes on Figs. 1.13(b) and 1.13(c).

To study the coupled modes from coupling of different azimuthal modes, another experiment was carried out by Yang and Astratov [76] with two size-matched dye-doped polystyrene microspheres of 5 μm nominal diameters assembled in touching position on a substrate. The spatially resolved spectroscopy was performed by an imaging spectrometer with the detection geometry sketched in Fig. 1.14(f). In this geometry not only the coupled modes at the equatorial plane but also all WGMs present in the bi-spheres are captured by the spectrometer. As seen in the spectral image in Fig. 1.14(c) the coupled modes display unusual kite-like shapes. The formation of such shapes can be explained by the coupling of multiple pairs of azimuthal modes with the same $m$ numbers [76]. The split components that originated from different pairs of azimuthal modes have different intensity maxima positions on spheres shown by lines $A$, $B$, and $C$ in Fig. 1.14. The splitting along central line $A$ represents the strength of coupling between fundamental WGMs located in the equatorial plane $m=l$ where the spatial overlap of modes is maximal, as shown in Fig. 1.14(c). On the other hand, the lower-order azimuthal modes $m<l$ have intensity maxima at points $B$ and $C$, some distance away from the equatorial plane. Due to the smaller overlap their coupled components have smaller splitting as indicated by the intersection of lines $B$ and $C$ with the kite shape in Fig. 1.14(e). Fig. 1.14(d) presents a



comparison of the three mode splitting spectra.

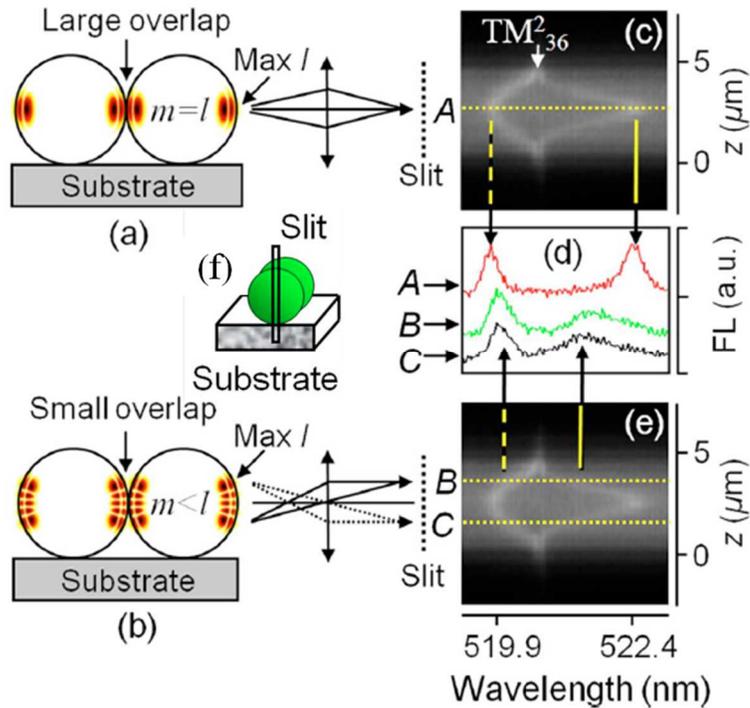

Figure 1.14: Schematics of intensity distributions for (a) fundamental and (b) lower-order azimuthal WGMs. Locations of intensity maxima are indicated by dotted line *A, B* and *C*. (d) Comparison of mode splitting spectra. (f) Geometry of spectroscopy detection. [76]

It is obvious that the coupling strength is dependent on the electric field overlap of the two coupled modes. In order to describe the properties of the inter-sphere coupling, it is desirable to introduce a mode overlap parameter. The parameter will be determined by the convolution of the modes' electric fields. One can expect strong couplings for the WGMs with the matching wavelengths whose orbital lay nearly in the same plane and are close to the contact point. Identical spheres with the same size and refractive index are ideal examples for the study of strong coupling, since they have exact same WGMs.



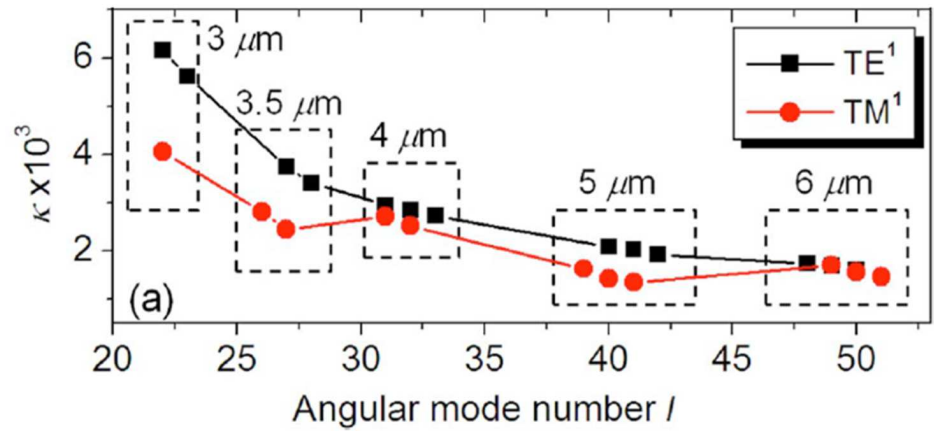

Figure 1.15: Coupling coefficient $\kappa$ for WGMs in size-matched bi-sphere pairs with nominal diameters of 3, 3.5, 4, 5 and 6 μm respectively in contact position. [76]

The inter-sphere WGMs coupling coefficients have been measured for fundamental modes ($q$=1, |$m$|=$l$) of size-matched fluorescent polystyrene bi-sphere pairs with nominal diameters of 3, 3.5, 4, 5 and 6 μm respectively, and the results for the spectral range of 510-560 nm are presented in Fig. 1.15 [76]. The coupling parameters were determined as half of the normal mode splitting normalized by the wavelength of a given WGM peak. It can be seen that larger sphere has higher mode number $l$ for the same spectral range, since its larger circumference can accommodate larger number of wavelength. For both polarizations the coupling coefficient $\kappa$ decreases with the increasing sphere diameter. It can be explained that for smaller spheres more electric fields would extend outside due to larger bending curvature. Thus larger electric fields overlap between two spheres leads to stronger coupling. In the experiment two spheres were placed in contact so the maximum available coupling coefficient was obtained for a given size. In all cases the obtained mode splitting amounts are larger than the resonance linewidths, indicating strong



coherent resonant couplings.

Identical spheres have strong inter-sphere coupling. However, identical spheres with exact same WGMs are difficult to obtain with current technology. Microspheres or other microresonators have inevitable size variations in fabrication [9, 100]. Thus the study of coupling between size-mismatched spheres without exact resonant conditions has been carried out. The coupling is not expected to be as efficient because both spatial and spectral overlap between WGMs of adjacent spheres would diminish. Since the modes of neighbor spheres have the maximal overlap when they are in contact position, the coupling strength is likely to be reduced by increasing the inter-sphere gap.

The coupling between two size-mismatched microspheres with various inter-sphere gap was simulated using finite-difference time-domain (FDTD) method modeling by Kanaev *et al*. [101]. The focus of this study was on coupling effects in a bi-sphere system formed by spheres with index 1.59 but of different diameters as 3 and 2.4 μm. Due to the fact that each microsphere possesses its own comb of WGM wavelengths, it is possible to study coupling phenomena for a number of situations with different resonant wavelength detuning (Δ) between the WGM wavelength of source sphere (S, 3 μm) and that of receiver sphere (R, 2.4 μm). Because coupling happens in the vicinity of closest eigenstates in the R and S spheres, three cases can be examined at different wavelengths, corresponding to Δ = -0.5, 5 and -7.5 nm [101].

The analysis began with the two spheres in touching position *d*=0 where the regime of strong coupling is evident in the near resonant case of Δ=-0.5 nm, demonstrating large normal mode splitting (NMS) of 5 nm. The strong coupling was defined as a regime where NMS exceeds the linewidths of the individual WGM resonances. In contrast in the



weak coupling regime, the positions of the resonances were not perturbed by the interaction between the two spheres. Increasing the separation between the two spheres leads to a gradual transition to weak coupling. In the nearly resonant case of $\Delta$=-0.5 nm the splitting between the coupled resonances is reduced to 0.5 nm for $d$=0.3 μm gap, as shown in Fig. 1.16(a).

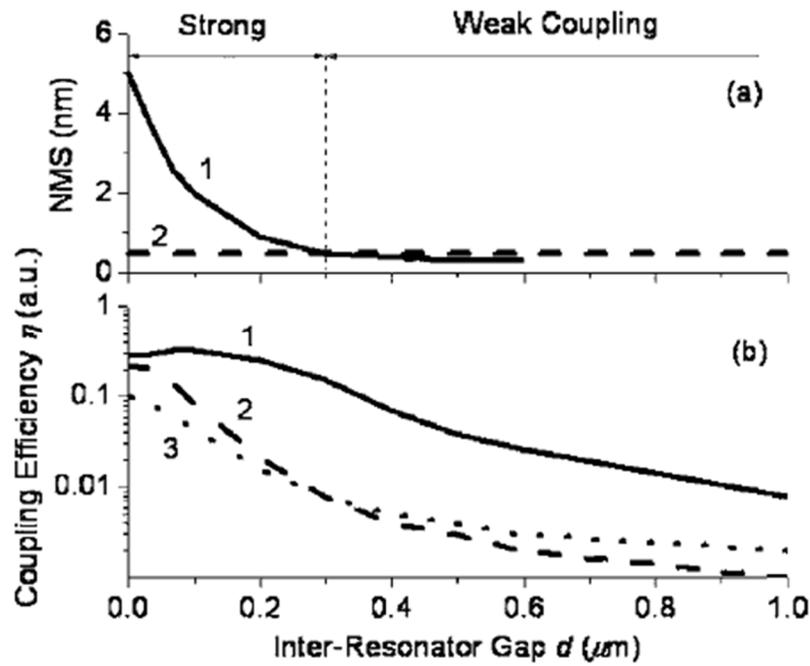

Figure 1.16: (a) NMS in nearly resonant case $\Delta$=-0.5 nm as a function of the inter-sphere gap size $d$. The dashed line indicates the linewidth (0.5 nm) of individual resonance. (b) The coupling efficiency ($\eta$) as a function of gap $d$. The solid line 1 represents the nearly resonant case $\Delta$=-0.5 nm, the dashed line 2 corresponds to $\Delta$=5 nm, and the dotted line 3 corresponds to $\Delta$=-7.5 nm. [101]

In this study, the estimate of coupling efficiency ($\eta$) was calculated by $\eta$=$E_R$/$E_S$, where $E_R$ is the energy deposited into the R sphere calculated by integrating the area under the WGM peak and $E_S$ is the total energy in the S sphere by integration with the



same spectral interval. The results of calculated coupling efficiencies are presented in Fig. 1.16(b). In the nearly resonant case ($\Delta$=-0.5 nm) the coupling efficiency shows a maximum of 0.33 at separations $d$~0.1 μm, which indicates a point of critical coupling where coupling loss matches internal resonator loss. With the gap larger than that the coupling efficiency decays monotonously. In the detuned cases ($\Delta$=5 and -7.5 nm) the coupling efficiencies are much smaller, with a fast decreasing rate at small gaps. The transitions to weak coupling occur at shorter separations and they show no approach to critical coupling point. An estimate of the energy transfer between the two cavities in a weak coupling regime using a coupled mode theory leads to even smaller efficiencies due to almost zero spectral overlap of the two neighbors' modes in detuned cases ($\Delta$=5 and -7.5 nm). Experimentally the enhancement of the coupling efficiency comes from excitation of modes with the distorted noncircular shape [101].

### 1.3.2. Coupling of WGMs in Linear Chains of Resonant Spheres

As discussed above, coherent resonant microspheres can be used as photonic atoms to build various structures referred to as photonic molecules. Bi-sphere is the simplest example of such molecules to demonstrate the resonant coupling and the coupled molecule modes. As a continuation of building photonic molecules, we can extend the bi-sphere in length to obtain one-dimensional molecules of linear chain structures.

In the visible or near infrared region, the resonance frequency of the microresonator has to be controlled with an accuracy ~ $1/Q$ for efficient coherent coupling. In the case of microdisk resonators, the size of these disks has to be controlled precisely down to a few nanometers. This is extremely challenging for current fabrication techniques. WGMs in dielectric microspheres possess high $Q$-factor and small mode volume. When



microspheres are placed close to each other, photons in one sphere can tunnel to the neighbor sphere due to the overlap of the adjacent evanescent cavity modes. Although identical spheres with exact same WGMs are also difficult to obtain for a current technology, it is possible to select a few coherent resonant spheres by comparing their fluorescence spectra and assemble them into various photonic molecules of interest.

Experiments performed by Kuwata-Gonokami *et al*. [102] used dye-doped Polystyrene spheres of 4.2 μm in diameter with refractive index of 1.59. From the resonance position of the florescence spectra, the size of the spheres can be sorted with an accuracy of $1/Q$, typically 0.05% in this experiment from a group of spheres with a size distribution of ~1%. Fig. 1.17(a) shows the typical spectra of spheres with 1% size difference. Then the select spheres were arranged in a linear chain structure on a silicon V groove substrate to ensure the straightness and reproducibility of the microsphere chain, as seen in Fig. 1.17(b). The gap shown in the inset between the spheres and the V groove can preserve a high $Q$ value for WGMs orbiting within the vertical plane. The spheres were in contact positions due to the electrostatic force.

Figure 1.17(c) presents the obtained fluorescence spectra from a single sphere and size-matched two to six-sphere chains in the vicinity of $TE_1^{29}$ resonance. The arrows indicate positions of coupled WGMs in each sphere chains. We can clearly see that the number of coupled modes increases as the number of spheres increases. The number of coupled modes tends to be equal to the number of spheres in the chain. Due to low signal to noise ratios some of the coupled modes are not visible. A peak at the resonance of uncoupled mode from single sphere was observed in all spectra due to large NA of the collecting fiber.



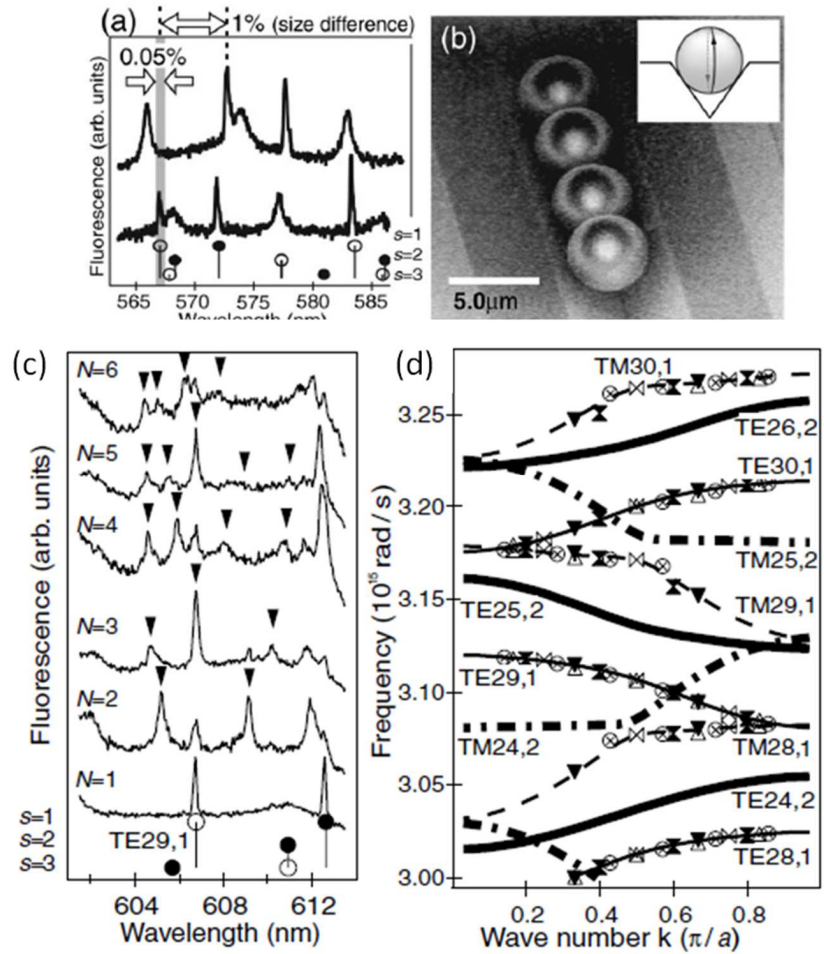

Figure 1.17: (a) Typical fluorescence spectra of spheres with 1% size difference. (b) Scanning electron microscope image of four spheres arranged linearly on a silicon V groove. (c) Fluorescence spectra from microsphere chain structures consisting of one to six spheres. (d) Energy band diagram obtained by plotting the experimental result shown in (c) and calculation results from the coupled oscillator model. [102]

The tight-binding model can be used to calculate the photonic band modes of one-dimensional chains [102]. With the tight-binding approximation, the dispersion relation for a linear chain made of an infinite number of resonators can be described as:

$$\omega(\mathrm{k}) = \omega_0 + 2\mathrm{g}\cos(\mathrm{ka}) \tag{1.7}$$



where $\omega_0$ and a are the resonant frequency and periodicity of structure (diameter of sphere), g is the inter-sphere coupling parameter, and k is the mode index corresponding to the wave vector. For a linear chain consisting of finite number (N) of spheres, tight-binding model calculations yield the eigenfrequencies as:

$$\omega_i = \omega_0 + 2g\cos(k_i a), \quad k_i = \frac{\pi i}{(N+1)a} \tag{1.8}$$

where i is an integer from 1 to N. With this relation, the photonic band dispersion can be obtained from the experimental results. Fig. 1.17(d) show the dispersion curves of calculated modes for a one-dimensional chain of $N$=30 microspheres using the tight-binding model for the first ($q$=1) and second order ($q$=2) modes of both polarizations. The interaction between TE and TM modes can be considered negligible due to their orthogonal electric fields. The inter-sphere coupling parameters used in the calculation was obtained from experiments.

Identical resonators with exact same resonant properties are difficult to obtain in fabrications. The idea of using microheater to control the resonance of individual microresonator by fine tuning its size through heating was proposed by Melloni *et al*. [103]. Microring resonators can be fabricated on silicon-on-insulator (SOI) technological platform. Fig. 1.18(a) shows scanning electron microscope image of the microring chain that was fabricated. Such structure is referred to as coupled resonator optical waveguides (CROWs) [104-107] since optical wave can propagate through the structure by resonant coupling between adjacent neighbors. A metallic heater consisting of high-resistive layer of nichrome was deposited on each ring and connected via low-resistance gold electrodes.



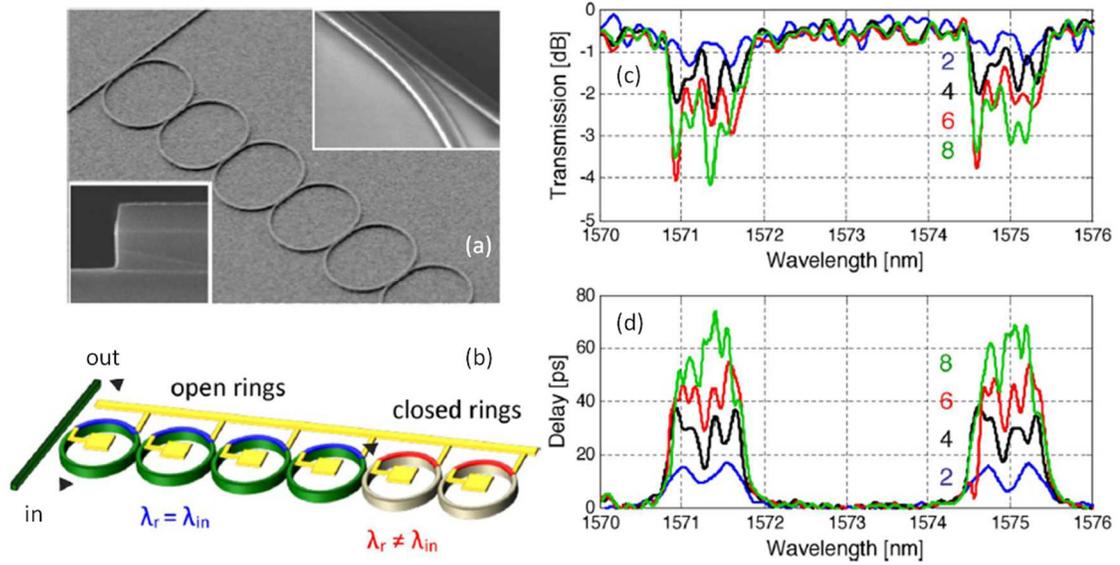

Figure 1.18: (a) Microring resonator chain without the cover layer and heaters. (b) Sketch of the tunable CROWs line. Each heater on the ring can be controlled individually. (c) Measured spectral intensity and (d) group delay for the coupled resonator chain with different number of open rings. [103]

With this set-up the realization of resonant coupling of various numbers of rings in a chain and the study of light delay in such structures are achievable. By controlling the WGMs of individual ring through heating, the resonant wavelength can be aligned and the coupling efficiency can be optimized. The number of open rings coupled in the system can be controlled as well as sketched in Fig. 1.18(b). The incoming signal at wavelength $\lambda_{in}$ couples to the CROW and propagates inside the open rings resonating at $\lambda_r$=$\lambda_{in}$, until it is reflected by the first closed (off-resonance) ring. When reflected back to the first ring, the signal couples to the waveguide and output with an accumulated delay. Fig. 1.18(c) shows the measured transmission spectra of the silicon CROW when 2, 4, 6 and 8 rings were tuned on-resonance. When 2 or 4 rings were coupled in the system the number of coupled modes is equal to the number of resonators. However, with further



increase of the number of ring resonators the mode splitting was saturated at four. This is due to the fact that fine tuning individual temperature to align resonances of more than 4 spheres is difficult in practice. Fig. 1.18(d) shows the corresponding group delay. The delay increases proportional to the number of coupled rings due to the increase of optical path length in propagation. So such coupled structure can be used as tunable delay lines. The group delay also shows the spectral characteristics of the coupled modes. Overall the tunable CROW with multiple heaters is bulky and not convenient to use. The structure can be miniature, however, the electronics controlling the heaters would require a board filled with the equipment for each photonics chip. In addition, the alignment of resonant modes of a large number of rings to realize efficient coherent coupling remains difficult in practice.

### 1.3.3. Coupling of WGMs in Arrays of Microresonators

As discussed above identical microresonators having the same resonant properties are ideal photonic atoms that can be used as building blocks for coupled photonic molecules, which is not limited to one-dimensional (1-D) linear structures but can also be extended to various two-dimensional (2-D) photonic arrays as well as three-dimensional (3-D) photonic crystals. However, due to the fact that large quantity of identical microresonators are difficult to obtain with current techniques, the optical coupling and transport properties of 2-D and 3-D photonic molecules have yet to be well demonstrated in experiments.



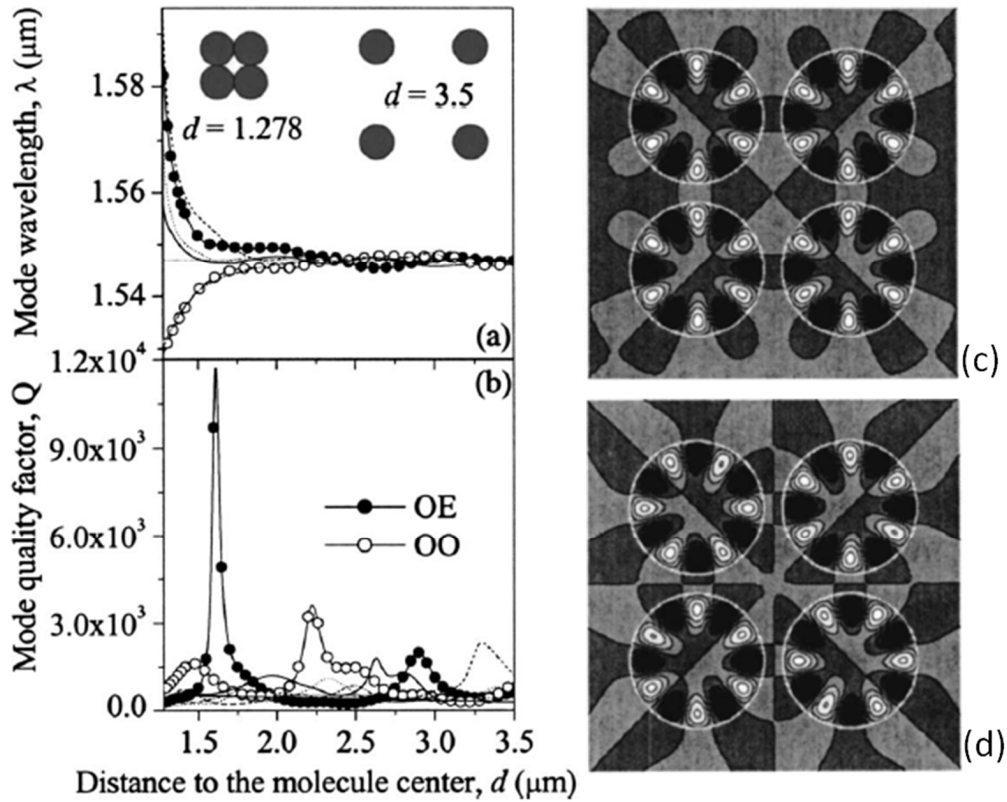

Figure 1.19: (a) Wavelength shift and mode splitting and (b) change of $Q$-factors of WGM in a 4-sphere square photonic molecule as a function of distance between the cavity center and the molecule center. Magnetic near-field pattern of (c) OE and (d) OO modes in the square molecule with $d$=1.45 µm. [110]

Several calculations and simulations have been carried out by Boriskina [108-110] and other researchers [94, 111-114] to study the coherent resonant coupling properties in size-matched 2-D photonic molecules. As an example, a symmetric 4-sphere square molecule consisting of microcavities with diameter of 1.8 µm and effective refractive index of 2.63 was considered by Boriskina [110] and presented in Fig. 1.19. A single microcavity with such parameters supports $TE_1^6$ WGM at λ=1.547 µm with $Q$=513. Figs. 1.19(a) and 1.19(b) show the wavelength shifts and $Q$-factor changes of the split molecule modes with an increasing distance between the centers of individual cavities



and the center of the square molecule. It can be seen that the degenerate WGM of single cavity split into four non-degenerate modes with even (E) and odd (O) symmetry along the square diagonals and the $x$ and $y$ axes and two double-degenerate modes. Among them OE and OO modes show much higher $Q$-factors. Their near-field mode patterns at $d$=1.45 μm are shown in Figs. 1.19(c) and 1.19(d), respectively. Especially for the non-degenerate OE mode, more than 20 times higher $Q$-factor was demonstrated with optimal distance of $d$=1.6125 μm. All other modes have significantly lower $Q$ at this distance. Potential $Q$-factor enhancement is expected to be seen in other symmetrical photonic molecule geometries such as triangle and hexagon shapes.

Experimental attempts, though not many, have also been made to study the resonant couplings among 2-D photonic molecules [100, 115]. Some qualitative observations have been made and interesting modes splitting were found. However, the results are still ambiguous and not precise due to the fact that the microresonators used as photonic atoms to build the molecules are not sufficiently monodisperse so that the size mismatch creates non-uniform coupling conditions among the molecule.

In the work presented in Fig. 1.20, GaInAsP microdisk arrays were fabricated by Baba *et al*. to form photonic molecules [100]. Each disk diameter $D$ was designed to be 3 μm. However, it had a fabrication error of ~50 nm. The inter-disk distance $d$ is small so that the evanescent coupling can occur. The fabricated microdisk arrays are shown in Figs. 1.20(a) and 1.20(b). The arrays were pumped at room temperature by 0.98 μm continuous wave laser, and the emission light was collected by a multimode fiber and analyzed by optical spectrum analyzer. Clear mode splittings were observed in all of the photonic molecules. For the simplest bi-disk molecule, the observed split of bonding and



antibonding modes agreed with that by FDTD calculation. In larger molecules the coupling between disks are more complicated and there are many degenerate modes. Generally the number of eigenfrequencies in the molecule is equal to the number of disks *N*. This relation was actually observed in the experiment for the linear chains. However, broken degeneracy caused by the slight disorder sometimes introduces subpeaks.

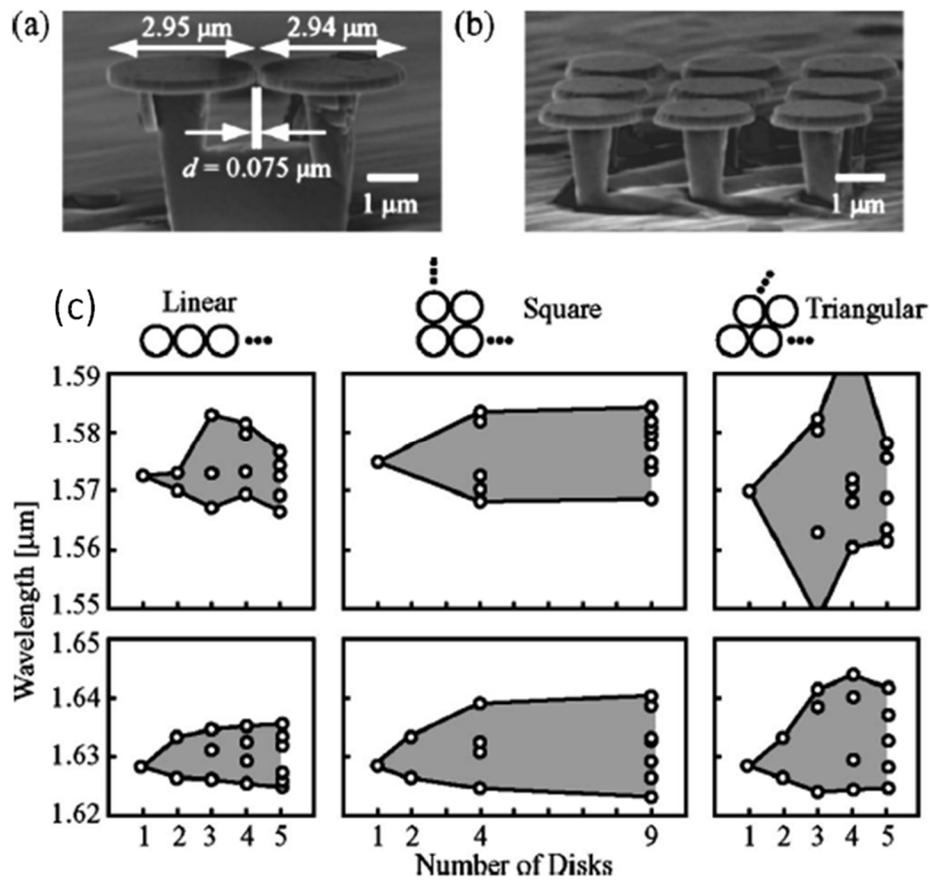

Figure 1.20: Images of photonic molecules composed of microdisk (a) bi-disks and (b) 3×3 square array. (c) Modal wavelength with number of disk *N*. Upper and lower are experimental and calculated results, respectively. Gray regions denote mini-bands. [100]



Modal wavelengths observed in the experiment and calculated by the FDTD are plotted with number of disks $N$ for various geometries, as shown in Fig. 1.20(c). It is seen that the number of coupled modes increases with $N$ and mini-bands are formed for larger photonic molecules. The experimental plots are not precisely corresponding to the calculation. However they qualitatively agree with the calculation and are expected from the analogy to the molecular chemistry. The amount of mode splitting increases rapidly in a smaller photonic molecule and saturates in a larger one. The results suggest that the coupled modes are dominated by the coupling between neighboring disks. The coupling strength is enhanced in a more close-packed structure and results in a wider split [100].

### 1.4. Radiation Pressure and Resonant Enhancement of Optical Forces

In this Section, we discuss the radiation pressure effects because they can be considered as a novel tool for sorting spheres with resonant WGM peaks [5, 6]. Radiation pressure, as the pressure exerted on any surface exposed to electromagnetic radiation, has been proved and known for over a century [116]. Such forces created by electromagnetic waves have been observed with comet motion in astronomy, now have been widely used in particles and cells manipulation in biology [117], and been applied in atoms trapping and cooling in particle physics [118]. However, only recently it was discovered that the WGMs in dielectric microspheres can significantly enhance the forces of radiation pressure usually induced by light reflections and refractions on the sphere's surfaces [119-121]. Since the WGMs and the associated force enhancement is highly sensitive to the sphere's diameter, such resonant light pressure effects can be used to develop sorting techniques which enables obtaining large quantity of coherent resonant microspheres needed for advancing microresonators based photonic devices.



### 1.4.1. Radiation Pressure

The concept of radiation pressure was first proposed by Kepler back in 1619 to explain the observation of a tail associated with a comet that always points away from the Sun [122]. The idea that light as a type of electromagnetic radiation possesses momentum so that it would exert a force upon any surface exposed to it was published by Maxwell in 1862 [122]. And then this phenomenon was first proven experimentally by Russian physicist Lebedew in 1900 [116].

By the methods of both classical wave electromagnetism and photon particle model we can deduce that if light was incident on a total absorbed object, as shown in Fig. 1.21(a), it would exert a force of $P/c$ pushing the object due to the total transfer of light momentum, where $P$ is the power of incident light and $c$ is the speed of light. If the light was incident on a total reflected surface with a normal angle, as sketched in Fig. 1.21(b), it would propagate along the reverse direction after the interaction with two times of momentum change, thus creates a force of $2P/c$ on the object. These two simple examples illustrate the classical limits for radiation pressure for light total absorption and total reflection cases.

The experiment of using optical forces to trap and propel particles of micrometer size was first demonstrated by Arthur Ashkin in 1970 [123]. Since then the applications involving optical forces have been quickly developed and has now been widely adapted in areas from physics to biology. One well-known example is the optical tweezers [124, 125], which becomes an useful tool in trapping and manipulating cells and bacteria [126] as well as atoms and neutral particles [127].



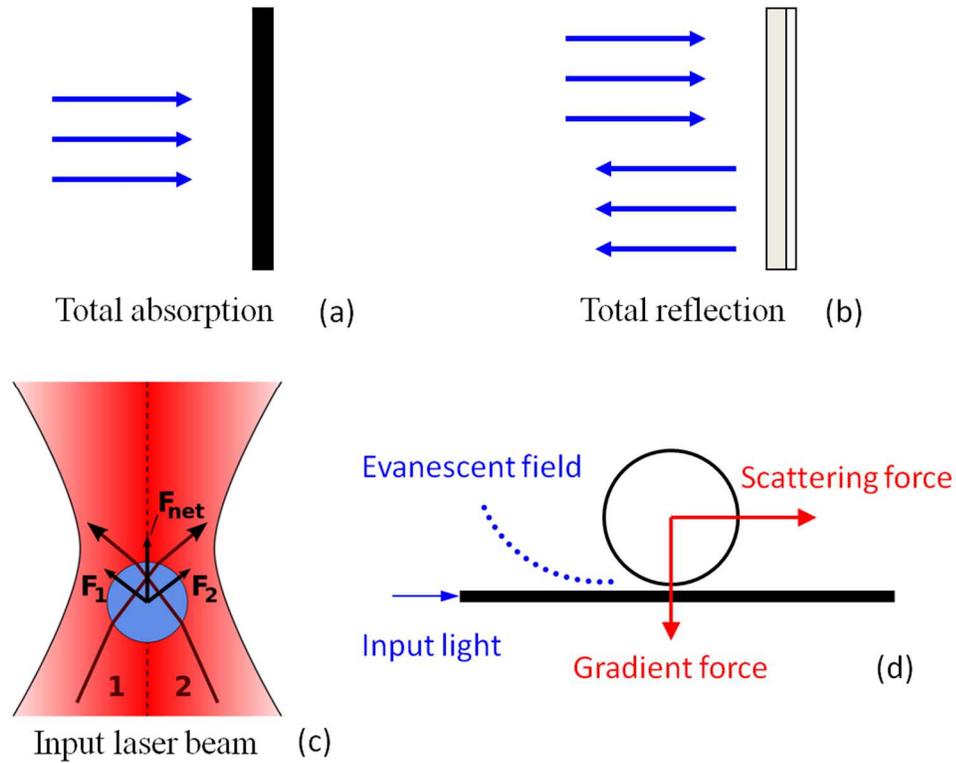

Figure 1.21: Sketches of light incident on total absorbed object (a) and total reflected surface (b). Illustration of sphere placed in focused beam (c) and near a waveguide (d).

Fig. 1.21(c) illustrates the principle of basic optical tweezers. When a spherical particle is placed near the focus of a highly focused beam, it will be trapped at the center of the focus due to refraction and reflection of beams with different magnitude. When the sphere is off the center, the net force will point to the direction of the highest light intensity and thus bring the particle back to the focus center. Such type of force is called gradient force since its direction is along the gradient of light intensity and tends to bring the particle to the place of the highest intensity. Another type of force is called scattering force which usually describes the force in the direction of light propagation. It is somewhat analogous to the force exerted on totally reflecting particle in a sense that the



force is determined by the fraction and by the directionality of the scattered light. It is the results of partial transfer of incoming light momentum. Fig. 1.21(c) shows the situation that the sphere trapped at the center of the beam waist experiences no transverse gradient force. The net optical force is the scattering force that pushes the sphere upward which can be balanced with its gravitational force and creates a stable trapping. In the case that the particle is placed in the evanescent field of an optical waveguide instead of in the free-space beam, similar effects of optical forces can be observed, as sketched in Fig. 1.21(d). The gradient force will attract the particle and trap it near the surface of the waveguide where the electric field is at maximum. There will also be a scattering force that pushes the particle in the light propagation direction. Combination of these two forces could create stable trapping and propulsion of microparticles or cells along an optical waveguide.

### 1.4.2. Optical Trapping and Propulsion of Microparticles

Optical trapping and propulsion of microparticles in the evanescent field have been demonstrated in experiments. The typical experimental set-up used to observe optical forces that evanescently exert on such particles usually build on surface slab waveguide [128-137] or subwavelength optical microfiber [138-145]. To ensure free movement of the particles the experiment is carried out in liquid medium and usually within integrated microfluidic devices. The optical forces can be used to trap and deliver biomedical cells and drugs [128, 130, 131], or trap and sort metallic or dielectric microparticles according to their different propulsion velocities [137, 141, 142].

Microparticles trapped and transported by optical forces was demonstrated by Schmidt *et al*. on a slab waveguide within a microfluidic device [133]. The device



consists of SU-8 polymer waveguides combined with polydimethylsiloxane (PDMS) microfluidic channels. The waveguide height is 560 nm and the width is 2.8 μm. The fused silica substrate has a refractive index of 1.453; the SU-8 film has an index of 1.554 at λ=975 nm; and the water as cladding has an index of 1.33 providing large refractive index contrast for light confinement and strong evanescent field gradients.

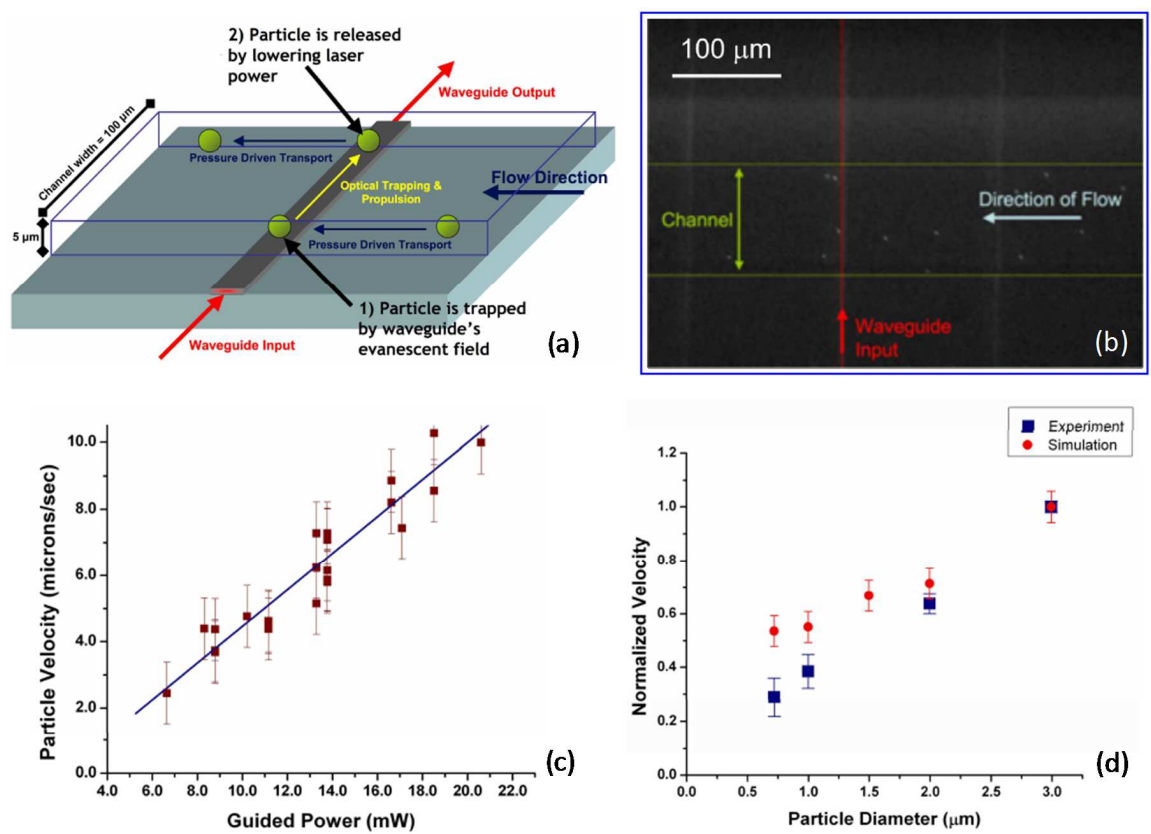

Figure 1.22: (a) Schematic of experiment set-up. The optical propulsion is perpendicular to the direction of water flow in the channel. (b) Top view from the microscope for the propulsion experiment. (c) Terminal optical propulsion velocity with output power for 3 μm diameter spheres on the same waveguide. (d) Experimentally measured and numerically calculated relative terminal velocity as a function of sphere diameter. [133]



The experimental set-up is shown as Fig. 1.22(a). Dielectric particles were driven by water flow along the microfluidic channel. When a particle came in contact with the optically excited waveguide it may be trapped in the evanescent field and start moving in the direction of light propagation. A larger portion of particles would be trapped with lower flow speeds and higher optical powers. A top view from the microscope is presented in Fig. 1.22(b). Due to the drag force on the particle in the solution, the particles would quickly reach a terminal optical propulsion velocity. Multiple particles trapping and propelling could happen. And particles could be knocked off the waveguide due to fluctuations in the fluid flow or physical irregularities in the waveguide. But in most cases they were transported to the wall of the channel and remained trapped until the trapping force was reduced by lowering the input power.

The optical guided power in the waveguide at the channel location was estimated taking into account the losses of the waveguides and the bends. The relationship between the guided power and the optical propulsion velocity was studied using a series of polystyrene spheres of 3 μm diameter. The linear dependence was found as shown in Fig. 1.22(c). Peak optical propulsion velocity of 28 μm/s was reported using guided powers of 53.5 mW. And optical trapping to pull the spheres from the water flows was found to be difficult at guided powers below 6.8 mW, when the flow fluctuations made optical trapping unstable. There also exists a dependence of the optical propulsion velocity on size of the sphere. Various measurements were taken to compare the propulsion velocities of several sizes of polystyrene spheres. The average velocities are normalized to that of the 3 μm sphere and are shown in Fig. 1.22(d). Smaller particles have slower optical propulsion velocities with the same guided power and the decrease in velocity



with size is slightly sharper than numerical calculations. However in both cases a near linear trend is observed. It should be noted that while this trend is valid over the limited range of sphere diameters studied here (0.7-3 μm), conclusions about the optical propulsion behaviors outside this region cannot be drawn.

Such phenomena of optical trapping and propulsion of microparticles in the evanescent field has been observed using subwavelength optical fiber as well by Brambilla *et al.* [139]. Subwavelength optical fibers have a considerable fraction of the propagating mode that lies outside the physical boundary of the fiber and the evanescent field extending into the medium can be exploited for optical trapping and propulsion. Particles with higher refractive index than the surrounding medium will be attracted to the high intensity region by the gradient force and propelled in the direction of light propagation by the scattering force.

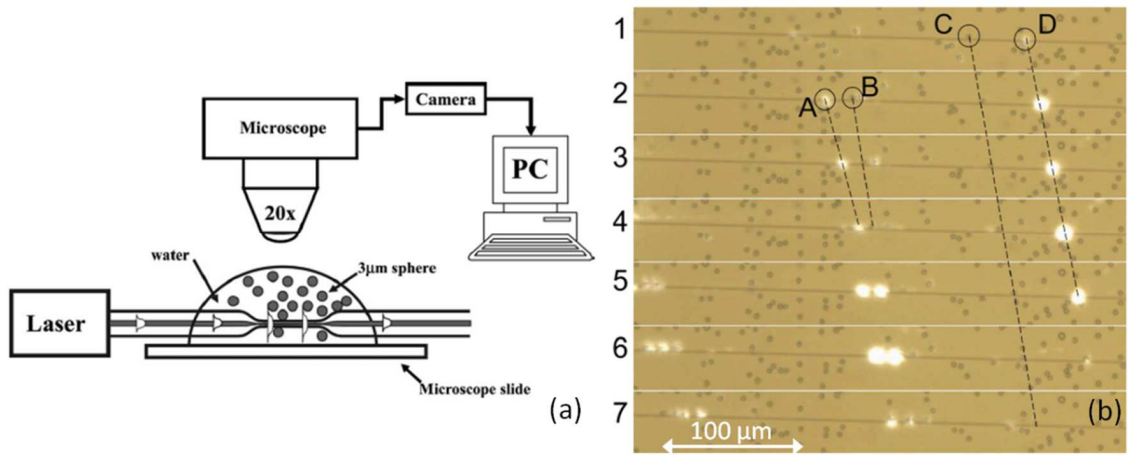

Figure 1.23: (a) Schematic of the experimental set-up: a CW laser launches light into a subwavelength optical fiber immersed in a water suspension of polystyrene microspheres of 3 μm diameters. (b) Consecutive photos taken at 1 s intervals for 3 μm polystyrene spheres propelled along an optical fiber with 0.95 μm diameter. [139]



The schematic of the experimental set-up is shown in Fig. 1.23(a). The subwavelength fiber was manufactured from single mode optical fiber using the flame brushing technique [146]. The transition regions at both sides are about 15 mm, and the minimum waist diameter is ~0.95 μm with a length of 6 mm. The tapered fiber was fixed on a microscope slide with one end connected to a CW laser. A drop of water suspension containing polystyrene microspheres of 3 μm diameter was laid on top of the microscope slide to surround the tapered fiber. While in the untapered region the power in the propagating mode is completely confined within the fiber, in the tapered section of submicrometer dimension a considerable fraction of the power is propagating in water. Real-time monitoring of particles motion was realized with a CCD camera mounted on the microscope and connected to a computer.

When the laser was switched on, particles near the taper were trapped and stably propelled in the direction of light propagation. Particles motions were observed in the region of the minimum waist diameter. A series of snapshots taken at 1 s time intervals were presented in Fig. 1.23(b). The particles' velocities were calculated from an average over a few seconds. The velocities of tracked particles A–D have been estimated from the snapshots to be 9.0, 7.0, 8.0, and 9.4 μm/s, respectively, with the taper input power ~400 mW. However, even though the taper loss in air and in pure water was estimated to be less than 1 dB, the output power of the taper immersed in the microsphere suspension was found to be below the detection limit. Therefore the optical guided power at the tapered region that is responsible for the optical trapping and propulsion of microspheres was difficult to estimate.



### 1.4.3. WGMs Resonant Enhancement of Optical Forces

Arthur Ashkin and Joseph Dziedzic observed resonant enhancement of the optical forces exerted on microdroplets by an optical beam in 1977 [1], which was the first known microcavity optomechanics effect. However, for a long time there was few studies been reported for this effect because the resonant force peaks were usually been weak and difficult to observe with free-space beam illumination [5]. In recent decade, evanescent field couplers such as prism [46, 80], surface waveguide [34, 81-83], and tapered fiber [28, 30, 45, 49, 84-86] have been developed and effective excitation of WGMs in microspheres by evanescent side-coupling has been demonstrated in experiments. Furthermore, stronger peaks of optical forces have been theoretically predicted for dielectric spheres positioned in evanescent field [119-121, 147-149]. Since the force peaks are associated with WGMs in microspheres, the forces have a significant dependence on the sphere sizes and thus are very sensitive to the size variation. Therefore this effect provides a possibility of optical sorting of microspheres with their WGMs positions overlapped at the wavelength of the laser source.

The resonances in optical propulsion force when a spherical particle placed in the evanescent field of an optical waveguide was predicted with Mie theory calculations by Jaising and Helles [119]. It illustrated strong resonant peaks of propulsion velocity with large peak-to-background ratio. The resonances in the transverse force was also shown resulting in potential optical repulsive force instead of trapping force under certain resonant conditions [119]. However, the authors suggested that at the maximal velocities the transverse force is attractive, thus the particle could be trapped near the waveguide while being propelled at high speed. Furthermore, size-selective optical forces for



microspheres using evanescent wave excitation of whispering gallery modes was demonstrated in calculations by Ng and Chan [120] and was later presented with the prism coupler by Xiao *et al* [121]. It was suggested that such resonantly enhanced optical propulsion force with a large peak-to-background ratio could be utilized for sorting resonant microspheres with an accuracy of $\sim 1/Q$.

As an example, a model of microsphere illuminated by an evanescent wave was considered in Ref. [120]. The geometry is depicted in Fig. 1.24(a). The evanescent wave is assumed to be decaying exponentially from the interface. The edge of the sphere is assumed to be $\lambda/2$ away from the interface. The light intensity at the interface is assumed to be $I_0 = 10^4$ W/cm$^2$.

The optical forces were calculated by the multiple scattering and Maxwell stress tensor formalism [150]. The parameters are incident wavelength $\lambda=520$ nm, dielectric constant $\varepsilon_{sphere}=2.5281$ for polystyrene, radius $r_s=2.317$-$2.375$ μm (size parameter $kr_s=28$-$28.7$). The radiation pressure spectrum as a function of the size parameter for a sphere in air under plane wave illumination is plotted in Fig. 1.24(b). As a comparison, the optical force in the wave propagation direction induced by TE polarized evanescent wave is shown in Fig. 1.24(c). In order to mimic the sphere's intrinsic loss and the coupling loss due to existence of the waveguide, a small imaginary part is added to the dielectric constant $\varepsilon_{sphere} = 2.5281 + 10^{-5}i$ for the evanescent wave incident case. The strongest peak in Fig. 1.24(c) is for the $\text{TE}_1^{39}$ mode with a force of $\sim 6$ pN. For comparison the weight of a 5 μm diameter polystyrene sphere is $\sim 0.7$ pN. Therefore this resonant optical force can cause significant acceleration to the particle.



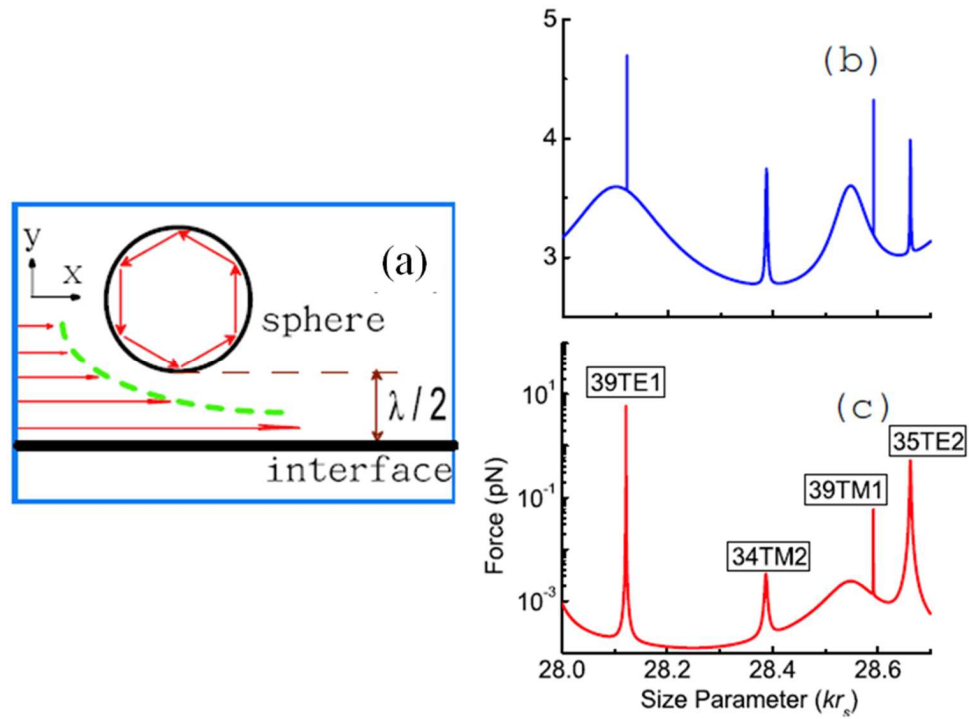

Figure 1.24: The optical forces acting on a microsphere along the propagating direction of the incident wave. (a) Schematic of the geometry. (b) The incident wave is linearly polarized homogeneous plane wave with a uniform intensity of $I_0 = 10^4$ W/cm$^2$. (c) The incident wave is TE polarized evanescent wave with $I_0 = 10^4$ W/cm$^2$ at the interface. [120]

Comparing Figs. 1.24(b) and 1.24(c), it can be seen that the contrast between the optical forces at on- and off-resonance is significantly higher for the evanescent wave. For the case of a plane wave illumination the peak-to-bottom ratio for the optical force is less than 2. On the contrary, for the case of an evanescent wave incident, the optical force at on-resonance can be enhanced by several orders of magnitude than that at off-resonance. The off-resonance optical force in the evanescent wave case is negligibly small. Since the evanescent wave decays exponentially, the average amount of incident field that interacts with the microsphere is small. However, when the WGM resonance is



excited, the evanescent coupling can be very efficient leading to a significant enhancement of optical forces by several orders of magnitudes. Furthermore, the linewidth of the WGM is usually very narrow proportional to $1/Q$, which provides the possibility of highly accurate size-selective microspheres propulsion and sorting. In a collection of microspheres, an evanescent wave can exert significant propulsion forces on those microspheres whose sizes are in resonance with the incident light while leaving those microspheres that are not at resonance basically untouched. With this method, one can sort the microspheres according to their size or resonant frequency with an accuracy of $1/Q$. With a typical $Q$-factor of $10^3$-$10^6$ for microspheres, the selected spheres will have much higher uniformity of resonances compared to that obtained by conventional fabrication techniques.

## 1.5. Summary

The literature review presented in Section 1.2 can be considered as a general introduction to physics and applications of microresonators. In a narrower sense, it can be considered as a review of the previous work related to Chapter 2 of this dissertation. From the review in Section 1.2, we know that dielectric microspheres are excellent 3-D microresonators that possess higher $Q$-factor WGMs resonances in a small mode volume. High quality microspheres are commercially available and they are easy to manipulate using optical tweezers [124, 125], self-assembly [151, 152] or a fiber probe under microscope equipped with micromanipulator [9]. Evanescent couplers such as prism, slab waveguide and tapered fiber can effectively excite high-$Q$ WGMs in microspheres. Among them tapered microfiber is the most commonly used one, due to the highest $Q$-factor it can potentially load and the readiness to be connected to fiber integrated



devices for experiments of light coupling and transmission measurement. Efficient evanescent microfiber-to-microsphere coupling has been demonstrated and was used in sensing applications [15]. In previous studies, the taper is usually fabricated by drawing a piece of fiber in flame [15, 49, 139]. The drawn fibers have the length of the tapered region on the scale of several centimeters. Taking into account that the tapered region has a diameter of a few microns, such tapers are fragile and thus difficult to be transported without broken. Therefore, there is a need to develop more compact tapers which ideally should be incorporated in a robust platform that can be used as a transportable tool for sphere characterization, sensor applications and advanced physical studies. Such platform should keep the microfiber fixed and protected, and also allow studies of different media surrounding microspheres on WGM resonant properties. Moreover, the microspheres investigated in these studied were mostly limited to those made from conventional materials such as silica glass and polystyrene, which have small refractive index contrast in a liquid. Although the WGM sensors have been developed, the effects of the surrounding medium on WGM resonances have not been well studied. The dependences of the spheres' $Q$-factors on the size, the refractive index, the medium, and the mode number have more to be explored.

In Chapter 2, we characterized WGMs in microspheres made from borosilicate and soda lime glasses, and especially from high index material barium titanate glass (BTG). The high refractive index of 1.9 and 2.1 is preferable for making high-$Q$ compact photonic devices. The tapered microfiber was fabricated with chemical wet etching with a length of several millimeters. A platform integrated with the tapered fiber was designed and fabricated to enable the characterization experiments with the same coupler both in



air and in liquid environments. Spheres of different sizes were studied for the evanescent side-coupling to the tapered fiber coupler in contact position. Coupling parameters were analyzed and WGMs quantum mode numbers were determined by fitting the analytical formula. The dependences of coupling parameters on the diameter of the taper were investigated and the coupling regime was determined. Near critical coupling condition was achieved. Transmission spectra were compared with air and water as the media. Substantially lower $Q$-factors were measured in water environment due to less sphere-medium refractive index contrast. However, it was also demonstrated that the water environment can facilitate the coupling to WGMs due to larger extent of the evanescent field and "healing" of the surface roughness. An exponential increase of the $Q$-factor with the sphere diameter was found for relatively small spheres ($D$<20 μm) in both air and water environments. The BTG spheres, even in water, demonstrate $Q{\sim}3{\times}10^3$ for 5 μm spheres, which increases to $Q{\sim}1.5{\times}10^4$ for 14 μm spheres. Therefore these compact spheres possess sufficiently higher $Q$-factors compared to conventional silica or polystyrene spheres for developing WGMs based devices.

The review presented in Section 1.3 can be considered as an introduction into physics and applications of coupled microresonators. In a narrower sense, it can be considered as a review of the previous work on coupled cavities related to Chapter 3 of this dissertation. As discussed in Section 1.3, coherent resonant coupling between two identical microspheres has been demonstrated in experiment [76, 99]. The individual WGM mode splitting into bonding and antibonding coupled modes were observed [76, 99]. And the concept of photonic molecule was brought out that the coupled photonic structures can be built using identical microresonators as photonic atoms [94, 95].



Individual microresonators such as microdisks, microtoroid, and microsphere are readily available, however, resonators with identical resonances are very difficult to obtain in fabrication since the high-$Q$ WGMs are very sensitive to the size variation. As we have seen in Section 1.3, size mismatched microspheres with detuned resonant wavelengths have substantially lower coupling strength [101]. Tuning of individual resonances in a coupled structure by controlling the microheaters attached to each microresonators was found to be bulky and difficult, and thus not practical in real applications [103]. Currently, a few microspheres with identical resonances can be selected via spectroscopic comparison [6, 102]. Coupled modes among linear chains built by up to 6 coherent resonant spheres have been demonstrated. And tight-binding model has been applied to explain the coupling in chains and predict the coupled modes positions [102]. However, in that work the WGMs were excited by dye fluorescence and all radiation was collected by the fiber probe. Thus the coupled modes were found weakly pronounced with strong presence of uncoupled modes and high noise. Also, only 1-D linear coupled structures have been studied in experiment. Attempt to construct 2-D photonic molecules was made with fabrication of GaInAsP microdisk arrays [100]. However, the fabrication error was too large to find a good agreement with the simulation results. Generally, the state-of-the art in this field reached a saturation level determined by the fabrication of coherent resonant coupled microcavities. Some functionality of the coupled cavity structures have been demonstrated. However, it has only been achieved using a few manually sorted or externally tuned building blocks. A need for a different technological approach of large-scale sorting photonic atoms with almost identical resonant properties became apparent.



In Chapter 3, we studied various 1-D and 2-D photonic molecules built by identical microspheres with overlap WGM resonances both theoretically and experimentally. Finite-difference time-domain (FDTD) method was used for numerical simulation of optical coupling and transport properties among photonic molecules. We have found commonly seen mode splitting patterns for each molecule configuration, and presented them as spectral signatures associated to each coupled photonic molecules. The spectral signature is represented by the number of split supermodes and their spectral positions which was found to be a unique and relatively stable property linked to each photonic molecule. The electric field maps of coupled molecules were also presented that showed interesting light coupling and transport phenomena. Experimentally, we selected size-matched polystyrene microspheres with mean diameter of 25 μm by spectroscopic characterization and comparison, which were assembled into various 1-D and 2-D photonic molecules in aqueous environment. Taper-to-molecule side-coupling experiments were performed and the obtained supermodes spectra were found to be in an excellent agreement with the spectral signatures predicted by the simulation. These experiments with simple 1-D and 2-D molecules provide a demonstration of potential complicated photonic devices that can be built by microspheres with identical resonant properties. However, such sphere selection method is time consuming and has low throughputs. Obtaining a large quantity of coherent resonant microspheres as building blocks of coupled cavity strucutures remains challenging.

The review presented in Section 1.4 can be considered as an introduction into physics of radiation pressure effects. It is related to the subject of this dissertation work because in Chapters 4 and 5 we studied these effects with novel approach of using



WGMs resonant spheres, and proposed sorting microspheres with uniquely uniform resonant properties by utilizing the resonantly enhanced optical forces. In Section 1.4 we discussed radiation pressure in the free-space as well as in an evanescent field. Microparticles such as dielectric microspheres placed in the evanescent field of a surface waveguide or a tapered fiber can be trapped near its surface and propelled along the light propagation direction [128-145, 153-156]. We already know from Section 1.2 that high-$Q$ WGMs in microspheres can be effectively excited with these evanescent couplers. And theoretical calculations predict that on-resonance optical forces can be several orders of magnitude higher than the off-resonance forces for the same size of sphere [119-121]. Use of resonant light forces opens up a unique approach of high-volume automatic sorting of coherent resonant microspheres with the uniformity proportional to $1/Q$ [120, 121]. Considering the typical $Q$ of $10^3$-$10^6$ for microspheres, such uniformity is much higher compared to that in coupled-cavity structures obtained by the conventional fabrication technologies [100]. However, the fundamental limits for the peak amplitudes of the resonantly enhanced propulsion force have not been studied, and the resonant transverse force has not been properly discussed. Moreover, such effects of resonant enhancement of optical forces have yet to be demonstrated in experiment. Small polystyrene spheres with diameters less than 5 μm used in most optical propulsion experiments can hardly confine light and sustain WGMs in a water medium. Thus the propulsion was only induced by the non-resonant optical scattering force. Also, there has been no discussion of practical designs of sphere sorting devices based on the resonant optical forces.



In Chapter 4, we characterized the WGMs in polystyrene spheres with a range of diameters from 3 to 20 μm, and found that only large spheres with diameters of 15-20 μm can support WGMs with $Q\sim10^3$ while resonance was barely seen for spheres less than 5 μm in aqueous environment [5, 157]. Optical trapping and propulsion phenomena were observed with a tapered fiber integrated microfluidic platform for all studied sizes of polystyrene spheres. The dependence of measured propulsion velocities on the sphere diameter was analyzed statistically [5]. Uniform Gaussian distributions of velocities were obtained for small spheres. However, some extraordinary high velocities were recorded and large scattering of velocity data were seen for large spheres ($D$=15-20 μm). These experimental observations serve as indirect evidence of resonant enhancement of optical forces due to pronounced WGMs in large polystyrene microspheres.

In Chapter 5, we proved our hypothesis of resonant enhancement of optical forces directly in experiments by individual manipulation of microspheres and spectroscopic control of the relative position of laser wavelength and WGM resonances. We integrated an optical tweezers into abovementioned platform to enable the precise trap and movement of each individual spheres [6]. In contrast to the work presented in Chapter 4, where the detuning between the fixed laser wavelength and WGMs in spheres was realized randomly by the 1-2% size deviation of the microspheres, here we can characterize the WGMs in a specific sphere while keeping it at the vicinity of tapered fiber and tune the emission wavelength of tunable laser to any desired positions [6]. Therefore we realized a precise control of the wavelength detuning. We investigated the spectral properties of optical forces in the optical propulsion experiments with polystyrene spheres of 10 and 20 μm mean diameters. The distribution of propulsion



velocities was found inversely correlate to the fiber transmission spectrum. A comparison of optical propulsion forces between strongly resonant 20 μm spheres and weakly resonant 10 μm spheres provides a direct demonstration of the effect of resonant enhancement of propulsion force due to excitation of WGMs in microspheres. The temporal properties during sphere propulsions were also studied and stable radial trapping was determined. Analytical calculations of optical trapping force and propulsion force were performed to explain the physical mechanism and to compare with the experimental results.

CHAPTER 2: EVANESCENT COUPLING TO MICROSPHERES USING
TAPERED MICROFIBERS

## 2.1. Introduction

In recent decade there has been a great deal of attention focused on the coupling to

whispering gallery modes (WGMs) in microspheres [17, 18, 76, 158] and on developing

evanescent couplers to spheres based on tapered fibers [28, 30, 45, 49, 84, 86, 92,

159-162], surface waveguides [34, 81-83, 163], and prisms [46, 80, 164]. The interests

arise from a number of optical properties determined by the high quality ($Q$) WGMs

resonances: (a) non-linear effects in spheres made from special materials with high

nonlinearity [86, 161]; (b) nanoparticle sensing applications [30, 159, 162, 165]; and (c)

resonant radiation pressure effects [119-121]. Because of a strong size dependence of the

WGM frequencies, the resonant optical forces can be used for size sorting the

microspheres with an unprecedented small standard deviation of the sphere diameters on

the level of $1/Q$. Such size-matched microspheres can be used as building blocks for

coupled resonator optical waveguides based on the tight binding [102] of WGMs.

Many of these applications, especially for biomedical cells or nanoparticles, require

a liquid medium as surrounding environment. However, a reduction of WGMs $Q$-factors

in a liquid is expected due to the decreased refractive index contrast. For example, in

order to possess $Q\sim10^4$ in water in the near-infrared range, conventional microspheres

made from silica glass with index $n_s$=1.45 or polystyrene $n_s$=1.59 should have diameters

larger than 30 μm [5, 157]. Such spheres are too bulky for building compact photonics



devices. The general solution to this problem is offered by utilizing higher index materials, such as chalcogenide glass [161], lead silicate glass [86], and silicon [83, 166] microspheres. However, their manufacturing process is less well established and these spheres are not currently available commercially.

In this Chapter, we developed a tapered fiber-to-microsphere platform for coupling light to microspheres made from a variety of materials with refractive indices of $1.47 \leq n_s \leq 2.1$ and diameters of $3 \leq D \leq 22$ μm in both air and aqueous environment [9]. Compact microspheres that can maintain high-$Q$ WGMs in an aqueous medium are required for many applications involving liquid immersion, such as sensing nanoparticles and biomedical cells. We have identified that high index microspheres made from barium titanate glass (BTG) are commercially available [70]. We observed $Q$-factors approaching $10^4$ in compact ($D \sim 5$ μm) BTG spheres with $n_s$=1.9 and 2.1 immersed in water, thus such spheres are perfect candidates for these applications. We also studied the dependence of coupling parameters on the fiber diameter by re-attaching the spheres at different positions along the transitional region of the tapered fiber. We demonstrated near critical coupling regime with intrinsic $Q \sim 3 \times 10^4$ for water immersed 14 μm BTG spheres. The difference of evanescent coupling in air and water media was compared with the same taper coupler and the $Q$-factors were summarized for all of the measurements. The work presented in this chapter was completed in collaboration with Dr. Oleksiy Svitelskiy.

## 2.2. Fabrication of Tapered Microfiber Integrated Platform

Evanescent coupling has been proven to be an efficient way to excite WGMs in the microresonators [28, 40, 45]. Among all of the evanescent couplers tapered fiber is the most commonly used one, due to the highest $Q$-factor it can potentially load and the



readiness to be connected to fiber integrated devices for light coupling and measurement [37]. In our experiments, the coupling to microspheres was realized in a tapered fiber integrated platform [9]. Tapered fibers are usually fabricated in two methods: flame pulling [146, 167, 168] and chemical etching [85]. The pulled fibers have the length of the waist on the order of several centimeters. Taking into account that the tapered region has a diameter of a few microns, such tapers are fragile and thus difficult to be transported without broken. We wanted to develop a robust platform that can fix and protect the tapered fiber. We also wanted an enclosed volume that allows studies of different media surrounding microspheres on WGM resonant properties with the same taper. The method of chemical etching has advantages in simple requiring equipment and short taper length on the order of a few millimeters It can also be integrated into a mechanically robust platform with an enclosed bath, following the fabrication procedure developed in this work [9].

We used the chemical etching method with hydrofluoric acid (HF) [85] to fabricate the tapered fiber that was integrated into a platform [9, 169, 170]. First we stripped off the plastic coating of a central short region of a piece of telecommunication single-mode silica fiber (SMF-28e, Coning), since the plastic will not interact with HF and we want to keep most of the coating for protection. Then we fixed the fiber piece in a rectangular plexiglass frame with the stripped region positioned at the center. We overfilled a plastic vessel with HF solution with 20% concentration by volume (diluted from 49% hydrofluoric acid, J.T.Baker®). The dilution will decrease the etching rate thus yielding a better control of the resultant waist diameter. The solution will form a cone-shape meniscus above the vessel. We attached the frame to a 3-D stage and positioned it to



immerse a small fraction of the fiber without coating inside the HF solution, as shown in Fig. 2.1(a). The etching length can be adjusted by the size of the vessel, usually ranging from several millimeters to several centimeters. Experimentally, by controlling the time of holding fiber inside the solution, one could etch the fiber down to any desired diameter. We found the minimum waist diameter of ~1 μm was achievable by utilizing such technique. Since the waist is so narrow approaching the resolution limit of the microscope with an objective lens of 0.6 NA, the diameter is an estimate with an error of ~0.25 μm. The transition between the etched and unetched region occurs from both sides gradually and smoothly over ~2 mm length.

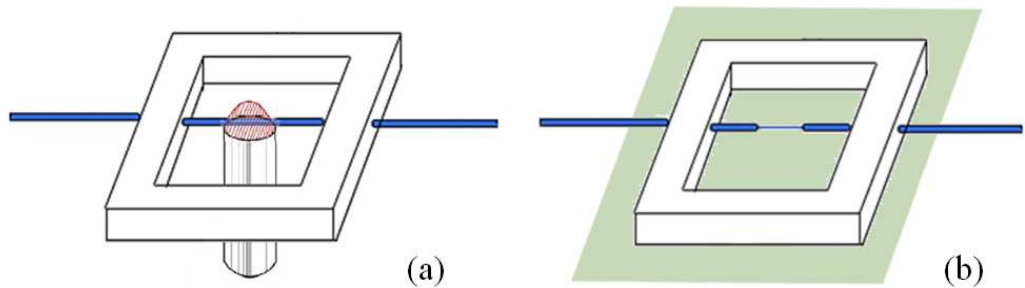

Figure 2.1: Schematics of experimental platform fabrication. (a) A piece of fiber was fixed in a plexiglass frame and was immersed in HF solution for etching. (b) The frame was sealed at the bottom with a glass plate to form a bath. [157, 170]

After completing the etching, the bottom of the frame was sealed with an optical glass plate, thus forming a bath that can be filled with liquid if needed, as sketched in Fig. 2.1(b). Therefore the platform can be used in the study of evanescent coupling both in air and in liquid environment. Both pigtails of the fiber were equipped with standard angled



physical contact (APC) optical fiber connectors, which allow easy connecting to various fiber integrated optical equipments.

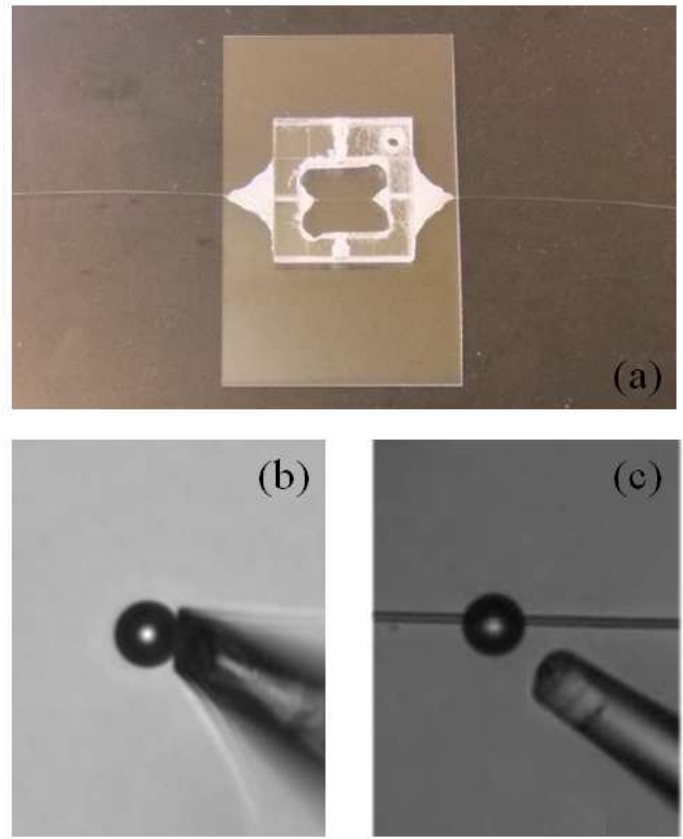

Figure 2.2: Photos of (a) tapered fiber fixed in a Plexiglas frame sealed with a glass plate at bottom, (b) a sphere of ~10 μm diameter picked up by a sharpened fiber tip, (c) sphere attached to the tapered region of the fiber with waist of ~1.5 μm. [157, 170]

An example of the fabricated platform integrated with the tapered fiber that was used in the experiments is presented in Fig. 2.2(a). The platform was placed under the microscope (IX71, Olympus). Using a sharpened optical fiber tip that was attached to a 3 dimensional hydraulic driven micromanipulator, dielectric microspheres can be picked up,



transported, and positioned at desired locations along the taped fiber due to electrostatic binding force. Fig. 2.2(b) shows the microscope picture of a sphere of ~10 µm diameter being picked up by a sharpened fiber tip. Fig. 2.2(c) shows a ~9 µm diameter sphere being attached to the tapered region (diameter ~1.5 µm) of the fiber, which is the situation when the experimental measurements were taken.

For the evanescent coupling experiments, the tapered fiber was connected to an unpolarized broadband white light source (AQ4305, Yokogawa) and an optical spectrum analyzer (AQ6370C, Yokogawa) in order to obtain the transmission spectrum in a broad spectral range. When the microsphere is placed in the near field of the tapered region or positioned in contact with the taper, WGMs in sphere will be excited and will be observed as dips in the transmission spectrum. Fig. 2.3 gives a schematic image of the experimental set-up in the study of taper-to-sphere coupling.

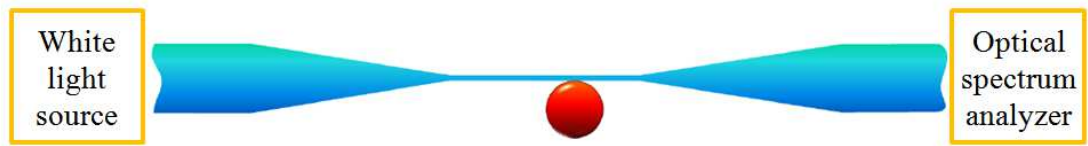

Figure 2.3: Schematic of broadband fiber transmission measurement when the sphere is in contact with the waist of tapered fiber.

### 2.3. Evanescent Coupling to Various Microspheres in Air and Water Media

Spheres made from three types of materials were used in these measurements. The spheres have different refractive index: borosilicate glass $n_s$ = 1.47, soda-lime glass $n_s$ = 1.5, polystyrene $n_s$ = 1.59, and barium titanite glass $n_s$ = 1.9. Examples of measured



transmission spectra with individual spheres attached to the tapered fiber at its waist with diameter of ~1.5 μm are shown in Fig. 2.4. Transmission spectra measure in air is presented in Fig. 2.4(a), while spectra shown in Fig. 2.4(b) were measured when the taper and sphere were immersed in water. Spectra were shifted vertically with respect to each other for clear comparison. It should be noted that the data in Figs. 2(a) and 2(b) was obtained using different spheres, however, we tried to select the spheres with similar diameters for such comparison.

It can be seen that in the air ($n_{air}$ = 1) all spheres with the refractive indices from 1.47 to 1.9 demonstrate prominent WGMs resonances, which were presented as narrow dips in the transmission spectra. In contrast, in water only the barium titanite glass sphere with the highest refractive index $n_s$ = 1.9 shows strong resonances. The $n_s$ = 1.5 soda-lime glass sphere shows weak and broad resonances, while the $n_s$ = 1.47 borosilicate glass sphere does not show any visible resonance. This observation is what we can expect, since the refractive index of water ($n_{water}$ = 1.33) is much higher than that of air leading to a significantly reduced index contrast. The less refractive index contrast between the sphere and the medium leads to less effective confinement of light. Especially if the index of medium is close to that of sphere, total internal reflection will hardly realize, and most light will leak through radiation thus the sphere can no longer maintain WGMs. Note that the resolution of optical spectrum analyzer is ~0.1 nm for our measurements. For the case of $n_{s3}$ = 1.9 sphere in air, individual resonances were not well resolved in the transmission spectrum.



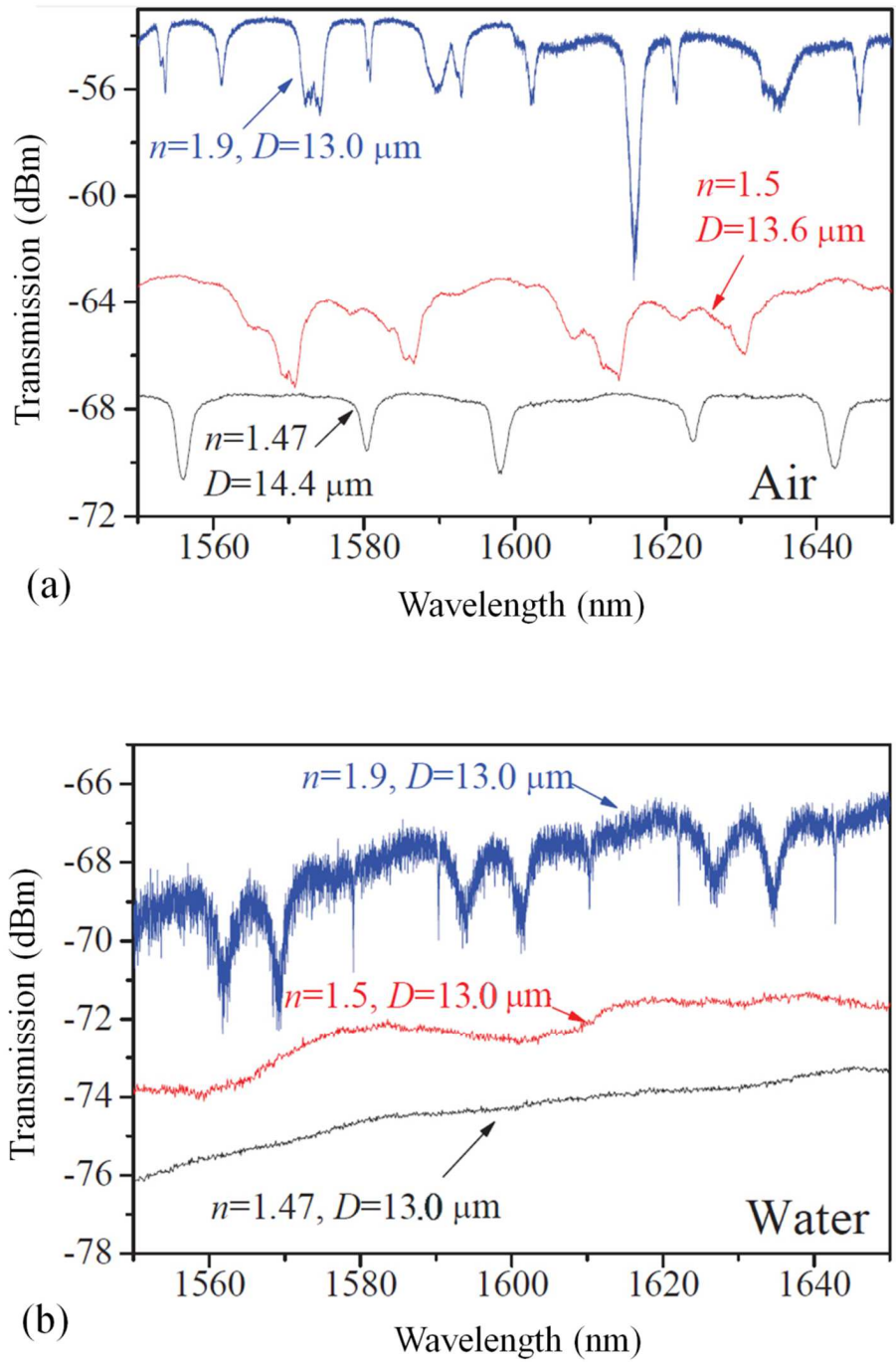

Figure 2.4: Fiber transmission spectra for coupling to spheres made from various materials with similar sizes when (a) in air, (b) immersed in water. [170]



Immersion of the structure in water generally leads to reduction of the signal transmitted through the fiber, which is evident from the smaller signal-to-noise ratio of the top spectrum ($n_s$ = 1.9) in Fig. 2.4(b) compared to that of Fig. 2.4(a). However, in the meanwhile, water medium may improve the coupling conditions between the fiber and the sphere with respect to the situation in air, due to larger extent of the evanescent field outside the fiber. The water immersion can also somewhat "heal" the fiber surface roughness due to substantially less refractive index contrast compared to the case of the same fiber surrounded with air. As seen in Fig. 2.4(b), some of resonances in the top spectrum ($n_{s3}$ = 1.9) appear as strong and narrow dips with large $Q$-factors.

The phenomenon of light scattering at the wavelengths resonant with WGMs in spheres can be observed using a tunable laser in visible regime. Different light scattering images were seen when a scan of wavelength was taken by a tunable semiconductor laser from 632 to 639 nm. The off-resonance and on-resonance (with WGMs) cases can be easily distinguished. The image of Fig. 2.5(a) shows a 14 μm diameter BTG sphere placed in contact with d~1.3 μm tapered fiber when there is no incident light. Figs 2.5(b) and 2.5(c) illustrate light scattering for off-resonance (634.8 nm) and on-resonance (635.5 nm) excitation of WGM in the sphere, respectively. It is obvious that light scattering from the sphere is much stronger when the laser emission is on-resonance and WGM is effectively excited, while the amount of scattering light is very small for off-resonance incidence.



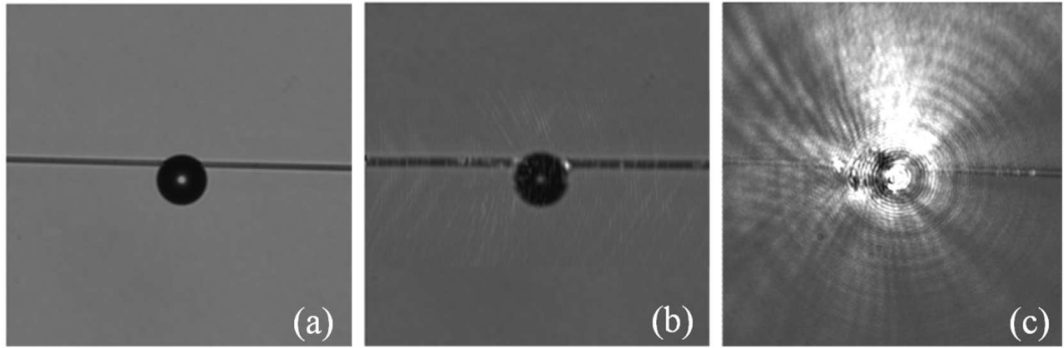

Figure 2.5: Images of a 14 μm diameter BTG sphere placed in contact with d~1.3 μm tapered fiber when (a) no incident light, (b) laser emission wavelength off-resonance with WGM in sphere, (c) laser emission on-resonance with WGM.

In the measurement of optical coupling, when a series of such resonances were obtained over a sufficiently long wavelength range, it is possible to identify all WGMs' quantum modal numbers based on the Mie scattering formalism [36, 171].

Assume size parameter $x = ka$, where $a$ is the sphere radius and $k$ is the wave number in vacuum. The position $x_{q,l}$ of WGMs with a radial mode number $q$ and an angular mode number $l$ in a sphere of refractive index $n$ can be expressed as a series in $v^{-1/3}$, where $v = l + 1/2$ [171]:

$$nx_{q,l} = v + 2^{-\frac{1}{3}}\alpha_q v^{\frac{1}{3}} - \frac{P}{(n^2-1)^{\frac{1}{2}}} + \left(\frac{3}{10}2^{-\frac{2}{3}}\right)\alpha_q{}^2 v^{-\frac{1}{3}}$$

$$-\frac{2^{-\frac{1}{3}}P(n^2-2P^2/3)}{(n^2-1)^{3/2}}\alpha_q v^{-2/3} + O(v^{-1}), \qquad (2.1)$$

where

$$P = \begin{cases} n & \text{for TE modes} & (2.2a) \\ 1/n & \text{for TM modes} & (2.2b) \end{cases}$$

and $\alpha_q$ are the roots of the Airy function Ai(-z).



One example of such mode numbers determination is presented in Fig. 2.6(a) for the transmission spectrum of 1400-1700 nm range for a $D = 14$ μm, $n = 1.9$ BTG sphere in contact with a $d = 2.5$ μm tapered fiber in water [9]. The dips in transmission are due to coupling to WGMs with orthogonal polarizations, labeled TE and TM, respectively [164]. The radial numbers ($q = 1$ or $2$) indicate the number of WGM intensity maxima along the radial direction. The angular number $l$ represents the number of modal wavelengths that fit into the circumference of the equatorial plane of the sphere. The observed dips are believed to be inhomogeneously broadened by the partial overlap of modes with different azimuthal numbers $m$, which are degenerate in a perfect sphere but will split in real physical spheres due to their slight (<1%) uncontrollable ellipticity [161]. Four subsets of WGMs (TE$^1$, TM$^1$, TE$^2$, TM$^2$) have been identified and labeled in a single spectrum, as illustrated in the diagram at the bottom of Fig. 2.6(a). The first order radial modes ($q = 1$) have narrower resonance widths and have a larger free spectral range (FSR) compared to the second order modes. As for different polarizations, TE and TM modes with the same radial mode number have similar widths as well as similar FSR.

For better demonstration of the resonance widths, an enlarged version of the region marked in red rectangle in Fig. 2.6(a) is shown in Fig. 2.6(b). It can be clearly seen that the second order modes (~1.3 nm width) are much broader than the first order ones (~0.14 nm width), which suggests that they experience a bigger loss due to the larger spatial extent of the modes. It should be noted, however, that the measured narrow linewidth of the first order modes may still be limited by the resolution of spectral analyzer (~0.1 nm).



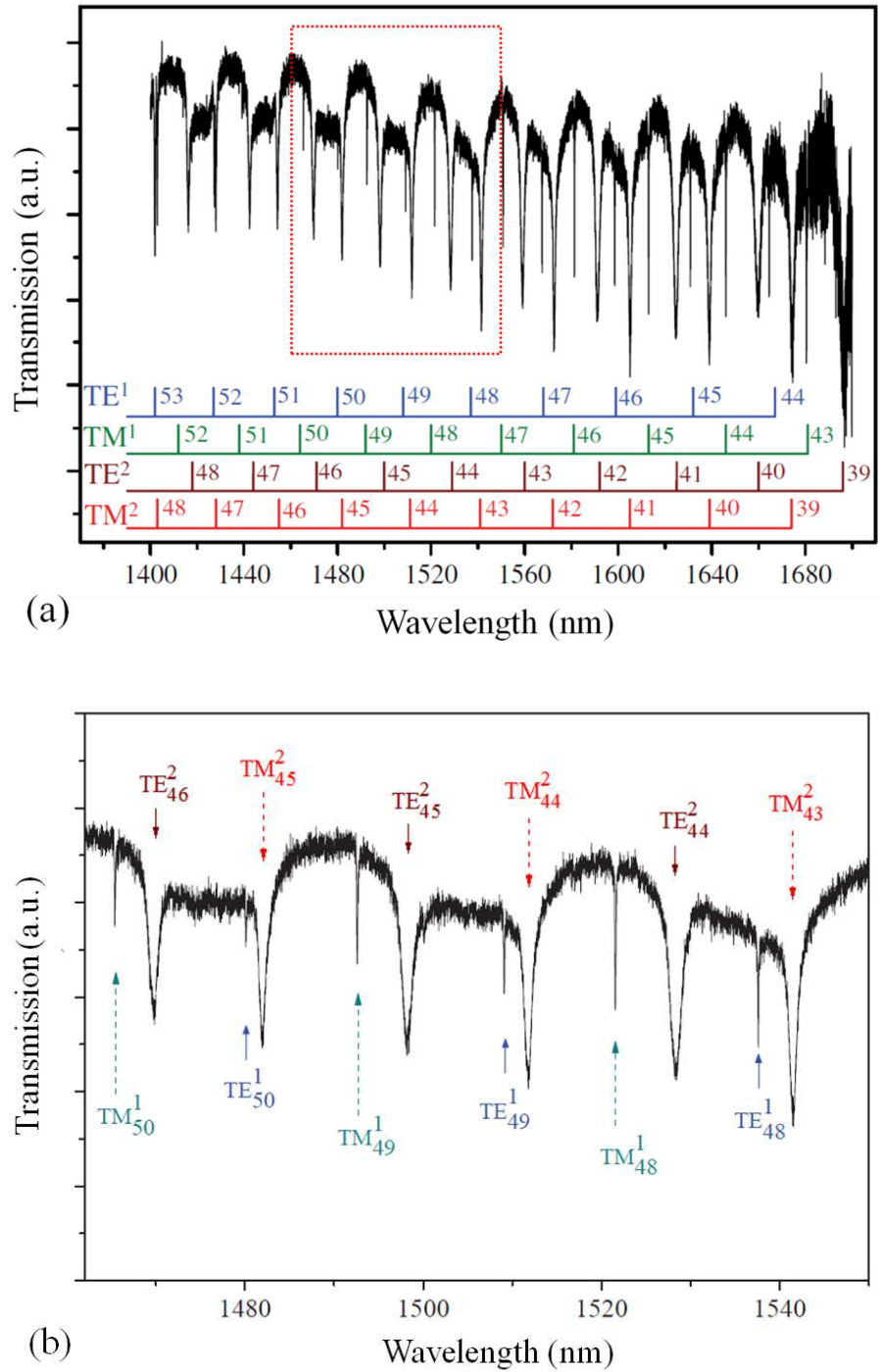

(a)

(b)

Figure 2.6: (a) Transmission of the $d = 2.5$ μm tapered fiber in contact with a BTG sphere ($D = 14$ μm, $n = 1.9$) in water. Bottom diagram shows calculated radial and angular modal numbers for four subsets of WGMs. (b) Enlarged portion of the spectrum marked in red rectangular in (a). [9]



We characterize the resonances using their $Q$-factors, defined as $Q = \lambda_0 / \Delta\lambda$, where $\lambda_0$ is WGM's resonant wavelength and $\Delta\lambda$ is the width of the resonance measured at half of the depth (in analogy to full width at half max, or FWHM). The $Q$-factors for $q = 1$ modes were found to be approximately an order of magnitude higher compared to $q = 2$ modes. In fact, the first radial order ($q = 1$) resonances are so narrow that our measured widths are likely to be limited by the resolution of the spectral analyzer.

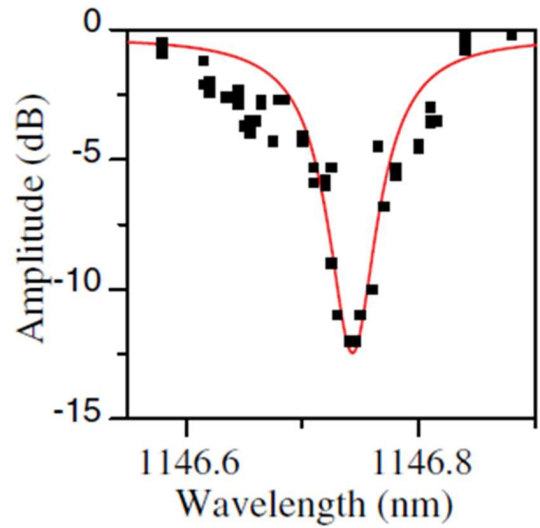

Figure 2.7: Example of a first radial order ($q$=1) WGM resonance measured with high-resolution tunable laser. The red curve is a fit with Eq. (2.3). [9]

For high-resolution measurements, we used a different technique based on using a laser with tunable emission wavelength. The detection is provided by a photo diode. By tuning the wavelength of the laser source by fine steps and measuring the transmission, a transmission spectrum can be obtained. We used the mode-hop-free intervals of a semiconductor laser tunable in the 1140-1250nm range (Toptica Photonics). The transmission at a certain wavelength was calculated by dividing the fiber transmitted



power with the sphere attached by the transmitted power through the taper without the sphere. A transmission spectrum measured with such method is shown in Fig. 2.7. A fit of the transmission data using a single-mode model [172, 173] demonstrates a first order ($q$ = 1) WGM resonance possessing an intrinsic $Q$-factor of $3{\times}10^4$, in comparison to $Q = 1{\times}10^3$ for second order mode ($q = 2$) presented in Fig. 2.6. The spectrum also shows a significant transmission dip with a depth of about 13 dB. Thus on resonance only 5% of incident optical power transmits through the tapered fiber while the most of light is coupled into the sphere, which demonstrates the realization of a near critical coupling condition.

## 2.4. Coupling Parameters and the Dependence on Taper Diameter

Understanding of the mechanisms of coupling light to WGMs in spheres usually required variation of one of the experimental parameters controlling the coupling constant. Often this is achieved by variation of the gap separating the tapered fiber from the sphere [45]. However, providing variation of the gaps with nanometric accuracy is a difficult experimental task. In our work, we found that instead of variation of the gap we can vary the thickness of the fiber with the sphere being always in a closed contact position. This study is much easier to realize experimentally due to the gradient nature of our microfibers and our ability to precisely move that sphere along the taper. In our experiment, the fiber diameter gradually decreases from the original ~125 μm to the waist ~1 μm. To characterize the coupling strength and to investigate the coupling dependence on the diameter of the tapered fiber, the same BTG sphere ($D = 11.2$ μm, $n = 2.1$) was moved along the taper from thicker to thinner parts in water and was repositioned at several places with different taper diameters [9, 169]. The measured transmission spectra



along with corresponding microscope images showing tape diameters are shown in Fig. 2.8. For clear presentation the neighboring spectra are shifted by 1 dBm with respect to each other.

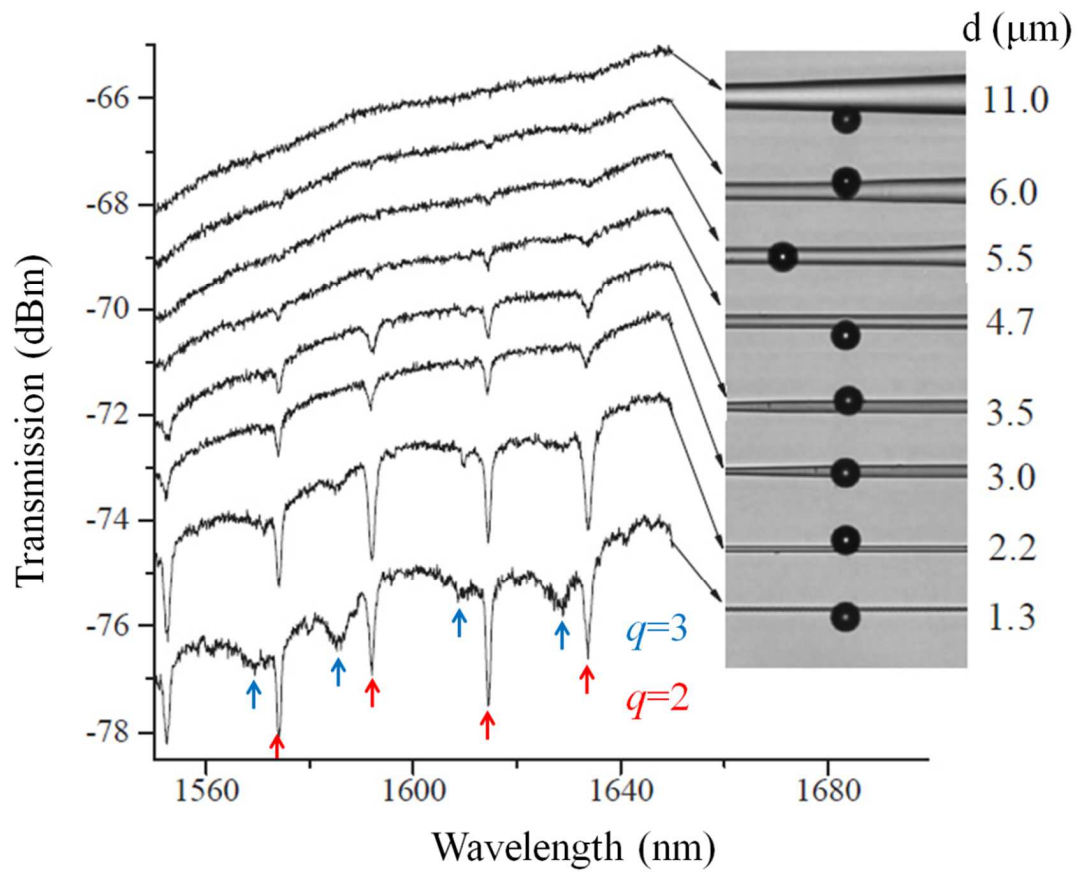

Figure 2.8: Resonances of BTG spheres ($D$ = 11.2 µm, $n$ = 2.1) attached at different positions along the taper in water, as shown by images to the right. For clarity, spectra are shifted by 1 dB with respect to each other. [9, 169]



The narrow and broad dips in the bottom spectrum ($d = 1.3$ μm) represent the second and the third radial order ($q = 2$ and 3) WGMs, respectively. The first order modes are too narrow to be detected with the limited resolution (~0.1 nm) of the system. An obvious trend can be seen that when moving the sphere from thicker to thinner region. The depth of the features observed in the fiber transmission monotonously increases as the sphere moves to thinner parts of the taper. In contrast, if the sphere continuously moves to thicker parts of the taper, the coupling feature gradually vanishes. It should be noted that for all of the coupling cases the resonant wavelengths remain approximately the same. Very slight variation may be induced by the unavoidable elliptic imperfection in the shape of the sphere. It can also be seen that the resonance widths are very similar for all curves with prominent WGMs dips.

To further understand the dependence of evanescent coupling on the taper diameter, we fitted the transmitted power near the resonance with coupling parameters using an approximated single-mode model [172, 173]:

$$\text{P} = e^{-\gamma} \times \frac{(\beta - \beta_0)^2 + \left(\dfrac{\gamma}{2S} + \alpha - \kappa\right)^2}{(\beta - \beta_0)^2 + \left(\dfrac{\gamma}{2S} + \alpha + \kappa\right)^2} . \tag{2.3}$$

In Eq. (2.3), $\beta = 2\pi n_s / \lambda$ is the propagation constant, $\beta_0$ is defined as $\beta_0 = 2\pi n_s / \lambda_0$, where $\lambda_0$ is the resonant wavelength. And $\gamma$ is the coupling loss, $\alpha$ is the field attenuation coefficient in the sphere, $\kappa$ is the coupling constant, and S is the circumference of the sphere.

It should be noted that the parameters used in this model represent effective average values characterizing coupling among several modes in the tapered fiber and a few



azimuthal WGMs [9]. Fig. 2.9 shows examples of fitted spectra using Eq. (2.3) for a second radial order ($q = 2$) WGM resonance near $\lambda = 1614.4$ nm with various taper diameters. The same BTG sphere ($D = 11.2$ μm, $n = 2.1$) was attached to the taper at four different positions with approximate diameters of $d = 1.3$, 2.2, 3.0 and 3.5 μm.

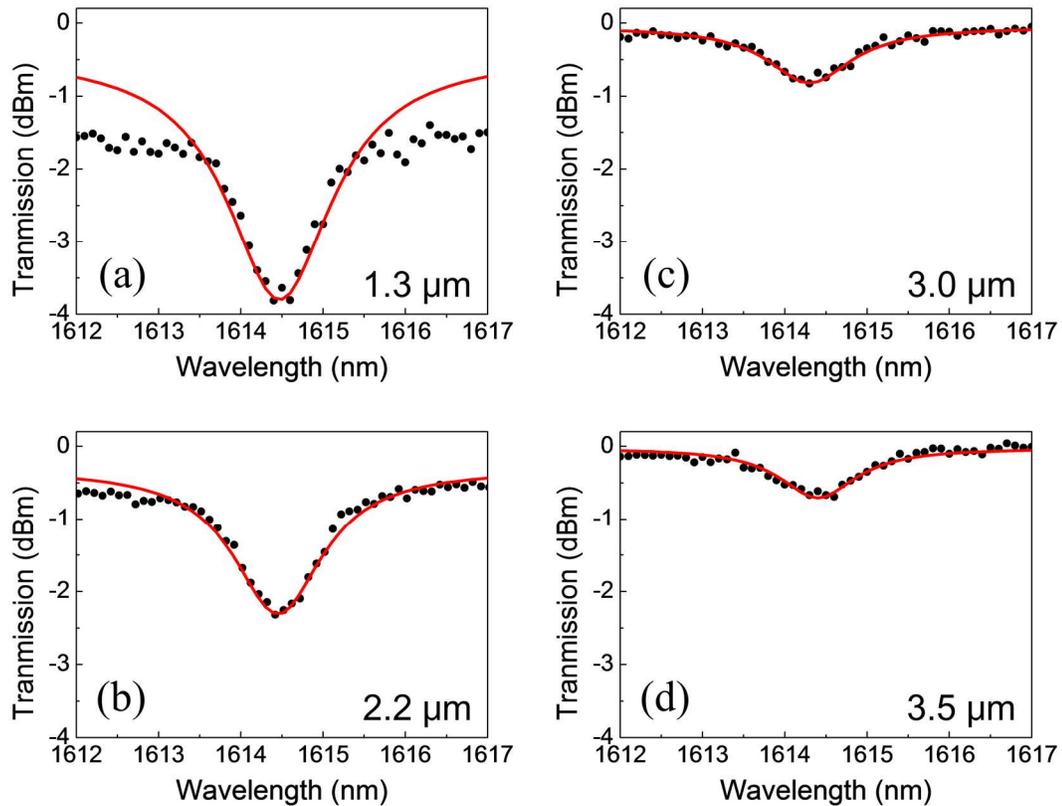

Figure 2.9: Examples of fitting with Eq. (2.3) for the transmission spectra of the same sphere positioned at different taper diameter as shown in the figure. [9]

It can be seen that for all of the curves good fittings can be obtained for the measured spectra except for the thinnest $d = 1.3$ μm case where some deviations are observed. Due to high flexibility of the extremely thin tapered region, contacting with a



sphere may cause some bending of the taper that would lead to a loss not related to the coupling. It can be seen in Fig. 2.9 that the resonant wavelength $\lambda_0$ and the resonance width $\Delta\lambda$ are generally independent of the taper diameter. And the depth of the resonance dip becomes deeper with decreasing taper diameter, but is limited to the level of a few dBm. That is an indication of the undercoupled regime [37]. In this regime $\kappa<\alpha$, where $\Delta\lambda$ is largely dependent on $\alpha$, and the depth is mainly determined by $\kappa$. With this trend critical coupling can potentially be achieved with further decrease of the taper diameter.

Let us take a more detailed look into these coupling parameters. Summary plots of the dependence on taper diameter for parameters $\gamma$, $\alpha$ and $\kappa$ are presented in Fig. 2.10. The coupling loss $\gamma$, which determines the level of the transmission away from the resonances, is shown to be rather small and even negligible for thick taper in Fig. 2.10(a). However, for the fiber diameter less than 3 μm $\gamma$ rises rapidly with decreasing diameter. The trend of reduced transmission away from resonances with a thinner taper can be attributed to the larger extend of evanescent field into the medium thus more light will be non-resonantly scattered by the microsphere placed in the near field. The effects of coupling to high-order WGMs with lower $Q$-factors could also reduce the transmission in a broad range.

The attenuation in sphere $\alpha$, as plotted in Fig. 2.10(b), is found to be around 26 cm$^{-1}$ for all the cases and almost independent of the taper diameter. It represents the loss, including both radiative loss and material absorption, which occurred inside the sphere. Since the same sphere was used for all the measurements, $\alpha$ is more or less constant and not dependent on the taper diameter. In the undercoupled regime $\alpha$ determines the resonance width and hence determines the $Q$-factor of resonance. Therefore similar



*Q*-factors were observed in the measured spectra regardless of the taper thickness. As a characterization of the coupling strength, coupling constant $\kappa$ was found to decrease exponentially with the taper diameter, as shown in Fig. 2.10(c). When the taper becomes thinner, a larger portion of electric field extends outside the fiber. The evanescent field decays exponentially into the medium. Thus thinning the taper diameter creates qualitatively similar effect as reduction of the fiber-to-sphere gap. A rough approximation can be made for the coupling constant through fitting as $\kappa = 19e^{-0.76d}$ cm$^{-1}$ where $d$ is represented in micrometers. It suggests that $\kappa$ approaches 19 cm$^{-1}$ for very thin taper, but will still be smaller than $\alpha$. This shows that the critical coupling condition $\alpha = \kappa$ cannot be met for the $q = 2$ WGMs in BTG sphere with $D = 11.2$ μm and $n = 2.1$ in water, likely due to the large mode loss and the limited *Q*-factors of the second order modes.

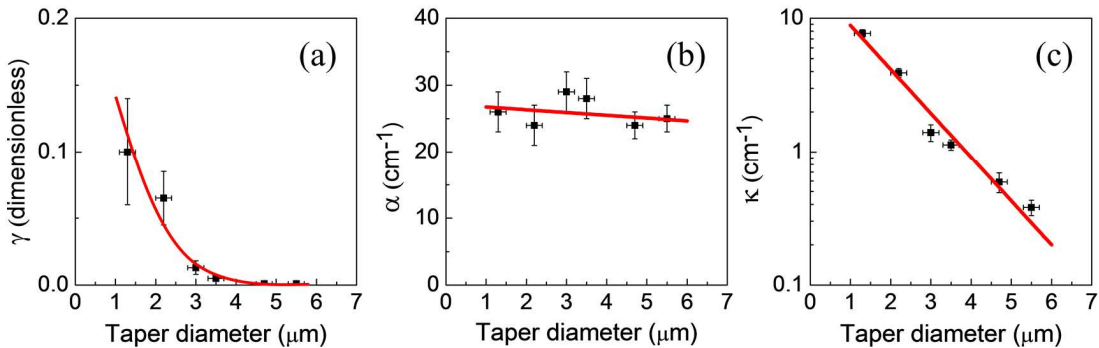

Figure 2.10: Dependences of coupling parameters $\gamma$, $\alpha$, and $\kappa$ on the taper diameter for a second order WGM resonance in BTG sphere with $D = 11.2$ μm and $n = 2.1$.

The first order WGM resonance is fitted in Fig. 2.7 using Eq. (2.3) for a BTG sphere with $D = 14$ μm and $n = 1.9$ in water. The values of coupling parameters $\alpha$ and $\kappa$ used in



the fitting are quite close as $\alpha = 1.3$ cm$^{-1}$ and $\kappa = 1.0$ cm$^{-1}$. The large depth of resonance (13dB) with only 5% fiber transmission at resonance is a clear demonstration that the coupling occurs close to critical regime. The small value of $\alpha$ indicates a small radiative loss inside the sphere for the first order high-$Q$ modes.

## 2.5. Dependence of $Q$-factors on Sphere Diameter

Quality factor is an important characterization of dielectric microspherical resonators. High-$Q$ microresonators are desirable for microlaser construction and various sensor applications. We have studied glass spheres with conventional refractive index made from borosilicate glass ($n = 1.47$) and soda-lime glass ($n = 1.5$), as well as high index spheres made from barium titanite glass ($n = 1.9$ or $2.1$ depending on proportion of composite). The $Q$-factors of all of the microspheres investigated in the evanescent coupling experiments were summarized in Fig. 2.11. The radial mode number ($q = 1$ or 2) were determined by comparison with mode numbers calculated by Eq. (2.1).

It can be seen that no matter what the refractive index of the sphere is and what the medium it is immersed in, for relative compact spheres ($D < 20$ μm) the $Q$-factor generally increases exponentially with a similar slope when the sphere size increases. WGMs in these compact spheres made from conventional index materials with n<1.5 practically vanish in water (Q<100). However, the BTG spheres with n=2.1, even in water, demonstrate $Q = 3 \times 10^3$ for small size of 5 μm diameter, which increases to $Q = 1.5 \times 10^4$ for 14 μm diameter. For the BTG sphere of similar size coupling in air, the $Q$-factor can approach to $10^5$. The size of the sphere mainly affects the mode attenuation in sphere ($\alpha$). Thus the size dependence of $Q$-factors is determined by the radiative loss, absorption loss, combined with the scattering from surface roughness [174]. The figure



provides us a reference to predict the $Q$-factors of microspherical resonators with various refractive indices and both in air and in water environment. It enables us to choose the size or material of the sphere for a required $Q$-factor for certain applications.

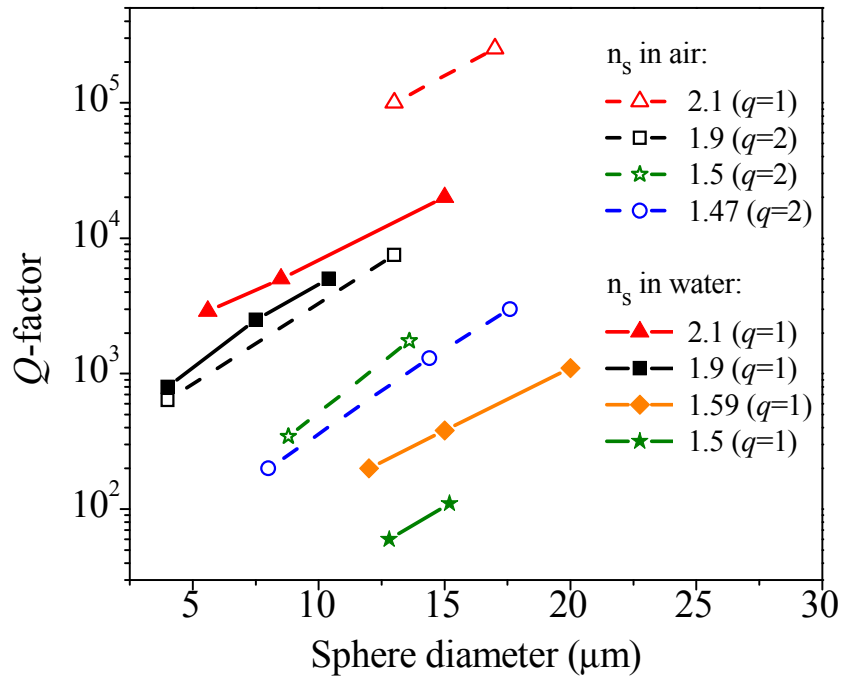

Figure 2.11: $Q$-factors of WGMs in the evanescent coupled spheres in air and in water, measured at ~1150 nm (for $q = 1$) and at ~1600 nm (for $q = 2$). [9]

## 2.6. Summary

In summary, we characterized WGMs in microspheres made from borosilicate and soda lime glasses, polystyrene, and especially from high index material barium titanate glass (BTG) both in air and in water environments. A mechanically robust platform integrated with a tapered fiber was fabricated to enable the characterization experiments with the same coupler for both environments. Spheres of a variety of sizes were studied



for the evanescent side-coupling to the tapered fiber coupler in a contact position. WGMs' quantum mode numbers were determined by comparison with the modes' resonant wavelengths calculated by the analytical formula. An impressive demonstration of good agreement with the experiment is represented by the match of positions of four subsets of WGMs calculated in two polarizations with different radial numbers for a BTG spheres with $D = 11.2$ μm and $n = 2.1$.

The dependence on the diameter of the tapered fiber was studied by moving the same sphere along the gradient region of the taper and positioning at locations with different diameters. Coupling parameters $\gamma$, $\alpha$, and $\kappa$ were fitted by a single-mode model and were analyzed for the dependence on the taper diameter. A weak to critical regime of coupling to compact BTG spheres in water was demonstrated. Near critical coupling condition was achieved with the selection of high index sphere ($D = 14$ μm, n = 1.9) and appropriate taper diameter ($d = 2.5$ μm). Transmission spectra were compared with air and water as the medium. For all studied spheres, substantially lower $Q$-factor was measured in water environment due to less sphere-medium refractive index contrast. However, the water medium was also found to help the coupling likely due to the further extending of evanescent field and effectively smaller surface roughness (in terms of the variations of the effective index) as a result of "healing" effect introduced by the liquid immersion.

An exponential increase of the $Q$-factor with the increase of sphere diameter was found with a similar slope regardless of the sphere index and the medium for relative small spheres ($D < 20$ μm). The study provides us a reference to predict the $Q$-factors of microspherical resonators with various refractive indices and both in air and in water



environment. It should be noted that our measured $Q$-factors were significantly smaller than the theoretically predicted $Q$-factors for a single-mode model [37, 40]. This is, generally, explained by a small (uncontrollable) elliptical deformation of spheres which leads to lifting degeneracy of the azimuthal WGMs. This factor, however, is typical for different physical spheres along with the surface roughness and surface contamination. In this sense, our measurements of $Q$-factors represent practical values that are accessible in real applications. WGMs in these small spheres made from conventional index materials practically vanish in water. However, the BTG spheres, even in water, demonstrate $Q = 3\times10^3$ for 5 μm spheres, which increases to $Q = 1.5\times10^4$ for 14 μm spheres. Therefore the high refractive indices of 1.9 and 2.1 for BTG spheres are preferable for developing compact high-$Q$ WGMs based photonic devices.

CHAPTER 3: SPECTRAL SIGNATURES OF PHOTONIC MOLECULES

## 3.1. Introduction

Microresonators supporting high quality factor ($Q$-factor) whispering gallery modes (WGMs) have been intensely investigated in recent decade with unique advantages as nanoparticle sensor, narrow-line filter, and low threshold laser [15, 17]. The terminology of "photonic atom" was first introduced by Arnold [93] in the context of understanding eigenmodes of spherical resonator based on an analogy between the solutions of Schrödinger's and Maxwell's equations in spherical geometry. The mathematical background of such analogy was described by Nussenzveig [175] and Johnson [176]. Using this analogy, the clusters consisting of coupled photonic atoms can be understood as "photonic molecules". It should be noted that in recent years this terminology has been used in a much broader context of cavities, not limited with microspheres. For example, photonic crystal cavities and semiconductor microcavities [177-180] can be also termed photonic atoms and corresponding coupled-cavity structures can be termed photonic molecules.

Single microresonators with indistinguishable WGMs positions can be used as photonic atoms to build photonic molecules. The ability to design the coupled structures provides another degree of freedom to engineer the optical density of states, resonant eigenfrequency, and the spatial distribution of the modes [181], which is attractive for microlaser design [100, 182-184]. Photonic molecules were proposed to be able to



increase the sensitivity in sensor applications due to the enhancement of electric fields in the coupling region [108]. Photonic molecules were also studied as a main approach to realize coupled resonator optical waveguides (CROWs) [106] which attracts great interests for applications as delay lines [185], optical storage [186], and to bend and switch light [109]. Optical gain and nonlinearity can also be tailored in such molecules which can lead to desired unidirectional propagation [187].

Strong WGMs-based coupling has been studied experimentally in bi-spheres [76, 99] and linear chains [102] constructed by almost identical microspheres. However, long chain of CROWs structure, large 2-D photonic molecules and possible 3-D integrated photonic lattice [188] built by size matched spheres are difficult to realize due to the 1-2% size disorder of available microspheres. This size mismatch will significantly reduce the coupling efficiency among the constituent atoms [101, 111, 158]. Resonance tunability of each photonic atom by controlling individual metallic micro-heater has been proposed [103], but it is not practical for large scale assembly. Very recently, we have demonstrated theoretically and experimentally a sorting technique that utilizes WGMs resonant enhancement of optical forces to select microspheres with overlapping resonances at any desirable wavelength [5-7, 189, 190], which is applicable in both air and liquid medium. Large quantity of spheres with resonant wavelength deviation of $\sim 1/Q$ can potentially be obtained from massive scale sorting techniques based on resonant optical propulsion. These pre-selected microspheres can be used as building blocks to construct complicated 2-D and 3-D photonic molecules that have not been well studied.

One property of photonic molecules that is not well noticed in previous studies is the fact that the spectra of photonic molecules with similar spatial configuration (including



number of resonators and their layout) have similar spectral properties [11, 12]. In what follows, we will call these common spectral characteristics "spectral signatures". A well-known example of such spectral signature is represented by the normal mode splitting in the bi-sphere molecules [76, 99]. In linear chain structures, the light tunneling from one resonator to the nearest neighbor can be described by 1-D tight-binding approximation [102]. The spectral signature of linear molecules can be understood by combining the tight-binding photonic dispersions with the wave-vector quantization due to periodic boundary conditions. It should be noted, however, that in 2-D microresonator clusters, the optical transport and coupling process is more complicated and poorly understood, which involves multiple paths for light propagation leading to interference, localization and percolation of light [188].

In this Chapter, we studied various photonic molecules built by microspheres with uniform resonant properties both theoretically and experimentally. The finite-difference time-domain (FDTD) simulations were performed in 2-D at the cross section of the 3-D structures to study the coupling properties of such molecules. Side-coupled fiber transmission spectra were calculated for 3-sphere and 4-sphere chains as well as 2-D arrays of 4-sphere square and 6-sphere ring. The coupling was investigated with three different sphere and medium index combinations. And the spectral signature for each molecule configuration was reproduced in all these cases. The spatial distribution of coupled modes, often referred to as supermodes [181], was also presented for 3-sphere chain, 4-sphere square and 6-sphere ring molecules revealing peculiar optical coupling and transport phenomena. Experimentally, photonic molecules were assembled with ultra-uniform polystyrene spheres of 25 μm mean diameter sorted by spectroscopic



characterization and comparison. The WGM resonances uniformity was controlled to be within 0.05% to ensure strong coupling and uniform coupling conditions. Various clusters of spheres such as bi-sphere, 3-sphere chain, 4-sphere square and 5-sphere plus were assembled in water environment. To the best of our knowledge this is the first experimental demonstration of resonant coupling effects in water-immersed photonic molecules. This is critical for nanoparticle and biomedical molecule sensing applications [57, 191]. Side-coupling experiments were performed with a tapered fiber and good agreement was found between transmission spectra obtained in experiment and spectral signature predicted in simulation. Also a new method of tuning coupling strength between adjacent spheres was experimentally demonstrated by adjusting the spatial excitation position, which can be controlled by the height of the tapered fiber.

### 3.2. Finite-Difference Time-Domain Modeling

Numerical simulation using finite-difference time-domain (FDTD) method was performed to study the optical coupling properties in microresonators [74, 192]. The software solutions used for the FDTD simulations are from OptiWave [193] and Lumerical [194]. The theoretical model is illustrated in Fig. 3.1(a) for the case of a single circular resonator side-coupled to a cylindrical waveguide. The models were constructed in three dimensional (3-D) spaces, however, all the simulations were performed in two dimensional (2-D) cases which describe properties of real 3-D structures at the equatorial plane of sufficiently large spheres.



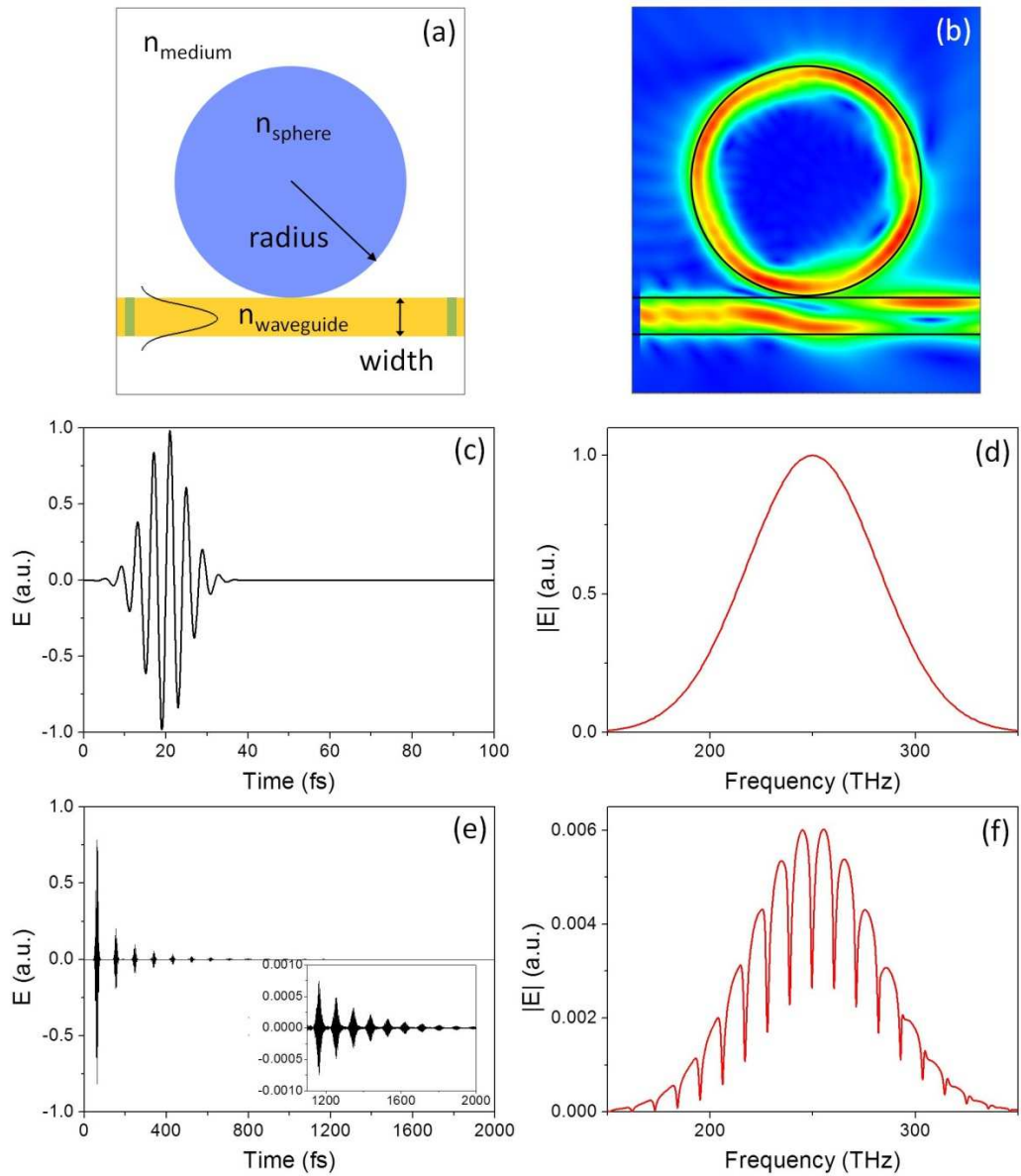

Figure 3.1: Results from FDTD simulation of a single circular resonator side-coupled to a cylindrical waveguide. (a) geometry and parameters considered; (b) map of E-field distribution; E-field of incident pulse in (c) time domain and (d) frequency domain; transmitted E-field in (e) time domain and (f) frequency domain. [11]



Fig. 3.1 shows an example of the simulation of fiber-to-sphere coupling [11]. A Gaussian modulated pulse of 10 femtosecond (fs) duration with central frequency at 250 THz was launched from the left into the waveguide, with only fundamental TE mode being considered. Figures 3.1(c) and 3.1(d) show the electric field of the incident wave in the time domain and the magnitude of such electric field in the frequency domain, respectively. At the end of the waveguide a time domain monitor was placed to record the transmitted electric field. The perfectly matched layer (PML) boundary condition was set to properly absorb propagating waves outside the computational space. The light can be trapped in a microresonator for a long period, thus the simulation time (2000 fs in this case) needs to be sufficiently long to allow multiple pulse circulations inside the sphere and to let the magnitude of the electric field sufficiently decays ($10^{-4}$ here), as seen in Fig. 3.1(e). Discrete Fourier transform of the time-domain signal was then performed to obtain the transmitted electric field in frequency domain, as presented in Fig. 3.1(f). The spectrum shows periodic sharp dips due to coupling of light to WGMs in the circular resonator. It also carries the spectral envelope coming from the input pulse. Normalization of such spectrum relative to the input yields transmission spectrum independent of the spectral shape of the source. Fig. 1(b) illustrates the electric field map results from the simulation, where coupling to the circular resonator and confinement of the light are clearly demonstrated.

### 3.3. Coherent Resonant Coupling in Bi-Sphere Molecule

We first studied the photonic molecule composed of two identical microspheres. The bi-sphere is the simplest configuration of photonic molecules. The spheres used in the simulation are of 7 μm diameter with a refractive index of 1.59 (polystyrene). The



medium is set to be air with an index of 1. A cylindrical waveguide of 1.5 μm diameter with index of 1.45, to mimic a typical tapered fiber in experiment, was placed at one side of the molecule perpendicular to the axis linking the centers of two spheres for evanescent side-coupling.

When the two spheres are positioned in contact with zero inter-sphere gap (g = 0) typical normal mode splitting (NMS) appear in the transmission spectrum [76, 99], as shown in the top spectrum of Fig. 3.2(c). The distribution maps of light intensity for shorter-wavelength antibonding mode and longer-wavelength bonding mode are presented in Fig. 3.2(a) and Fig. 3.2(b), respectively.

It should be noted that for the antibonding mode (Fig. 3.2(a)) the light is confined in two individual spheres and is expelled from the coupling region. There is no electric field presenting at the contact point. On the contrary, for the bonding mode (Fig. 3.2(b)) the evanescent field from two spheres is extended into the coupling region between the spheres. These two split modes can be analogous to the chemical molecular states described in quantum mechanics. For example, when two indistinguishable hydrogen atoms form a hydrogen molecule, the two identical electronic states of the atoms will split into a lower-energy bonding state and a higher-energy antibonding state, which are coupled molecule states, with electronic cloud distributions similar to the bonding and antibonding modes in bi-sphere photonic molecule.



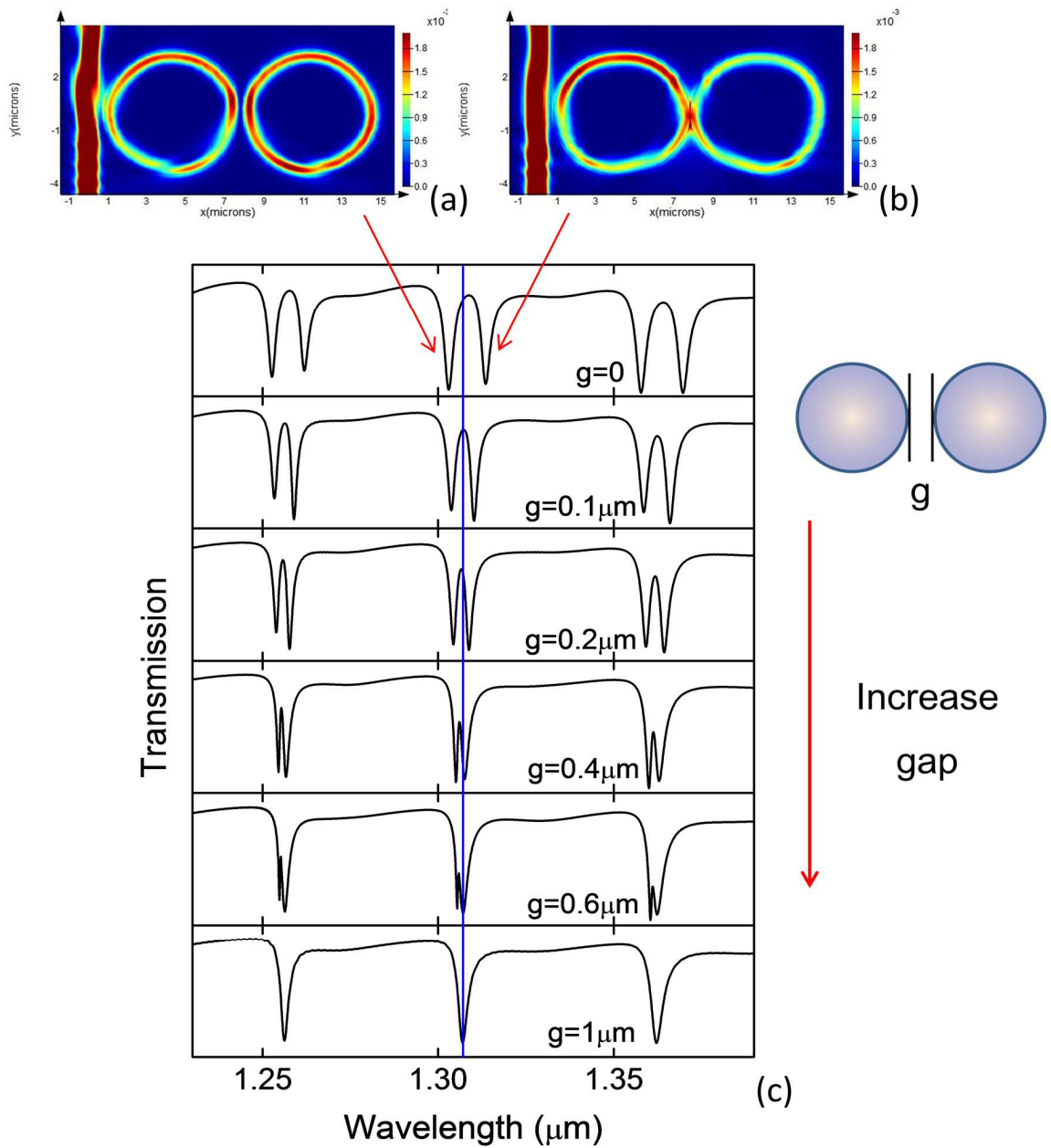

Figure 3.2: Simulation of normal mode splitting for bi-sphere photonic molecule composed of two identical 7 μm diameter spheres with index of 1.59. Light intensity maps for (a) antibonding mode and (b) bonding mode. (c) Dependence of mode splitting amount on the inter-sphere gap.



The amount of mode splitting represents the strength of the resonant inter-sphere coupling [76, 111]. For the case of touching bi-sphere (zero gap) the splitting between bonding and antibonding modes is 10.6 nm, which is much larger than the individual resonance width of 3.1 nm. It is an indication of strong resonant coupling between spheres. When the inter-sphere gap (g) gradually increases, the amount of mode splitting monotonously decreases, as shown in Fig. 3.2(c). The coupling strength depends on the mode overlap, both spectrally and spatially. For identical spheres the individual WGMs resonances are exact the same thus they have the maximum spectral overlap. When placed in contact the two individual modes have the maximum spatial overlap as well. As the two spheres separated, such spatial overlap decreases resulting in the reduction of the coupling strength and the mode splitting amount. When the gap increases to 1 μm the mode splitting no longer occurs and only individual WGMs from the single spheres present, as seen in the bottom spectrum of Fig. 3.2(c).

The blue vertical line in Fig. 3.2(c) indicates the wavelength of individual WGM resonance near 1.3 μm. As seen from the evolution of changing the gap, the double mode splitting is not symmetric. The wavelengths of bonding and antibonding modes for bi-sphere molecule with different inter-sphere gaps are presented in Fig. 3.3(a). The left and right black curves represent antibonding and bonding modes, respectively; and the blue line indicates the uncoupled WGM position, as noted in Fig. 3.3(b).



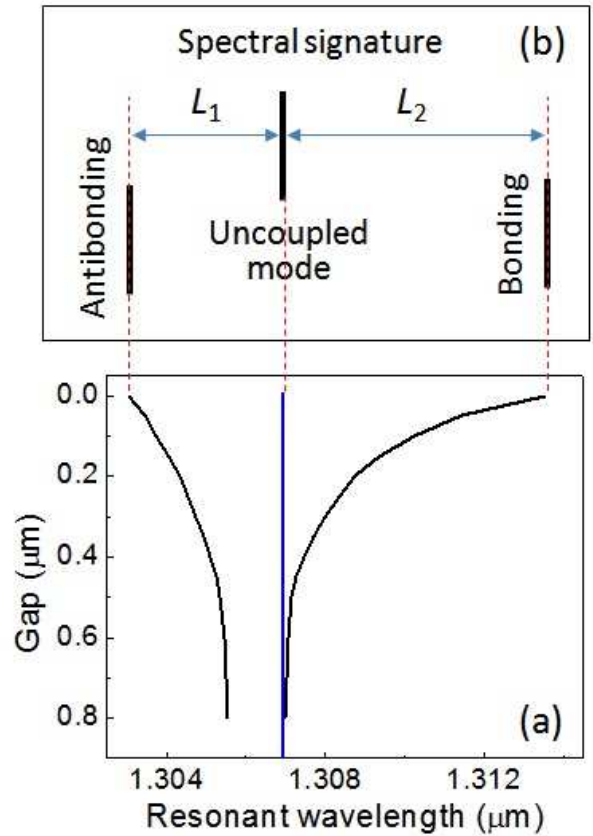

Figure 3.3: (a) Wavelengths of bonding and antibonding modes for bi-sphere molecule with different inter-sphere gaps. Blue line indicates the uncoupled WGM position. (b) Schematics illustrate the spectral signature of bi-sphere in touching position.

It can be seen in Fig. 3.3(a) that the mode splitting resulting from resonant coupling does not occur at an inter-sphere gap larger than 0.8 μm. When the gap decreases to 0.8 μm a shorter-wavelength antibonding mode emerges far away from the uncoupled WGM resonant wavelength, while the bonding mode remains close to the original WGM position. With further decreasing of the gap both spit modes move away slowly from the uncoupled mode and the splitting amount increases. This is an indication of enhancement



of resonant coupling strength. Since the evanescent field of WGMs exponentially decays in the medium the spatial modes overlap from two spheres is small with such a large gap. Thus the increase of splitting amount is slow until the gap reduces to less than 0.5 μm. After that the splitting of the modes increases rapidly due to greatly increasing mode overlap. It can be noted that the bonding mode moves to longer wavelengths faster than antiboding mode shifts to shorter wavelengths. Such phenomenon can be explained by the fact that the bonding mode concentrates at the vicinity of the region where the spheres touch (as seen in Fig. 3.2(b)) thus the change of coupling distance has a larger effect compared to the antibonding mode characterized with the electric field localization inside spheres (as seen in Fig. 3.2(a)). Since the region near the contact point between the spheres is characterized with higher effective index of refraction with decreasing gap, this would mean larger long-wavelength shift for the bonding mode, and thus the bonding mode is more sensitive to the change of the inter-sphere gap in comparison to the antibonding mode.

We have simulated bi-sphere molecules consisting of two identical spheres positioned in contact using various refractive indices and sizes, and found the mode splitting pattern is similar for all of the cases. As sketched in Fig. 3.3(b), the resonant coupling of bi-sphere will always result in two split modes; and when in touching position the wavelength shift of bonding mode is always larger than that of antibonding mode ($L_2 > L_1$). We propose such commonly seen split pattern as a characteristic spectral property associated with size-match resonant coupled bi-sphere photonic molecule and therefore can be treated as a spectral signature of bi-sphere molecule.



3.4. Spectral Signatures of Various Photonic Molecules

We continue to study the resonant coupling among photonic molecules of more complicated configurations than the bi-sphere. Due to WGMs' nature of trapping light, the simulation involving multiple microresonators needs more time to allow proper light transport and coupling and sufficient decay of the coupled light. For the following simulation we chose the shut-off criteria as when the total electric field in the computational space is less than $10^{-5}$ of the input field, and we found the results converged well.

Four configurations of photonic molecule were studied, including linear chains of 3-sphere (Fig. 3.4(b)) and 4-sphere (Fig. 3.4(c)), and 2-D structures of 4-spheres square (Fig. 3.4(d)) and 6-sphere ring (Fig. 3.4(e)). In order to find the spectral signature commonly seen in each configuration, the simulation was performed multiple times with different constituent microspheres in both air and water media. The refractive indexes of the medium and the microspheres as well as their diameters are shown respectively for these three cases in the top row of Fig. 3.4. Appropriate diameters were chosen for each case to ensure WGMs of first radial order are well pronounced while second order modes are suppressed. This approach enables unambiguous observation of supermodes in the fiber transmission spectrum that arises from inter-sphere coupling of the first-order WGMs.



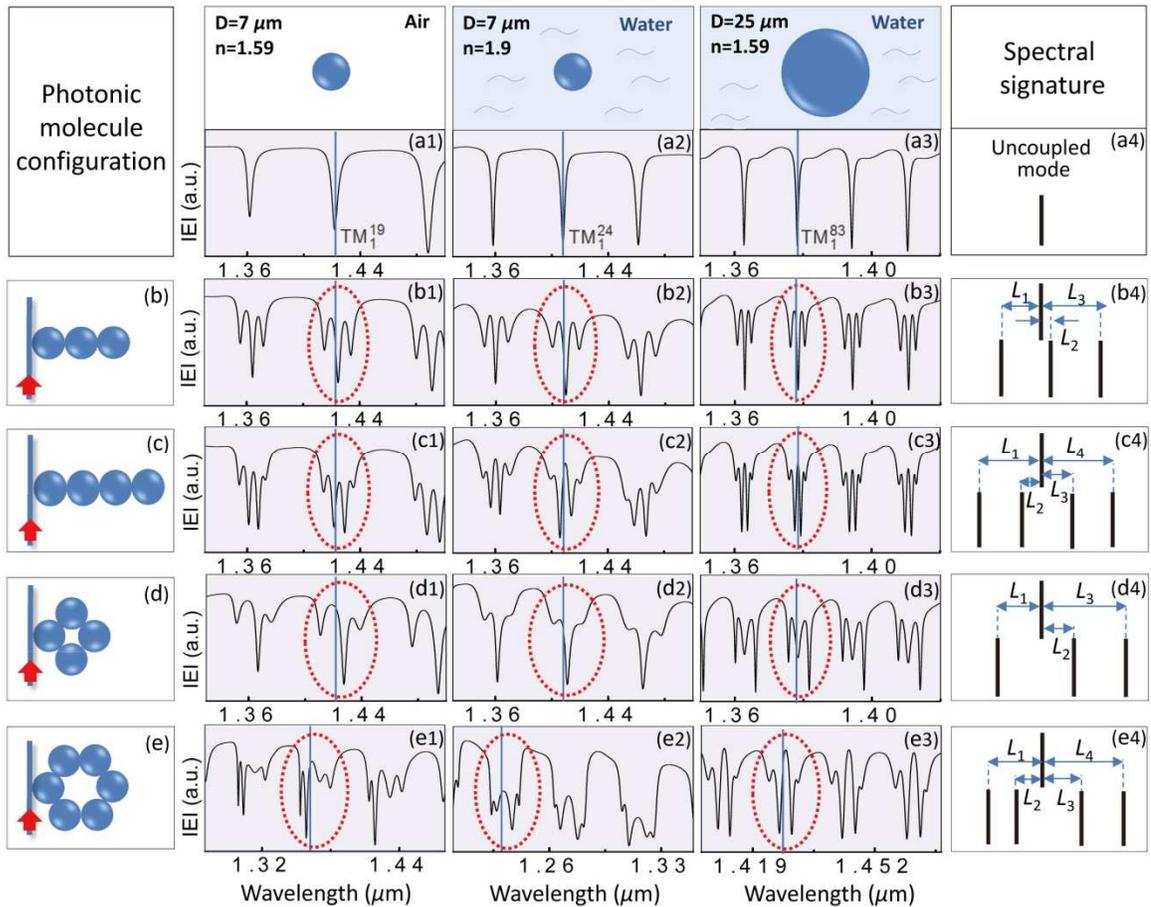

Figure 3.4: Simulated transmission spectra of single resonators and various molecule configurations (b-e), with constituting atoms of (a1-e1) 7 μm 1.59 index sphere in air; (a2-e2) 7 μm 1.9 index sphere in water; and (a3-e3) 25 μm 1.59 index sphere in water. (a4-e4) Summarized spectral signatures of corresponding photonic molecules. The FDTD simulation was performed in collaboration with Farzaneh Abolmaali.

The transmission spectra of side-coupling to single spheres were presented in Figs. 3.4(a1-a3) respectively for these three cases. The spectra show periodic dips due to coupling of light to WGMs in spheres. The WGMs of single resonator are referred to as uncoupled modes here to be distinguished from supermodes in coupled photonic



molecule. Simulation results for molecule configurations in Figs. 3.4(b-e) were presented for three combinations of parameters, diameter and index of circular resonator, along horizontal lines. Strong coupling and mode splitting are observed in every case. Blue vertical line indicates the wavelength of uncoupled mode, which helps to compare the relative positions of the split supermodes. Red ellipse marks a set of supermodes that arises from inter-sphere coupling of one WGM, representing a reproducible unit that periodically appears in the spectrum separated by the free spectral range. As one can see, the splitting patterns of supermodes are quite similar even though the size varied from 7 μm to 25 μm; index varied from 1.59 to 1.9; and medium changed from air to water. Let us consider, as an example, the ellipse enclosed parts in Figs. 3.4(b1-b3). The uncoupled WGM from single sphere is split into three supermodes. These modes are almost equally separated, and the central one has much higher magnitude. A slight red shift compared to the uncoupled mode is also noticed in all cases. As a characteristic representation of the splitting pattern commonly seen in all three cases, we plotted supermodes' positions relative to the position of uncoupled mode in Fig. 3.4(b4). Relative distance from each supermode to the uncoupled one is shown as $L_N$ where N is the number of the split spectral component. Thus, the spectral signature of three-sphere linear chain is characterized with N=3 and with a certain ratio of split components ($L_N/L_M$) schematically illustrated in Fig. 3.4(b4).

Spectral signatures summarized for each photonic molecule are presented in Figs. 3.4(b4-e4). As we can see, the signatures of linear chain molecules are qualitatively reproducible for a certain range of variation of parameters. Individual WGMs will split into bonding and antibonding modes with the number of supermodes equal to the number



of constituting atoms. The central supermode coincides with the uncoupled mode with a slight red shift [113]. The splitting is almost symmetric with approximately equal separations between adjacent supermodes. These observations are in a qualitative agreement with the spectrum of Bloch modes formed in coupled linear structures [111]. However, for 2-D molecules the symmetric splitting is no longer present. Transmission spectra in Figs. 3.4(d1-d3) show uncoupled mode split into three supermodes when 4-sphere square molecule is formed. In comparison to 3-sphere chain which also yields three supermodes, the modes in square molecule have wider splitting and larger red shift, and the separations are no longer equal. Although the appearance of supermodes in square molecule varies with constituent atoms, the number of split modes and their relative positions remain stable. Therefore it can still be considered as a signature in the spectrum to distinguish such molecule from others including the 3-sphere chain, as seen in Figs. 3.4(d4) and (b4). Similar properties can be noticed when comparing 6-sphere ring molecule to 4-sphere chain. While the ring molecule gives the same number of supermode as four, the separation between the two central modes ($L_2$ and $L_3$) is substantially larger than the other two separations ($L_1$ and $L_2$) thus enabling differentiation of this molecule from the 4-sphere chain, as seen in Figs. 3.4(e4) and (c4). It should be noted that the degeneracy may not be completed lifted due to symmetry of such molecules, resulting in less observed modes compared to the total number of constituting photonic atoms. However that is also a distinct property of a given photonic molecule associated with its configuration. Each photonic molecule with a particular configuration has unique coupling properties that give rise to its distinct spectral signature. The spectral signature of 4-sphere square (Fig. 3.4(d4)) is very different from



that of 4-sphere chain (Fig. 3.4(c4)) and 3-sphere chain (Fig. 3.4(b4)). Therefore we are able to identify the molecule configuration based on its spectrum, giving us the ability to potentially utilize such signatures for geometry or position sensing. We also observed that some of the coupled supermodes have much higher $Q$-factors compared to the uncoupled WGMs, as seen in Figs. 3.4(c3, d3, e1). The $Q$-factor increase may relate to the symmetry of the molecule configuration that was reported before [110, 114]. Hence the configuration could be designed for potential applications such as high order filters and multi-wavelength sensors. It should be noted, however, that the optical transport and coupling process is more complicated in large 2-D molecules formed by multiple spheres, which provide multiple propagation paths and WGM coupling possibilities. As a result, the spectral signature for complex cluster can be considered as a stable property only in a certain range of variation of parameters. Its appearance may also vary along a long spectral range. In this work, we established a similarity of spectrum for a certain parameter range (diameters from 7 to 25 μm and index from 1.59 to 1.9) and within a certain spectral range (hundreds of nanometers). These ranges turn out to be rather wide so that these properties should be common for the majority of real physical microsphere made from different glass materials.

## 3.5. Spatial Distribution of Supermodes in Photonic Molecules

To better understand the coupling and transport properties of coupled photonic molecules, we mapped the spatial distribution of electric field (E-field) for each supermode, as presented in Fig. 3.5. The simulation was performed using $n = 1.59$, $D = 7$ μm spheres in air, the same as in Fig. 3.4(a1-e1). Photonic molecule configurations and their corresponding transmission spectra are shown in Fig. 3.5(a-c), where vertical lines



indicate the position of uncoupled WGMs. Spatial E-field distribution was obtained by launching contiunous wave (CW) source at each supermode eigenwavelength.

A couple of interesting phenomena can be observed in the E-field map. For shortest wavelength antibonding mode (Fig. 3.5(a1-c1)) the electric field appears to be distributed uniformly among all constituent atoms, which is not seen for other modes. Also, the pattern of calculated E-field contains a large fraction outside the resonator, which contributes to the lower effective index of antibonding mode. The E-field fraction outside the resonators is especially noticeable in 6-sphere ring molecule. As shown in Fig. 3.5(c1) there are E-field standing waves present in the medium connecting adjacent resonators, constituting a large outer ring. These phenomena are believed to be related to the antibonding nature of such modes. The E-field is expelled from the region where the circular resonators touch due to WGMs phase mismatch, thus the light tunneling is not efficient. In the meanwhile, part of scattered light (in the external medium) incident on the adjacent sphere at a grazing angle can be coupled inside, which may contribute to the supermode coupling. As a result, relatively large portion of mode is presented in the medium leading to a higher effective index for such antibonding mode, which results in its shortest wavelength among all split supermodes. As discussed in Fig. 3.4, this antibonding mode in 6-sphere ring (Fig. 3.5(c1)) has much higher $Q$-factor compared to the uncoupled mode ($Q$ increases from 350 to 800). Similar $Q$-factor increase for antibonding mode was reported previously for symmetric photonic molecules and the $Q$-factor of some supermodes can be tuned by changing the gap between coupled resonators [110].



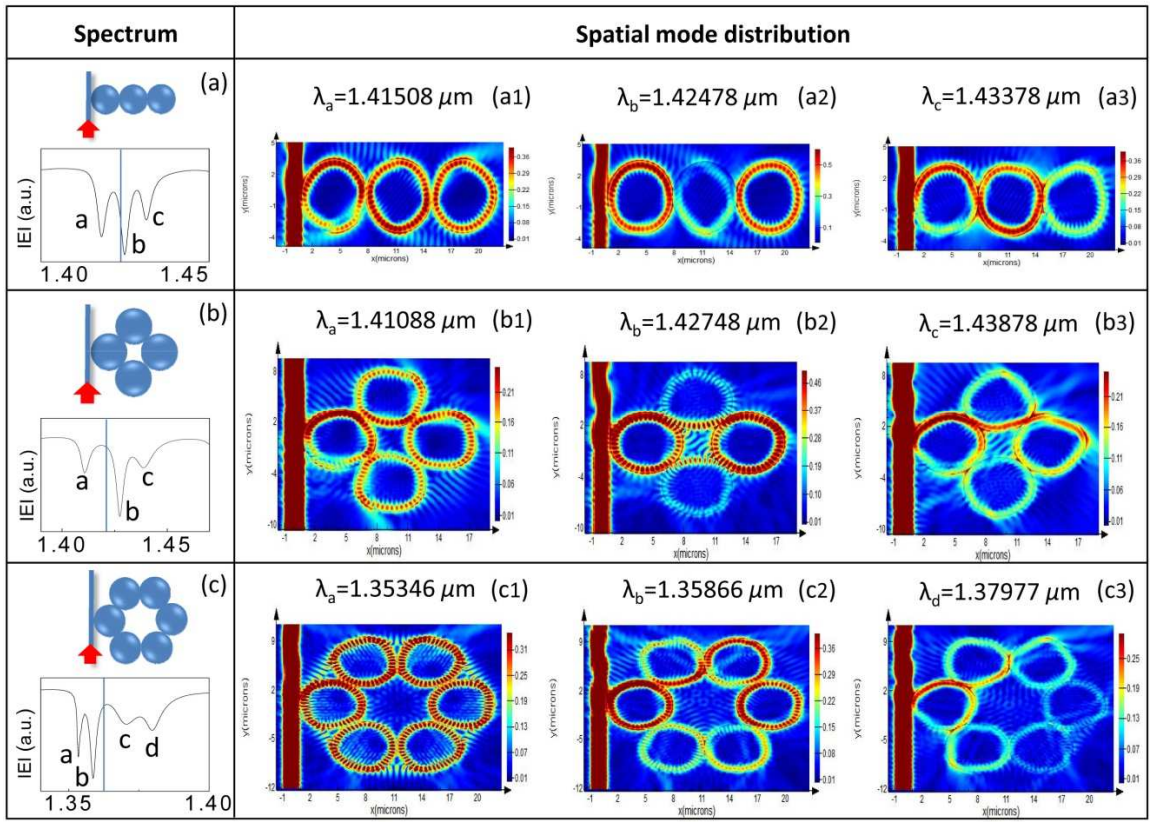

Figure 3.5: Simulated electric field map at each supermode eigenwavelength for three different molecule configurations. The FDTD simulation was performed in collaboration with Farzaneh Abolmaali.

For bonding mode, in contrast, the E-field tends to concentrate in the regions where the circular resonators touch due to better phase matching. Thus, the tunneling between neighboring cavities is more efficient in this case. As one can see in Figs. 3.5(a3-c3), the intensity distribution indicates that light is more likely to tunnel to the neighbor resonator than to stay in the original one and circulate. Therefore a large portion of the light propagates through the entire molecule making "arches" inside each resonator from one contact point to the next contact point. The 6-sphere ring molecule is likely to operate in



the over-coupled region, due to the fact that bonding mode (Fig. 3.5(c3)) is broader and shallower and has lower $Q$-factor compared to the antibonding mode (Fig. 3.5(c1)) with a stronger inter-sphere coupling.

For photonic molecule supermodes whose coupled WGMs resonances are close to their uncoupled positions, the E-field patterns inside individual resonators appear more circular-shaped similar to the WGMs E-field map inside a single resonator, as seen in Fig. 3.5(a2-c2). It is due to better phase matching between such supermodes and individual uncoupled WGM. If the supermode resonant wavelength is far away from that of uncoupled mode, large phase mismatch will cause broken circles (left resonator in Fig. 3.5(a1,b1)) or distorted uncircular shapes (Fig. 3.5(b3,c3)) for E-field distributed in individual resonators. It appears that light is forced to choose an optical path other than the perfect circle in the propagation in order to maintain the resonant condition of constructive interference. Such non-circular forced modes have been observed previously in size-mismatched bi-spheres [101]. Another noticeable point in Figs. 3.5(a2) and (b2) is that the E-field is concentrated at two side resonators, while the central one is almost dark. For linear chain this can be explained by the Bloch modes formation due to coupling. Ref. [111] showed in calculation and experiment that among the three split modes in a coupled 3-sphere chain, the center mode has maxima in the first and the third resonators whereas in the second (central) resonator the mode has minimum. For 2-D molecules the coupling is found to be more complicated; however, similar effects can be observed in Fig. 3.5(b2). By introducing one more sphere to connect two side ones, enhanced intensity is shown at two connecting "arches" while the rest region of two central spheres still remains dark. Enhanced E-field is also observed in the medium region enclosed by four resonators due



to formation of strong standing waves at the center of the E-field distribution in Fig. 3.5(b2). The E-field enhancement seen in the medium at the enclosed center of 4-sphere square and at the outer circle of 6-sphere ring molecules is advantageous in sensing application due to potentially stronger light-particle interaction within a larger detecting volume, not limited to the surface of the resonator in typical WGMs based sensors. The basic principle of sensing is based on a slight change of the effective index for WGMs causing one or several spectroscopic effects [30, 56, 195-200]: i) Spectral shift of coupled WGMs, typically to longer wavelengths for high-index nanoparticles, ii) Broadening of the corresponding coupled WGM resonance due to increased scattering losses, and iii) Splitting of WGMs caused by the formation of standing waves. The sensitivity of nanoparticle detection can be increased for some of the supermodes providing stronger overlap with the embedded nanoparticles such as in the case illustrated in Fig. 3.5(b2, c1). More detailed analysis of the potential sensor applications goes beyond the scope of this work; however, in principle it can be performed by FDTD modeling similar to the calculations presented in this work.

### 3.6. Experimental Apparatus and Procedure

In order to explore the coupling properties of photonic molecules and to verify the spectral signatures predicted in Section 3.4, we experimentally assembled various molecule configurations with pre-selected almost identical polystyrene spheres of 25 μm nominal diameter and measured transmission spectra with side-coupled tapered fiber in aqueous environment. Liquid environment enables particle movement and is critical to biomedical molecules, thus most of WGMs based sensors operate in liquid medium [57, 191]. To the best of our knowledge the present work is the first experimental



demonstration of coupled photonic molecule in aqueous environment. The general consequence of liquid immersion is connected with reduced $Q$-factors of WGMs in the constituting microresonators. On the other hand, high $Q$-factors are required for sensitive sensor action. For this reason, we have to select enlarged microspheres with diameter of 25 μm for this work due to small index contrast for polystyrene sphere in water. The selection of size-matched microspheres was achieved by micromanipulation followed by the spectroscopic characterization and comparison of individual spheres. The experimental platform for sphere selection is depicted in Fig. 3.6(a). A piece of single mode optical fiber (Corning SMF-28e+) fixed to the sidewall of a plexiglass frame was wet etched in hydrofluoric acid to achieve a waist diameter of ~1.5 μm with several millimeters in length [5, 9]. The fiber was connected to a broadband white light source (AQ4305; Yokogawa Corp. of America) and an optical spectral analyzer (AQ6370C-10; Yokogawa Corp. of America) for transmission measurement.

The frame was sealed with a microscope slide at the bottom and then filled with distilled water. The spheres used in our experiments are polystyrene microspheres (Duke Standards 4000 Series, Thermo Fisher Scientific, Fremont, CA, USA) with refractive index of 1.57 at 1.3 μm wavelength where transmission spectra were measured. Monodisperse polystyrene microspheres were chosen due to their good size uniformity of 1-2% standard deviation. Because water as medium significantly reduces the index contrast, sphere with large diameter is desirable to preserve sufficiently high $Q$-factor [9, 201]. We found 25 μm diameter sphere is a good candidate. It shows well-pronounced first-order WGMs dips in water ($\text{TM}_1^1, \text{TE}_1^1$) with $Q$-factor over $10^3$ while second-order



modes are not present, as seen in Fig. 3.6(b), which enables unambiguous modes interpretation.

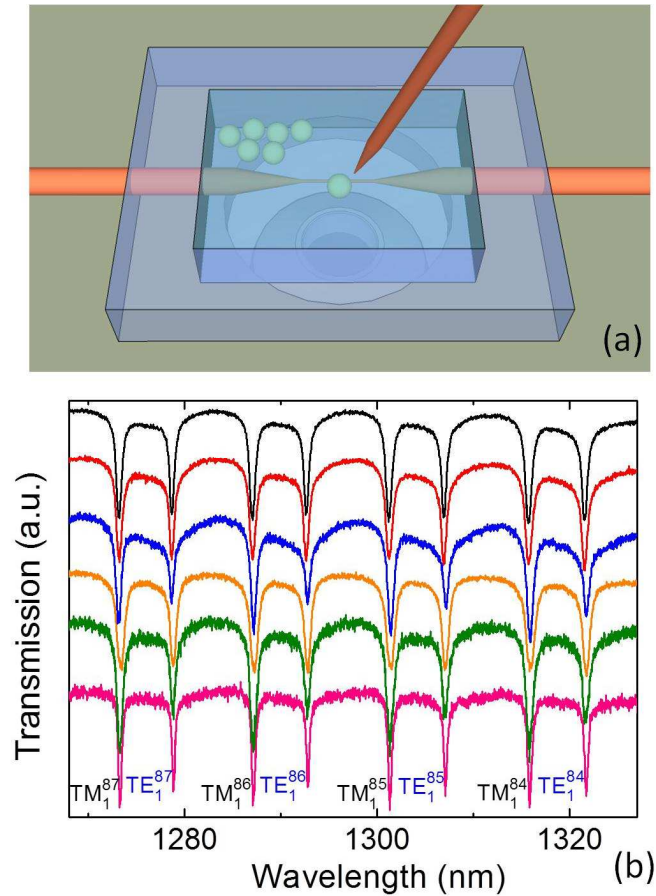

Figure 3.6: (a) Experimental platform for selection of size-matched microspheres by spectroscopic characterization and comparison. (b) Transmission spectra of six selected size-matched spheres with deviation of 0.05%.

The large sphere size also facilitates individual sphere micromanipulation. As shown in Fig. 3.6(a), a fiber tip was used to pick up sphere from the bottom and attach it to the tapered region of the microfiber. For each sphere broadband transmission spectrum was recorded and compared with others'. Resonant wavelengths of WGMs are highly



sensitive to the size of resonator, thus well-matched resonance positions in a broad range indicate almost identical sphere diameters. Such spheres were collected in a designated area that can be used as building blocks for assembling photonic molecules. Fig. 3.6(b) shows transmission spectra of a set of six selected spheres in a broad wavelength range, demonstrating extraordinary high size uniformity with a standard deviation of 0.05%.

These pre-selected spheres with almost identical size were assembled into certain molecular configurations on a microscope slide by micromanipulation. It should be noted that a very thin layer of adhesive needs to be deposited on the surface of the microscope slide first in order to fix the position of the spheres. A tilted plexiglass frame integrated with tapered fiber was fabricated for the molecule spectroscopic characterization, as sketched in Fig. 3.7. Droplets of distilled water were deposited on the glass plate to create an aqueous environment for the spheres. Then position of the assembled molecule (4-sphere square shown in Fig. 3.7) was adjusted by the 3-D stage to get in contact with the tapered region of fiber for the transmission measurement.

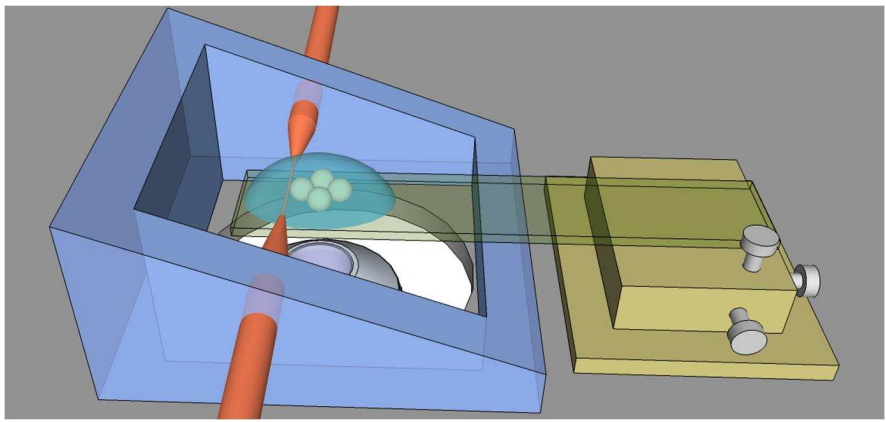

Figure 3.7: Experimental set-up for the tapered fiber side-coupling to assembled photonic molecules in water immersion.



### 3.7. Experimental Results and Discussions

We began our experimental studies with the bi-sphere photonic molecule due to its simplicity. With the procedure presented in Section 3.6, two polystyrene microspheres of almost identical size were selected and assembled on a glass plate in touching position. The spheres were immersed in water and brought in contact with the tapered fiber in perpendicular direction, as sketched in Fig. 3.8(a). The height of the sphere was carefully adjusted while fiber transmission spectrum was continuously recorded.

The height of the taper was roughly controlled using the focusing depth of the microscope objectives. The accuracy of this characterization was typically on the order of 1-2 μm. When the contact position of the fiber was close to the equatorial plane of the bi-sphere molecule, normal mode splitting (NMS) was observed and bonding mode and antibonding mode were well pronounced. However, the splitting amount was found to be highly sensitive to the relative height of the fiber. We repeatedly moved the spheres up and down and kept the fiber in contact while monitoring the transmission spectrum in the meantime. A position yielding maximum amount of mode splitting can be located. We assume that the fiber was aligned with the equatorial plane of the bi-sphere at this position due to the geometric symmetry. Then we slowly moved the spheres downwards (fiber moved upwards relatively) to monotonously increase the height difference (h) between taper and the equatorial plane, as illustrated in Fig. 3.8(a). Several examples of fiber transmission spectrum during this procedure are presented in Fig. 3.8(b) showing the decrease of splitting amount. It is well known that the amplitude of NMS is a representation of the coupling strength, and the coupling coefficient can be derived from the splitting amount as $\kappa = 1/\sqrt{3} \times \Delta\lambda/\lambda$ for bi-sphere [111]. The calculated coupling



coefficients are noted on corresponding spectra, as large as $1.6 \times 10^{-3}$ at the bi-sphere equatorial plane. When the fiber was detuned far away from that plane mode splitting was no longer present thus only the spectrum of uncoupled single sphere's WGM was observed. The way to model this effect in 2-D simulation is to assume that the change of the height of the taper can be understood by the variation of an effective gap between two coupled spheres. The position of the tapered fiber on the surface of the sphere defines a trajectory (and the plane) for fundamental WGM, indicated by the orbits shown in Fig. 3.8(a). If the sphere is elliptically deformed, the degeneracy of azimuthal WGMs is lifted. In this case, tapered fiber can be coupled not to a single, but to several azimuthal WGMs with trajectories close to the point where the fiber touches the sphere. The number of the azimuthal modes involved is usually not high and it depends on the sphere diameter. For sufficiently small spheres, it can be assumed that the taper is coupled to just single fundamental WGM in sphere. This mode can be coupled to a symmetric WGM in the second sphere. Since the coupling strength is directly related to the spatial overlap of two modes, the modes located in the same equatorial plane containing the bi-sphere axis will have the strongest coupling due to the maximal overlap which results in the largest spectral splitting. On the other hand, the WGMs located in the tilted planes relative to the bi-sphere axis will be less overlapped which should result in smaller spectral splitting.

In order to examine our hypothesis, we performed 2-D FDTD simulations for coupling between two identical circular resonators with varying inter-resonator gaps. To mimic our experiments we used resonators of 25 μm diameter with index of 1.57, medium index of 1.32, and cylindrical waveguide of 1.5 μm width with index of 1.45. We



performed 2-D simulation due to limited computational power, however, it provides a close analogy to our 3-D experimental situation.

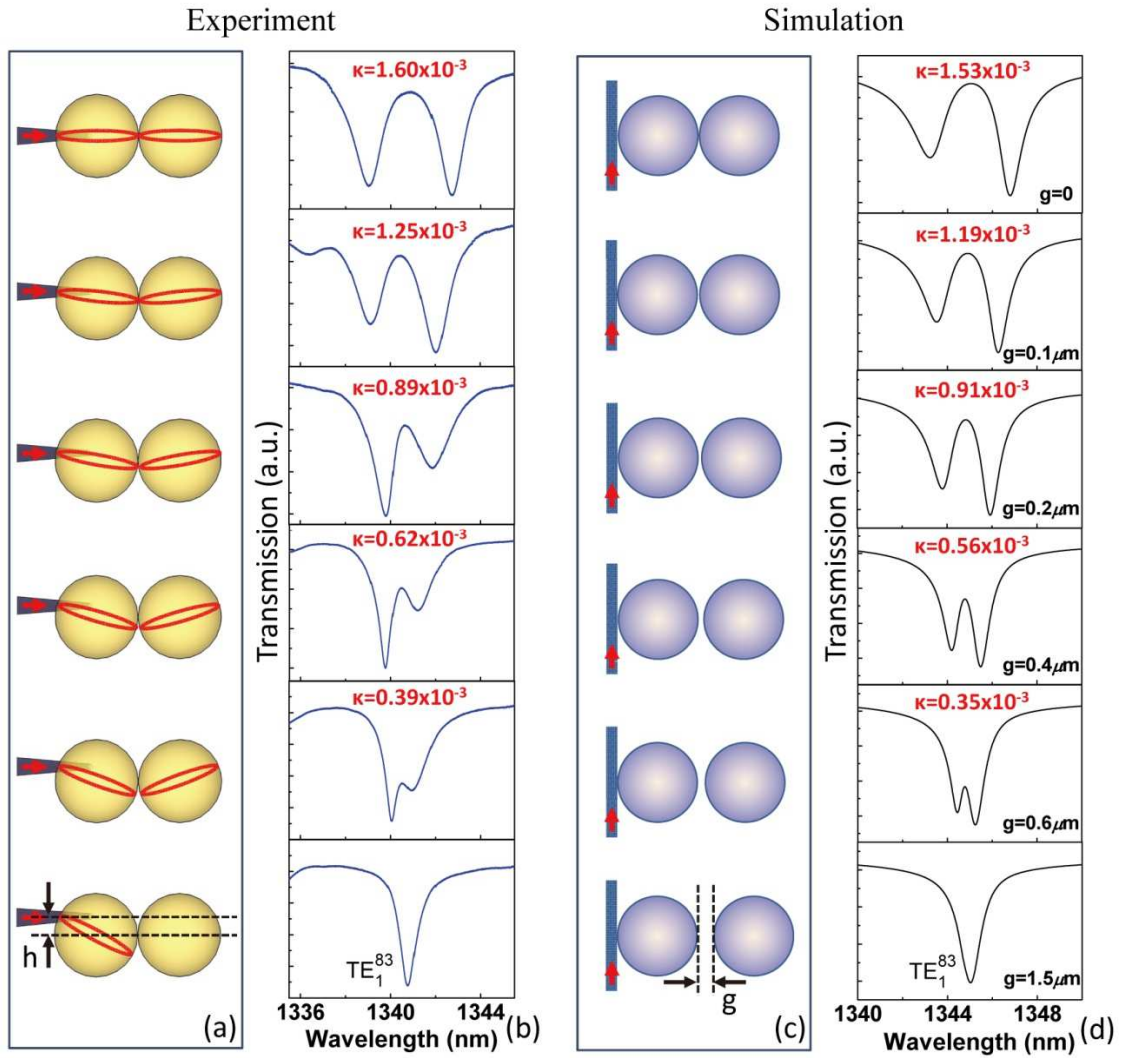

Figure 3.8: Increase of height difference (noted as h) between equatorial plane of bi-sphere and tapered fiber (a) with corresponding transmission spectra (b) in experiments. Increase of inter-sphere gap (noted as g) in bi-sphere (c) with corresponding transmission spectra (d) in simulation. Calculated coupling coefficients κ noted in corresponding spectra.



The configurations of simulation are sketched in Fig. 3.8(c). One circular resonator was placed in contact with the waveguide where femtosecond pulse with TE polarization was launched from the bottom and broadband transmission was recorded at the top. The gap (g) between two spheres was increased with 0.1 μm step in simulation. Several transmission spectra showing similar amount of mode splitting as seen in experiments are presented in Fig. 3.8(d). Coupling coefficients can be derived by the same method, which match well with those of corresponding experimental cases. Therefore the results support our hypothesis of off-axis coupling that the effective gap created by tuning the relative height of fiber in contact with two touching spheres can be viewed analogously to the physical gap between two separated circular resonators. Adjusting the gap between microspheres is a commonly used method to tune coupling coefficient [80, 101, 110, 202]. Here we demonstrated an alternative way of coupling strength tunability by controlling the plane of mode excitation. In this case we have full tunability of coupling coefficient from the maximum of $1.6 \times 10^{-3}$ to zero in bi-sphere molecule, which is equivalent to changing the physical inter-sphere gap from zero to the amount that eliminates coupling. This method is especially convenient when changing coupling coefficient simultaneously for all spheres in a complicated photonic molecule is required, where adjusting gaps for all of the spheres is not practical. Large number of spheres can form a closely packed array via self-assembly [203], and full tunability of coupling coefficient among all spheres can be realized by controlling only the position of the tapered fiber.

More complicated configurations were also assembled and studied using pre-selected spheres with almost identical size, including 3-sphere chain, 4-sphere square and 5-sphere cross, as presented in Fig. 3.9. First column (Figs. 3.9(a-d)) shows



microscope images for the molecules assembled and brought in contact with the tapered fiber at their shared equatorial planes. Positions were located by carefully adjusting the taper height while monitoring the transmission spectrum to find the place with the largest mode splitting. The rough surfaces appeared in these images are caused by a thin layer of adhesive on the glass plate to fix the position of spheres. Due to the large size of spheres (25 μm diameter) there is no contamination near the equatorial plane that would affect the resonant coupling effects.

Fiber transmission spectra with side-coupling to single spheres as well as various photonic molecules measured in water immersion are presented in Figs. 3.9(a1-d1). It should be noted that in order to obtain the spectral signature of a large structure we need to record the transmission in a long wavelength range. Since the excitation is provided by the unpolarized white light source both TE and TM WGMs are present in the fiber transmission spectra. The WGMs' radial and angular mode numbers are labeled in Fig. 3.9(a1). The recorded transmission spectrum is a superposition of coupled TE and coupled TM modes of the molecule. The coupling between two orthogonal polarizations is usually negligible. Corresponding to the experiment, the simulation was performed with both TE and TM polarized modes injected with equal amplitude, as shown in Figs. 3.9(a2-d2), for comparison. Both experimental and simulated spectra have spectral width approximately equal to two free spectral ranges (FSRs) which allows displaying periodically repeated patterns of the WGM splitting.



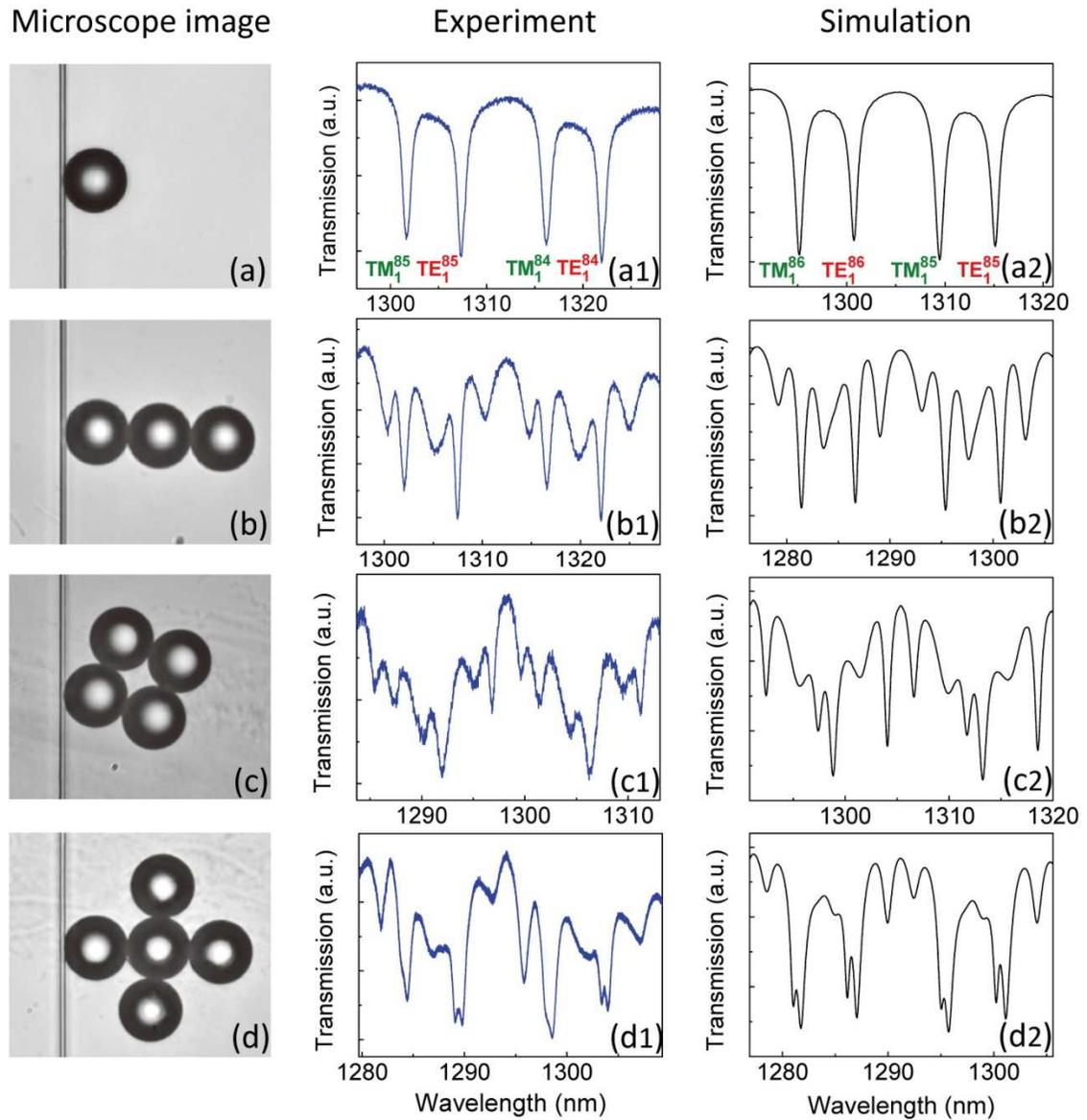

Figure 3.9: (a-d) Microscope images for various photonic molecules assembled with size-matched polystyrene microspheres of 25 μm mean diameter side-coupled to a tapered fiber with 1.5 μm diameter in water immersion. (a1-d1) Measured and (a2-d2) simulated fiber transmission spectra for corresponding molecule configurations. The FDTD simulation was performed in collaboration with Farzaneh Abolmaali.



Previous attempts were made to observe the coupled modes spectrum in photonic molecules consisting of GaInAsP microdisks [100], however, the large fabrication error in the size uniformity (>1%) resulted in a merely satisfactory comparison to the calculated spectrum. Here, using microspheres with size deviation of only 0.05% we demonstrated an excellent agreement between experimentally measured and FDTD simulated spectra for both 1-D chain and 2-D array molecules. The numbers of observed supermodes (dips) and their spectral positions match well in all cases. It is interesting that in some cases the fine mode splitting can be difficult to observe, but they still match the theoretical predictions. This is illustrated by the fine splitting of double-dip features shown in Fig. 3.9(d1) which are not perfectly resolved. In some cases, these double-dip features manifest themselves as distorted shoulders. Still, by comparing measured spectrum in Fig. 3.9(d1) with calculated spectrum in Fig. 3.9(d2) an unambiguous conclusion about similarity of the WGM splitting pattern can be reached. These results demonstrate experimental feasibility of resonant sphere sorting and photonic molecule assembling with high accuracy, and also provide strong evidence for our proposed concept of spectral signature discussed in Section 3.4.

### 3.8. Summary

In this Chapter we investigated coupling properties of photonic molecules consisting of size-matched coupled resonators theoretically and experimentally. We first studied the formation of bonding and antibonding modes in the bi-sphere molecule with FDTD simulation. The amount of mode splitting much larger than the individual resonance width demonstrates strong resonant coupling of two spheres in a touching case. The



dependence of mode splitting on the inter-sphere gap was studied, and the tunability of the coupling strength was realized by adjusting the gap.

Through modeling of bi-spheres with various sizes and refractive index we found that the coupled WGM spectra displays some characteristic patterns which are common for the entire range of variation of parameters used in our modeling. We proposed the concept of spectral signature as a set of spectral characteristics which are stable for a given spatial configuration of the photonic molecule in a sense that it is preserved with varying parameters of the system (such as index and size of the consisting spheres) in a relatively wide spectral range. Such characteristics include the total number of split supermodes and the ratio of splitting for each component. Experimentally we demonstrated the existence of such spectral signatures by assembling various molecules with pre-selected polystyrene microspheres with 0.05% size deviation in aqueous environment and measuring the transmission with tapered fiber positioned in contact at the molecules' equatorial planes. Characterized spectral signature can be used as a unique pattern to recognize molecule's spatial configuration in spectroscopy and find applications in position sensing as well as counterfeit technology.

Spatial distribution of the supermodes was also studied in simulation. Interesting coupling and transport phenomena such as Bloch mode formation, non-circular field pattern, coupling through medium, and peculiar propagation routes were observed and discussed. Further fundamental study of phase matching and coupling properties is required to give more rigorous explanations. Photonic molecules also show $Q$-factor enhancement for some of the antibonding modes and a tendency for mini-band formation with increasing number of photonic atoms. The latter tendency manifested itself due to



the appearance of a series of closely-spaced supermodes. Therefore the spectrum of the molecule can be engineered, providing additional freedom in design of lasing devices, narrow-line filters and multi-wavelength sensors. The E-field enhancement in the medium surrounding certain molecules provides larger detecting volume and higher sensitivity comparing to typical WGMs based sensors of single cavity.

Experimentally we demonstrated an alternative method of tuning coupling coefficient between coupled microspheres. By adjusting the plane of mode excitation, which was realized by controlling the position of tapered fiber, we obtained full tunability of coupling coefficient from the maximum to zero in two touching spheres. By comparing to the results of 2-D FDTD modeling we showed that it is equivalent to controlling the physical gap between two circular resonators from zero to the amount that eliminates coupling. This method is especially useful when coupling coefficients need to be tuned simultaneously among all spheres in a complicated molecule.

Out recent results show that the WGMs resonant enhancement of optical forces can be used to develop sorting technique to obtain resonant microspheres with overlapping WGMs at ~$1/Q$ precision in a large quantity [5-7, 189, 190], which can be realized in both air and liquid medium. These pre-selected microspheres can be used as building atoms to construct complicated 2-D and 3-D photonic molecules whose properties have not been well studied yet. The present work proved the experimental feasibility of assembling a few spheres in various configurations in an aqueous environment that still maintain strong coupling and high $Q$-factor. Large scale array can be formed via liquid assisted self-assembly [203]. Liquid environment is critical for particle movement and biomedical sample survival, however, compact photonic devices based on coupled



microresonators can also be developed in air using high-index (n>1.9) spheres which can possess $Q\sim10^4$ with sufficiently small sizes (D<5 μm) [9].

CHAPTER 4: OPTICAL PROPULSION OF DIELECTRIC MICROSPHERES WITH
EVANESCENT FIBER COUPLER

## 4.1. Introduction

Since the invention of optical [123, 124] and holographic [204] tweezers the use of optical forces for trapping and propelling microparticles has become widely accepted in areas from physics to biology. In recent years this field experienced a tremendous development due to the observation of novel mechanisms of light-matter coupling in optically bounded structures and broad use of integrated optical devices for particle trapping. Engineered optical beams such as Bessel [205, 206] and Airy [207] beams have been utilized. Novel trapping devices such as optical lattices [208] and miniaturized fiber-optics tweezers [209-212] have been developed. And the use of electromagnetic fields in optically bounded structures [213-215] has been explored. These approaches stimulated observations of optical pulling forces [216-218], optical tractor beams [219-221], and stable optical lift [222, 223]. Recently, near-field optical forces have been studied in the chip-scale optical devices integrated with microfluidic systems. Designs based on plasmonic structures [224-226] and photonic crystal cavities [227-229] have been proposed and demonstrated for microparticle trapping.

Studies of optical propulsion have attracted great attention for potential applications in sorting microparticles according to their size, index, or other properties [132, 135, 137, 141, 142, 230]. The propulsion of dielectric microspheres has been studied in liquid-immersed evanescent couplers based on dielectric waveguides [128-137], optical



microfibers [138-145], and prisms [231]. The radiation pressure exerted on particles in such structures is determined by the conservation of the total momentum along the propagation direction. However, due to small reflection and absorption of dielectric particles their propelling forces are greatly diminished in comparison with the classical examples of totally absorbing or mirror-like reflecting particles. The propulsion velocities normalized by the incident power have been found to be below ~1 mm/(s×W) for dielectric microspheres with diameters ($D$) from 1 to 10 μm in previous studies [128-136, 138-145, 231]. For strongly absorbing particles [232] or metallic particle [136, 233-235] the propulsion efficiencies can be increased.

Despite these advancements in this area one of the most important resources of optical manipulation still remains unexplored. It is connected with the use of internal optical resonances in microparticles for enhancing optical forces. Recently, interesting experiments on manipulating polystyrene nanoparticles in a circular motion around silica microspheres have been performed by Arnold *et al*. [58]. The optical forces have been resonantly enhanced due to whispering gallery modes (WGMs) in the microspheres. However the recipient of the optical force, the polystyrene nanoparticle, has been too small to possess internal resonances. The subject of the present work is connected with a reverse situation when the force is enhanced by the resonance in the moving microsphere. Due to inevitable ~1% microsphere diameter variations and the size-dependent nature of WGMs resonances, this effect can be used for sorting microspheres with WGM peaks overlapped at the wavelength of the laser source. This is a highly attractive property for fundamental studies and applications of coupled resonant cavity devices [76, 102, 158, 236, 237]. Although the resonances in optical forces have been observed in microdroplets



by the pioneer Arthur Ashkin more than 30 years ago [1], these effects were weakly pronounced. Some evidence of optical force enhancement due to resonances has been obtained in waveguide couplers [238] and in the case of off-axially shifted focused beams [239]. The notable advance in this area is the theoretical demonstration of high peak-to-background ratios of resonant optical forces in evanescent prism couplers [120, 121]. However, the theoretical limit of resonant optical forces has not been discussed and the realization of these resonant effects has yet to be explored.

In this Chapter, we first present calculations of the magnitudes of the resonant optical forces exerted on circular cavities in a simplified 2D model of surface electromagnetic waves. We show that the magnitudes of resonant propelling forces can approach and even exceed the limits established for totally absorbing particles. After that we present our experimental observations of optical propulsion of polystyrene microspheres in the vicinity of tapered microfibers in water environment. For some of the spheres with $D$ = 15-20 μm we observed giant power normalized propulsion velocities of ~10 mm/(s×W) that exceed the previous measurements in various evanescent couplers by an order of magnitude. These extraordinary high propulsion efficiencies approach estimations at the total absorption limit, indicating that a significant fraction of the guided power is utilized for creating such giant light pressure. In addition, we performed statistical analysis of a large number of optical propulsion events for microspheres with mean diameters from 3 to 20 μm and with ~1% size variations. We explained these results with the concept of resonant enhancement of optical forces due to WGMs excitation in spheres. These effects can be used for sorting microspheres with WGM peaks overlapping in the vicinity of the laser source wavelength $\lambda_0$: $\Delta\lambda/\lambda_0 \delta 1/Q$, where $\Delta\lambda$



is the detuning of WGM resonant wavelength. Taking into account that $Q\sim10^3$-$10^4$ are common for liquid-immersed microspheres [9], it opens up a unique way of selecting the building blocks of chip-scale structures with resonantly coupled WGMs for applications in coupled resonator optical waveguides and coupled cavity devices.

## 4.2. Theoretical Calculation of Resonant Optical Forces

To study the forces that can act on WGMs resonators we use the physical model illustrated in Fig. 4.1(a) [5]. An initial surface wave with frequency $\omega$ is guided by the boundary of the lower half-space with dielectric constant $\varepsilon_m<0$. The upper half-space has refractive index $n_b$. A dielectric cylinder with refractive index $n_s$ is located at a distance $d$ above the boundary. When the surface wave interacts with the cylinder, it can excite the WGMs in the cylindrical resonator. The material parameters used in the simulations are $\varepsilon_m=-2$, $n_b=1$, $n_s=1.4$. Since the material parameters are frequency independent the solution depends only on two dimensionless size parameters $kR$ and $kd$, where $k=\omega/c$ is the wavenumber and $c$ is the speed of light in vacuum. This 2D model captures the essential phenomena such as the excitation of WGMs by an evanescent field, the interaction of the excited mode with the guiding wave, and the creation of the scattered field.

The scattering of a guided wave by a resonator is a complicated diffraction problem. This problem is usually solved by expanding the initial guided wave in terms of the modes of the resonator in free space [14]. However, such an expansion is only an approximation and an accurate solution would require the use of the modes of the combined system, i.e. the waveguide structure and the resonator. This approximation is expected to become less accurate as the distance between the resonator and the boundary decreases. In the propulsion experiments such separation is difficult to control and may



vary significantly. Therefore we adapted a more rigorous approach that would allow us to obtain accurate results in a wide range of distances.

Our solution is based on the surface potential method [240-242] applied to the model shown in Fig. 4.1(a). We reduce the Maxwell equations to the wave equations for the magnetic field which has only one component oriented along the cylinder axis. Then we represent the scattered field outside the cylinder in terms of the single layer surface potential on the surface of the cylinder and Green's function for the two half-spaces. The surface potential is expanded in terms of the angular exponential functions. The total field inside the cylinder is expanded in terms of the cylindrical functions. We match the expansions inside and outside the cylinder by using the continuity condition for the magnetic field component and the tangential electric field component. This matching yields an infinite system of linear algebraic equations for the expansion coefficients. By truncating the system and solving it numerically we find these coefficients, and thereafter obtain the electromagnetic fields inside and outside of the cylinder. In particular, we find the amplitudes of the transmitted and reflected surface waves, as well as the far field radiation. The electromagnetic forces follow directly from the fields.



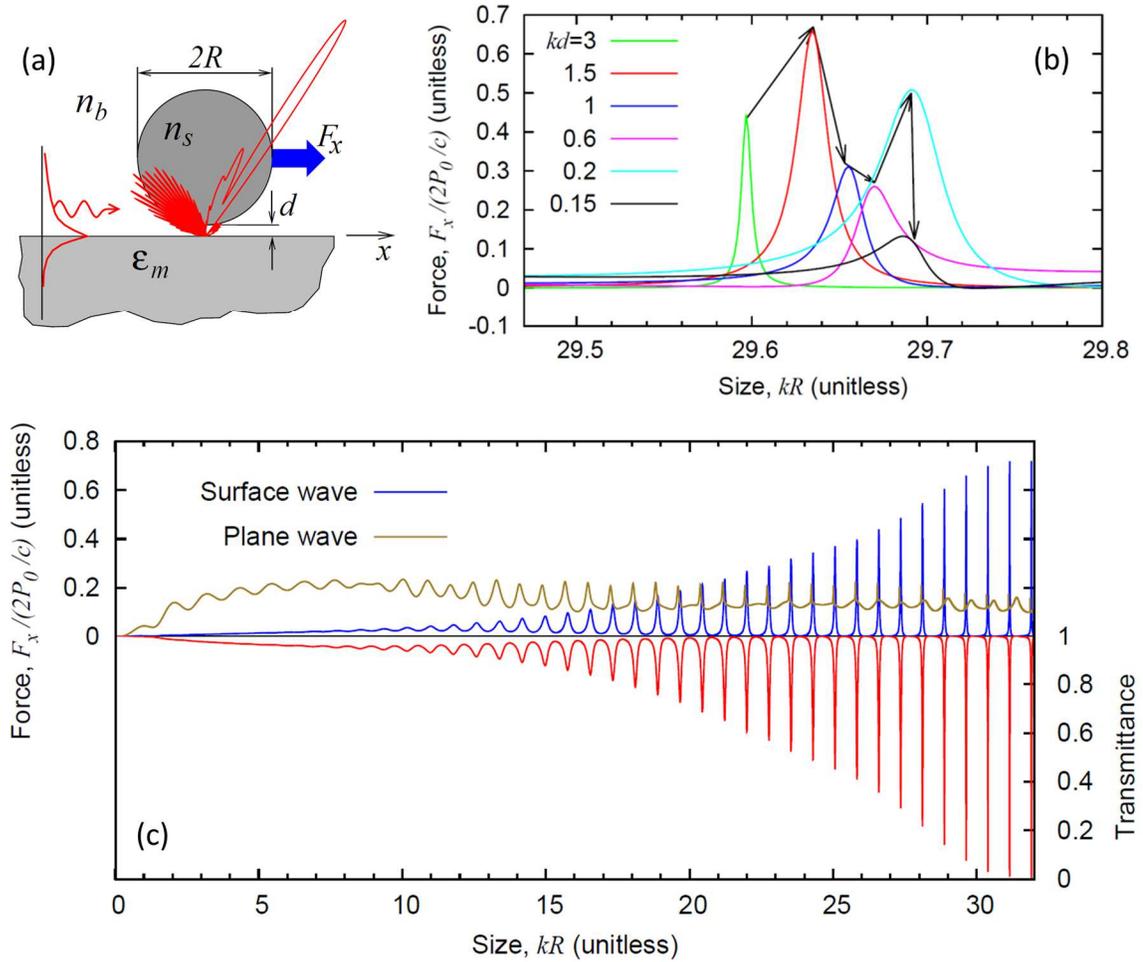

Figure 4.1: (a) Schematic of the illumination of a cylinder by a surface wave with the frequency $\omega$ guided by the boundary of a half-space with $\varepsilon_m < 0$. The scattering of the surface wave (the typical far field directionality of bulk radiation at resonance is shown in red) creates the propelling force $F_x$ along the surface. (b) Size-dependence of the resonant force on the cylinder for various values of the cylinder-boundary separation. (c, top frame) Size-dependence of the propelling force on the size parameter $kR$ for the excitation by a surface wave (olive curve corresponds to $kd$=1.5) and by a plane wave (blue curve). (c, bottom frame) Transmittance for the surface wave (red curve corresponds to $kd$=1.5). [5] The calculations were performed in collaboration with Dr. Alexey Maslov.

We verified the numerically calculated fields by checking the balance between the power of the initial surface waves and the sum of powers of the transmitted and reflected waves and the bulk radiation in the far field. The calculation of forces was verified by



obtaining an agreement between two approaches: by integrating the Lorentz force (with electric and magnetic components) over the cylinder cross section and by integrating Maxwell's tensor outside the cylinder.

A plane wave propagating in vacuum and totally reflecting from a mirror creates a force $2P_0/c$, where $P_0$ is the power incident on the surface. In the case of a partial reflection the force will be smaller, for example, becomes $P_0/c$ for a compete absorption. For surface waves, it is therefore instructive to investigate the ratio of the force and the quantity $2P_0/c$, where $P_0$ is the power of the surface wave, as an indicator of the efficiency of using the surface wave to propel microresonators with WGMs.

A comparison of the force created by a surface wave of power $P_0$ and a plane wave that has power $P_0$ per area of size $2R$ in the transverse direction is presented in Fig. 4.1(c) [5]. In both cases the resonances in forces for sizes $kR > 10$ are present. For the plane wave the peak amplitudes and peak-to-background ratios are limited, while for the surface wave both the peak amplitudes and peak-to-background ratios increase with $kR$ reaching extremely high values. The resonant peaks of forces correlate well with the dips in the transmission spectrum for the surface wave. When the transmission almost vanishes at resonances for large values of $kR \sim 30$, the normalized force can reach a value of around 0.7. This means that the surface wave can propel a transparent cylinder by WGMs excitation more efficiently than a plane wave can propel a totally absorbing cylinder. For a plane wave such a large force would correspond to a significant reflection. For the excitation of WGMs the reflected surface wave is practically negligible and the incident power is distributed between the transmitted surface wave and the bulk radiation. A typical example of the far field directionality of the bulk radiation is illustrated in Fig.



4.1(a). It shows a lobe at ~57° with the direction of the initial wave propagation. However, there is also a significant backward scattering in a range of ~120-150°.

The resonant propelling forces near a selected resonance for various values of distances from the sphere to the surface are presented in Fig. 4.1(b) [5]. Starting from a large separation $kd >> 3$ (not shown in Fig. 4.1(b)), the peak force increases with the decrease of $kd$. The maximum force is obtained at $kd$ ~1.5. The non-monotonic behavior of the magnitude of the optical force at $kd < 1.5$ may be related to interference effects, however this requires a more detailed analysis. An important consequence for optical propulsion experiments consists in an overlap of the calculated force peaks for a range of separations. A similar peak overlap takes place in the spectral domain for a sphere with $kR = 29.6\text{-}29.7$. Once the laser source is tuned into this resonance, the moving particle would experience an enhanced propelling force for a range of separations from the surface that should simplify the experimental observation of this effect. In principle, similar results can be obtained in other evanescent couplers including dielectric waveguides and tapered fibers.

### 4.3. WGMs Excitation in Polystyrene Microspheres by Tapered Fiber

Observation of resonant propelling effects requires the presence of a strong evanescent field in a liquid environment containing microspheres. Tapered microfibers provide a number of advantages compared to other evanescent couplers in such experiments. These include small optical losses on the level of a few decibels, natural integration with fiber-optics based light sources and spectrometers, and the capability to trap and propel spheres in 3D symmetric geometry. The tapered fiber integrated microfluidic platform is schematically illustrated in Fig. 4.2. We obtained adiabatically



tapered fibers by chemical etching of a single mode fiber SMF-28e in a droplet of hydrofluoric acid [9]. This technique allows obtaining tapers with ~1.5 μm diameters and millimeter-scale lengths. The tapered fibers were integrated with a microfluidic platform fabricated using a plexiglas frame as shown in Fig. 4.2. The frame was fixed at the top of a microscope glass plate to create a microfluidic cell with unrestricted optical access from the top. We selected polystyrene microspheres (Duke Standards 4000 Series Monosized Particles, Thermo Fisher Scientific, Fremont, CA) for propulsion experiments because of their ability to float in water due to the fact that the mass density of polystyrene is 1.05 g/cm$^3$ which is very close to that of water.

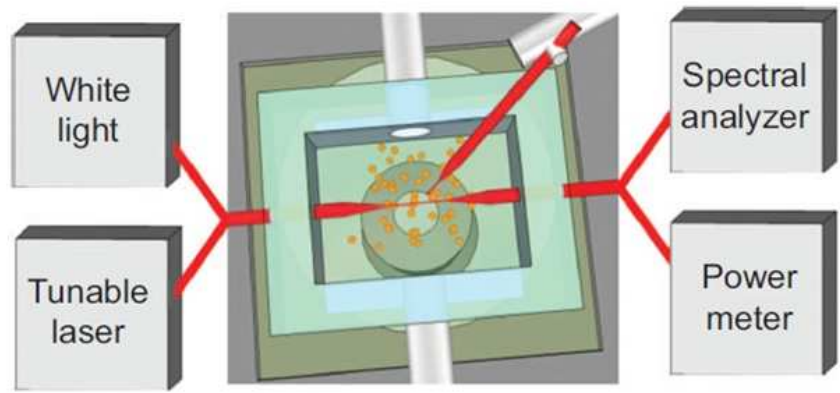

Figure 4.2: Tapered fiber integrated microfluidic platform for optical propulsion. [5, 157]

Observation of resonant propelling effects requires efficient evanescent coupling and excitation of WGMs in the microspheres. The fraction of incident optical power coupled to the sphere can be determined by the depth of the dip in the transmission spectra. The refractive index contrast in water is limited for polystyrene (1.59 vs. 1.33),



therefore spheres with large diameters may be required to obtain pronounced WGMs.

The evanescent coupling between the tapered fiber and polystyrene microspheres of various sizes were studied in the above described fiber-integrated microfluidic platform filled with distilled water [157]. Single microsphere of select size was pickup by a sharpened fiber tip and deposited to the waist of the tapered fiber, as shown in Fig. 4.3(a). The precise positioning of microsphere is possible under a microscope with a hydraulic micromanipulator. We have studied coupling of polystyrene spheres with diameters from 3 to 20 μm and found that for spheres smaller than 5 μm the fiber transmission remain flat thus demonstrating no WGMs coupling.

Figure 4.3(b) shows examples of several measured transmission spectra for spheres of different diameters [157]. The periodic dips in each spectrum correspond to the WGMs resonances excited in the spheres. At the wavelength of the resonance part of the optical power guided through the fiber is transferred into the sphere and eventually get lost. Therefore, the depth of the resonance dip characterizes the fraction of power transferred from the fiber to the sphere. The evolution of spectra in Fig. 4.3(b) clearly shows that the coupling strength increases with the sphere size: while the 7 μm sphere barely shows some resonances, the 20 μm sphere shows pronounced dips with the depths of about 3.5 dB, which indicates that ~ 55% of the guide optical power was coupled into the sphere. We also characterized the WGMs resonances using their $Q$-factors defined as $Q = \lambda/\Delta\lambda$, where $\lambda$ is the resonant wavelength and $\Delta\lambda$ is the resonance's width measured at half of its depth.



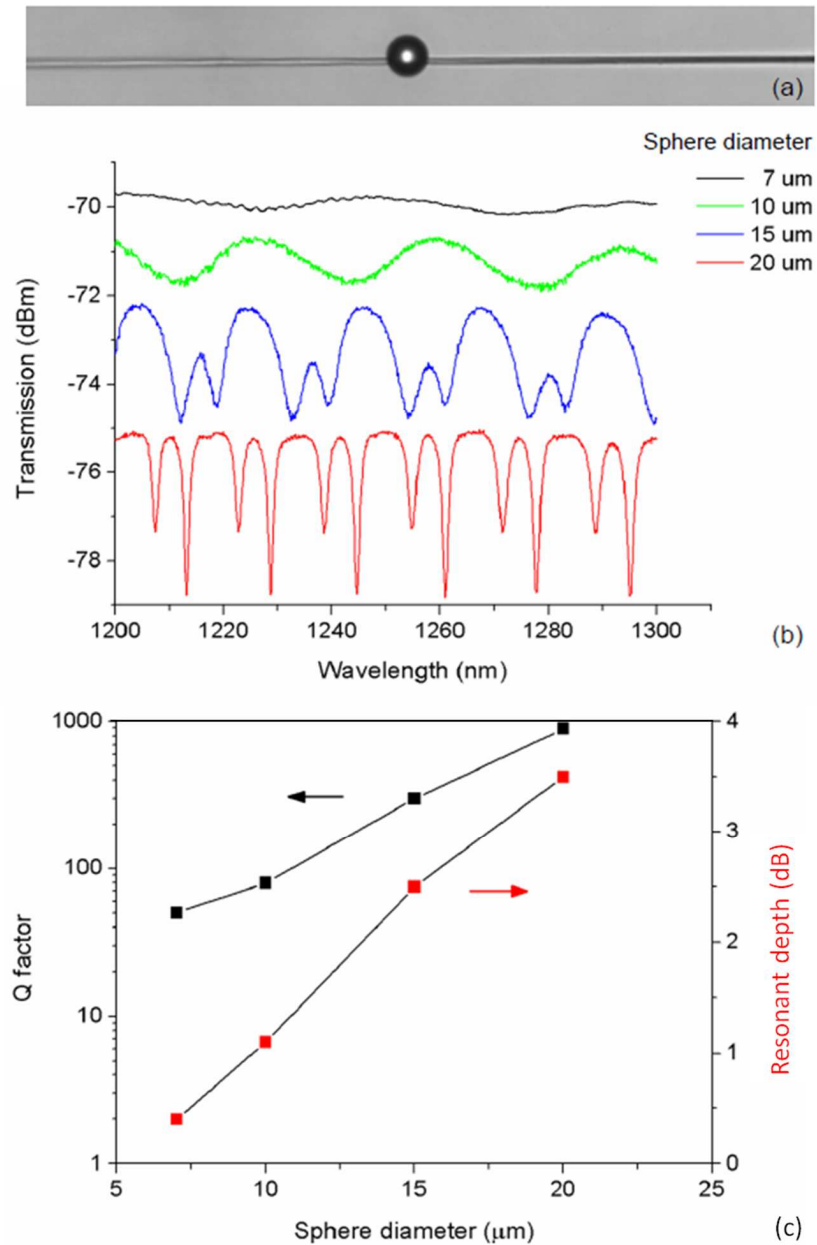

Figure 4.3: (a) Image of a microsphere attached to the tapered fiber for the light coupling investigations. (b) Tapered fiber transmission spectra with 7, 10, 15 and 20 μm polystyrene microspheres attached to the taper waist of ~1.5 μm, respectively. (c) *Q*-factors and resonant depths for 7, 10, 15 and 20 μm polystyrene spheres, derived from the transmission spectra shown in (b). [157]



The coupling depths and the *Q*-factors of the measured resonances as a function of the sphere size are presented in Fig. 4.3(c). It can be seen that the resonance depth (in dB) follows linear dependence on the sphere diameter (*y* axis at right); whereas the *Q*-factors increases exponentially with the sphere size (*y* axis at left). The results of evanescent coupling studies show that although small polystyrene spheres (*D* < 7 μm) cannot possess WGMs in water due to small refractive index contrast, strong WGMs excitation can be observed with spheres of large sizes (*D* > 15 μm). Therefore resonant effects may be present in optical propulsion of large polystyrene microspheres in water.

A more detailed analysis of the WGMs coupled in spheres can be performed by fitting the coupling parameters of a single-mode approximated model [9, 173] described in Chapter 2.

$$P = e^{-\gamma} \times \frac{(\beta - \beta_0)^2 + \left(\dfrac{\gamma}{2S} + \alpha - \kappa\right)^2}{(\beta - \beta_0)^2 + \left(\dfrac{\gamma}{2S} + \alpha + \kappa\right)^2}. \tag{2.3}$$

In Eq. (2.3), $\beta = 2\pi n_s / \lambda$ is the propagation constant, $\beta_0$ is defined as $\beta_0 = 2\pi n_s / \lambda_0$, where $\lambda_0$ is the resonant wavelength. And $\gamma$ is the coupling loss, $\alpha$ is the field attenuation coefficient in the sphere, $\kappa$ is the coupling constant, and $S=\pi D$ is the circumference of the sphere. In a weak coupling regime ($\kappa<\alpha$) which usually takes place for various compact (*D*<20 μm) water-immersed spheres the depth of the resonant dip in transmission spectra increases with the coupling constant $\kappa$.

To identify the range of sphere diameters most suited for observation of resonant propelling effects we determined how both parameters, $\kappa$ and $\alpha$, depend on *D*. Figs. 4.4(a-c) display a typical evolution of WGMs coupling features observed in fiber



transmission spectra for $D$ = 12, 15, and 20 µm spheres, respectively [5]. Due to spherical symmetry, WGMs in microspheres are characterized by radial $q$, angular $l$, and azimuthal $m$ mode numbers [17]. The radial number, $q$, indicates the number of WGM intensity maxima along the radial direction. The angular number, $l$, represents the number of modal wavelengths that fit into the circumference of the equatorial plane. Such waves propagate inside the sphere close to its surface so that they traverse a distance of roughly $\pi D$ in a round trip. The condition of constructive interference in the spherical cavity can be approximated as $\pi D \approx l(\lambda/n_s)$, where $\lambda/n_s$ is the wavelength inside the microsphere. In an ideal free-standing sphere with a perfect shape, the azimuthal modes represented by $m$ numbers are degenerate. However, in practice this degeneracy is usually lifted by small deformations of the microspheres such as uncontrollable ellipticity (~1%) of the real physical spheres. The dips observed in Figs. 4.4(a-c) are inhomogeneously broadened by the partial overlap of different azimuthal modes. Determination of $m$ numbers is not possible in this situation, however $q$ and $l$ numbers as well as the WGMs polarizations, $TE_q^l$ or $TM_q^l$, can be identified for each resonance dips, as labelled in Figs. 4.4(a-c). In order to determine mode numbers and polarizations, the positions of the WGMs resonances were compared in a broad range of wavelengths using the Mie scattering theory [5, 9, 10].

As illustrated in Fig. 4.4(d), the taper-coupled microsphere system operates in a weak coupling regime for spheres with diameter in 12-30 µm range. Both parameters $\kappa$ and $\alpha$ decrease exponentially with the increase of sphere size, with $\alpha$ decreasing at a faster rate. The extension of these two parameters long the trend suggests that a potential critical coupling ($\kappa=\alpha$) may be achievable around $D\sim44$ µm.



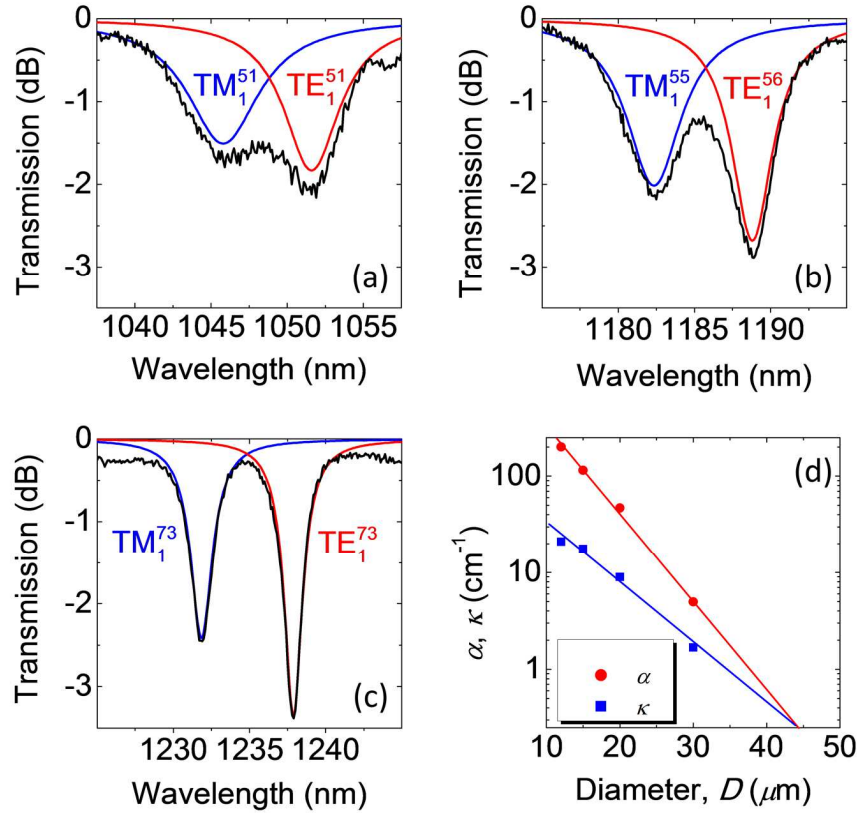

Figure 4.4: (a-c) Transmission spectra of taper with 1.5 μm diameter in contact with single polystyrene spheres with $D$ = 12, 15, and 20 μm in water, respectively. The WGMs numbers and polarizations are labeled. Red and blue curves represent results of fitting with Eq. (2.3). (d) Size dependence of the phenomenological coupling parameters $\alpha$ and $\kappa$ illustrating weak coupling regime with a critical coupling expected at $D{\sim}44$ μm. [5]

Although the maximal resonant optical forces are expected to be seen at critical coupling, these large spheres are too bulky to float and move; and their narrow first order ($q$=1) resonances with estimated $Q{\sim}10^5$ are difficult to use in practical optical propulsion experiments. On the other hand, polystyrene spheres with $D$=15-20 μm and $Q{\sim}10^3$ provide a better trade-off between their compact sizes and efficiency of WGM coupling. The 20 μm sphere shows dips with the depths of about 3.5 dB, which means that ~ 55% of optical power was transferred into the sphere. Assuming approximately uniform



directionality of light scattering, the peak of the resonant force can approach the total absorption limit ($\sim 0.55 \times P_0/c$) in this case.

## 4.4. Optical Propulsion of Microspheres with a Wide Range of Diameters

The conventional approach of studying optical propulsion effects is based on using a laser source and an imaging system to visualize light-induced motions of manipulated particles. The spheres in the micron-scale vicinity of the tapered fiber experiencing the evanescent field of the guided waves are attracted to the fiber by the optical gradient force and are propelled along the fiber due to the optical scattering force. The particles will quickly reach a terminal velocity ($v$) when the scattering force ($F_x$) is equal to the dragging force from the liquid, $C=6\pi\mu Rv$, where $\mu$ is the dynamic viscosity [128, 129, 133, 139].

Due to the focus of our study on resonant light pressure effects, we modified the conventional approach of experimental measurements for optical propulsion by evanescent fields. First, the range of sphere diameter was increased to $3 \leq D \leq 20$ μm including large spheres that could possess sufficiently strong WGMs. In comparison, previous studies mostly dealt with small spheres with $D<10$ μm, which only demonstrated non-resonant radiation pressure. Second, instead of the average propulsion velocities we measured the maximal instantaneous propulsion velocities ($v_{max}=\max(v)$), which represents the strongest force the sphere ever experienced. Moreover, all spheres are inevitably different and thus have random size deviations resulting in different positions of WGMs resonance. Instead of observing the motion of only a few spheres, we studied a substantial amount of spheres for each chosen diameter and the statistical distribution of $v_{max}$ was analyzed.



Optical propulsion of spheres was realized by adding a suspension of microspheres with certain mean $D$ values and ~1% diameter variation to the microfluidic platform illustrated in Fig. 4.2. The suspension of spheres of only one size was used for each experiment. A slow flux (~10 µm/s) was produced perpendicular to the taper by a micropump circulating in a closed microfluidic loop. Since the parameter of individual tapered fiber such as the thickness of the tapered region is difficult to precisely control, all propulsion events were recorded at the same section (~300 µm length) of the same taper. After completing measurements of a given sphere diameter, the microfluidic platform was emptied and cleaned and then filled with another suspension containing spheres with different $D$.

The tapered fiber was connected at one end to a single mode semiconductor laser tunable in the 1180-1260 nm range (TOPTICA Photonics AG, Gräfelfing, Germany) [243]. Due to small absorption and scattering losses (~3dB) in the tapered region we were able to control the total guided power ($P_0$) at the waist of the taper with ~5% precision. The propulsion velocity is expected to have linear dependence on $P_0$ for small spheres with $D$<10 µm [6]. In order to study the dependence of optical propulsion as a function of $D$ we fixed the power at the taper region for all measurements at a modest level of $P_0$=43±2 mW. The laser emission linewidth was narrower than 0.1 nm and is smaller than the width of any WGM resonances studied in this work. The laser wavelength was fixed at around $\lambda_0$=1200 nm, and the results for this study did not strongly depend on the selection of $\lambda_0$. In our experiments the wavelength detuning, $\Delta\lambda=\lambda-\lambda_0$, between the laser emission line and the WGMs resonances ($\lambda$) was realized due to random ~1% deviation of the sphere diameters.



The spheres tend to be separated from the fiber by a nanometric gap during the propulsion attributed to the double layer repulsion between the similarly charged particle and fiber [133]. This gap has been studied in experiments on a WGM based carousel [58], where polystyrene nanoparticles were trapped and propelled in a circular motion around silica microspheres. It has been shown that the particle is radially trapped due to a combination of a long-range attractive force and a short-range repulsive force. The attractive force is similar to the gradient force in optical tweezers. The repulsive force is thought to be from the repulsion between ionized silanol groups on the bare silica surface, and the negatively charged polystyrene nanoparticles. The average gap has been estimated to be about 35 nm. In this work we used significantly larger polystyrene spheres. It is likely that in the course of propulsion the gap sizes varies in a certain range which leads to a variation of the optical force. In addition, such gap size should depend on $P_0$, $D$, and the concentration of ions in a suspension. It is likely that the average size of the gap in our experiments was on the scale of few tenths of nanometers [58, 133]. However, additional studies are required for more precise characterization of that. It plays a critical role in achieving steady propulsion along the fiber since physical contacts with the fiber would retard spheres' motion. The small nanoscale gaps expected in our experiments mean that the results of spectroscopic characterization of WGMs in microspheres obtained with the spheres in contact with the tapered fiber would be applicable for a qualitative understanding of the possible role of WGM coupling in the course of propulsion.



## 4.5. Measurement of Optical Propulsion Velocities

The radiation pressure effects were studied by recording movies [5, 157] of individual propulsion events for each $D$ by an CCD camera (Olympus MicroFire, Olympus America Inc.) equipped with an inverted microscope. The movies consist of a series of snapshot frames with ~5 ms exposure time separated by ~160 ms time intervals. The propulsion velocities can be calculated from these recorded movies. Typical propulsion events for spheres with $D = 7$, 10, and 20 μm are represented by consecutive photos in Figs. 4.5(a-c), respectively [5].

The sequence of snapshots for the 7 μm spheres shows a steady motion with constant velocity, $v_{max} \approx v_{av}$, as illustrated by dashed construction lines in Fig. 4.5(a). We found that optical propulsion with constant velocity was typical for sphere sizes in the range $3 \leq D \leq 7$ μm. The propulsion velocity increases linearly with the size of the sphere in this range. After being propelled over a distance of several hundred microns the spheres eventually depart the taper. This can happen due to fluctuation in the liquid flux or because the spheres moved to a wider section of the taper where the evanescent field is weaker.

The motion of 10 μm spheres is less steady showing deviation of the instantaneous velocity measured between each neighbouring frames from the velocity averaged over a long period of time, as seen in Fig. 4.5(b). The particles can also spiral along the taper in some cases. As shown in Fig. 4.5(b), the sphere was on top of the taper at the beginning where the enlarged image of the taper can be seen in the first snapshot. Then the sphere moved to the side of the taper as seen in the third snapshot. And in the last snapshot the sphere was seen below the taper. The sphere may also rotate accompanying the



longitudinal propulsion and the spiral motion. However, the rotation cannot be observed

with the unmarked spherical objects in this experiment.

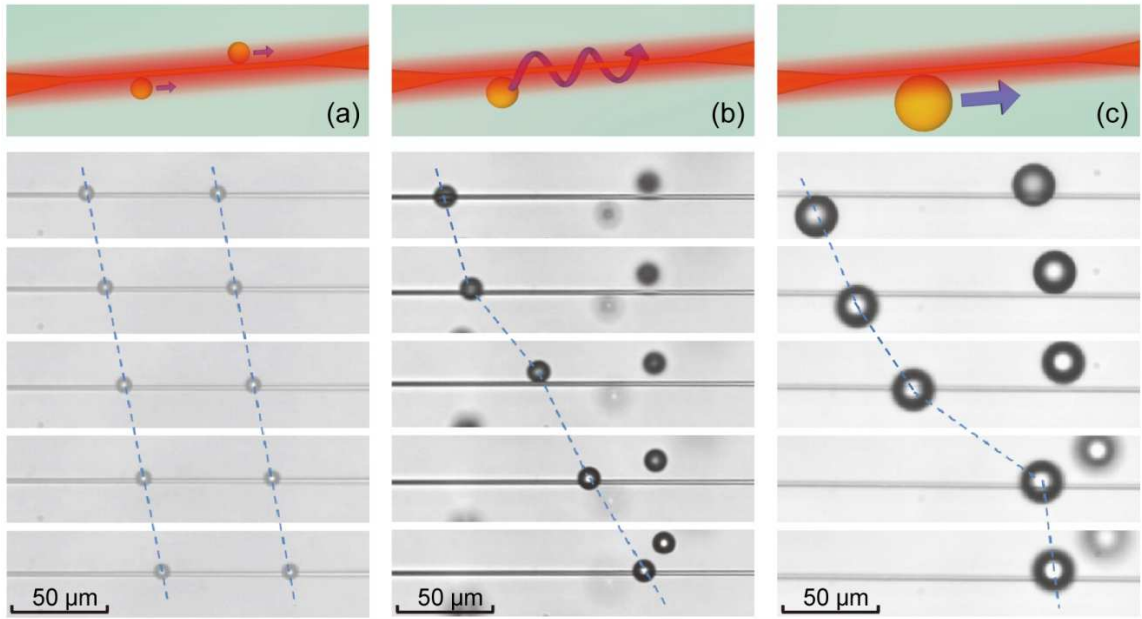

Figure 4.5: Sequences of snapshots taken with 160 ms time intervals illustrating optical propulsion of polystyrene spheres with different diameters: (a) 7 μm, (b) 10 μm, and (c) 20 μm, respectively. Laser light propagates in the tapered fiber from left to right. Inserts at the top of (a-c) schematically show the type of sphere motion represented by the corresponding consecutive photos. Propulsion of 7 μm spheres in (a) is very steady with $v_{max} \approx v_{av}$. Propulsion of 10 μm sphere in (b) shows some variations of the particle velocity. Propulsion of 20 μm spheres in (c) demonstrates giant instantaneous velocity between the third and fourth snapshots reaching $v_{max} \approx 0.45$ mm/s. [5]

For spheres with $15 \leq D \leq 20$ μm the variation in the instantaneous velocity becomes a

dominant factor, as illustrated for $D$=20 μm in Fig. 4.5(c). The explanation of this

phenomenon is connected with the fact that larger particles are more massive thus having

an increased probability of touching the fiber surface, which will cause the sphere to

brake and slow down. There may be other reasons for the seemingly discontinuous



motion of the larger spheres based on rapidly varying resonant effects. It is likely that the spheres were rotating along their own axis during the propulsion. This can lead to coupling to azimuthal modes with varying *m* numbers which can be split due to uncontrollable ellipticity (~1%) of the real physical spheres. The variation of the gap size can be another reason for the discontinuous motion of the larger spheres.

Since we are interested in unrestricted motion of spheres where the light pressure effects are maximally presented, we analyze recorded propulsion movies to find the maximal velocity measured between each neighbouring frames, $v_{max}$, for each propulsion event [5, 244]. In the example shown in Fig. 4.5(c) such maximal velocity is evident with ~70 μm jump for 20 μm sphere between the third and the fourth frame, leading to an extraordinary high velocity of $v_{max}$~0.45 mm/s. Such $v_{max}$ reaches ~60% of the terminal velocity estimated at the total absorption limit: $v=P_0/(3\pi c\mu D)\approx 0.76$ mm/s. In unit normalized by the optical power the measured velocity corresponds to ~10 mm/(s×W) which exceed previously published results for different evanescent couplers [128-136, 138-145, 231] by more than an order of magnitude. Taking into account that conventional optical forces on transparent microspheres cannot exceed a few percent of the force estimated at the total absorption limit, the only plausible explanation for the observed extraordinarily high velocities is based on the mechanism of resonantly enhanced optical force.



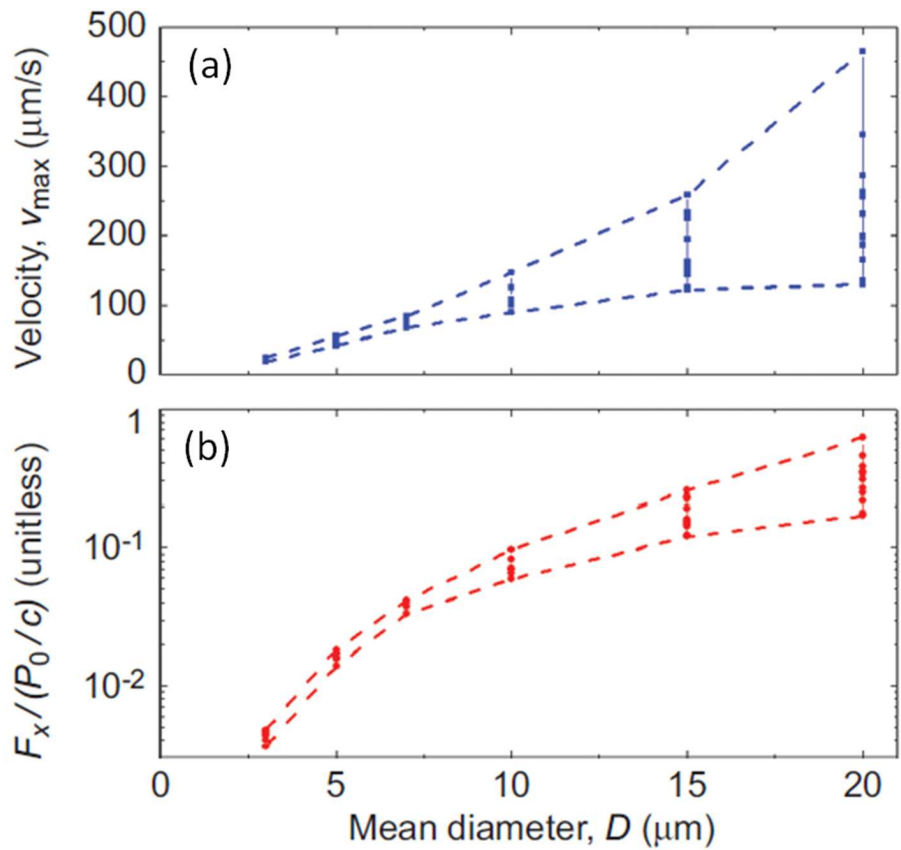

Figure 4.6: (a) Maximal instantaneous propulsion velocity ($v_{max}$) and (b) normalized propelling force $F_x/(P_0/c)$ for polystyrene spheres with various mean diameters, $D = 3, 5, 7, 10, 15,$ and 20 µm. For each mean diameters multiple measurements were performed using 20-30 spheres with random ~1% size variation. [5]

The dramatic difference in optical propulsion of small, $3 \leq D \leq 7$ µm, and large, $15 \leq D \leq 20$ µm, microspheres is demonstrated in details by $v_{max}$ measurements over a broad range of sphere diameters presented in Fig. 4.6(a) [5]. For each mean diameter the measurements were repeated for many spheres with ~1% size variations in the suspension. The purpose of these studies was to see how this size disorder would translate into the difference of propulsion velocities. It is seen that for small spheres the velocity is well



reproducible for each mean diameter ($D$ = 3, 5, and 7 μm) irrespective of the ~1% size variation. The propulsion was steady for all cases and the measured values of $v_{max}$ were very close for different spheres with the same mean diameter. In this range of sphere diameters $v_{max}$ was found to increase almost linearly with the sphere size in agreement with the results of previous studies [128-136, 138-145, 231]. The linear dependence is expected since the non-resonant optical scattering force is proportional to the volume while the drag force is proportional to the cross-sectional area.

For large spheres with $15 \le D \le 20$ μm multiple measurements revealed extremely scattered data with a broad distribution of $v_{max}$ in striking contrast with the case of small spheres. Such behaviour is expected with resonantly enhanced optical forces. As shown in Figs. 4.4(b) and 4.4(c), the WGM resonances with $Q \sim 10^3$ are well pronounced for such spheres. If the laser wavelength matched the positions of their WGM resonances, the propelling force would be resonantly enhanced due to a mechanism demonstrated for a simplified 2D model as discussed in Section 4.2. On the other hand, the non-resonant propelling forces (when the laser line is between WGM resonances dips) tend to be less observable for sufficiently large spheres. Random ~1% diameter variation should lead to a broad distribution (~10 nm) of wavelength detuning between the laser line and WGMs in different spheres. Only a small fraction of spheres with their WGM resonances position overlapped with the laser line are expected to be propelled along the fiber with a resonantly enhanced force. For such spheres the optical forces are expected to display dramatic variations from sphere to sphere depending on the precise amount of small detuning (below ~1 nm) between the laser and the WGMs positions. This would lead to a broad distribution of measured velocities $v_{max}$ and scattering forces $F_x$ for large spheres,



consistent with the results presented in Fig. 4.6(a) and 4.6(b), respectively.

## 4.6. Statistical Analysis of Propulsion Velocities

To study the transition from non-resonant to resonant propulsion effects in a greater detail we analyzed the probability distribution of $v_{max}$ values measured for multiple spheres with ~1% size variation, as illustrated by the histograms in Fig. 4.7 [5]. The maxima of the distribution histogram are normalized. For small spheres with $3{\leq}D{\leq}7$ µm the histograms present relatively narrow Gaussian-like distributions with ~15% standard deviation, as the Gaussian fittings shown in Figs. 4.7(a-c). It indicates that the optical propulsion is only based on the non-resonant mechanism. For 10 µm spheres the distribution becomes much broader which can be interpreted as the transition from non-resonant to resonant propulsion effects.

For 15 and 20 µm spheres the distributions become extremely wide demonstrating velocities varying for different spheres by a factor of four and six in Figs. 4.7(e) and 4.7(f), respectively. It should be noted that, due to limited experimental statistics and limited capability of determining the maximal propulsion velocities from the experimental movies, the shapes of these distributions are not precisely defined in Figs. 4.7(e) and 4.7(f). However, it is apparent that the width and shape of these distributions is significantly different from the narrow Gaussian-like distributions observed for small spheres, indicating that there is a new propulsion mechanism based on the resonantly enhanced optical forces.



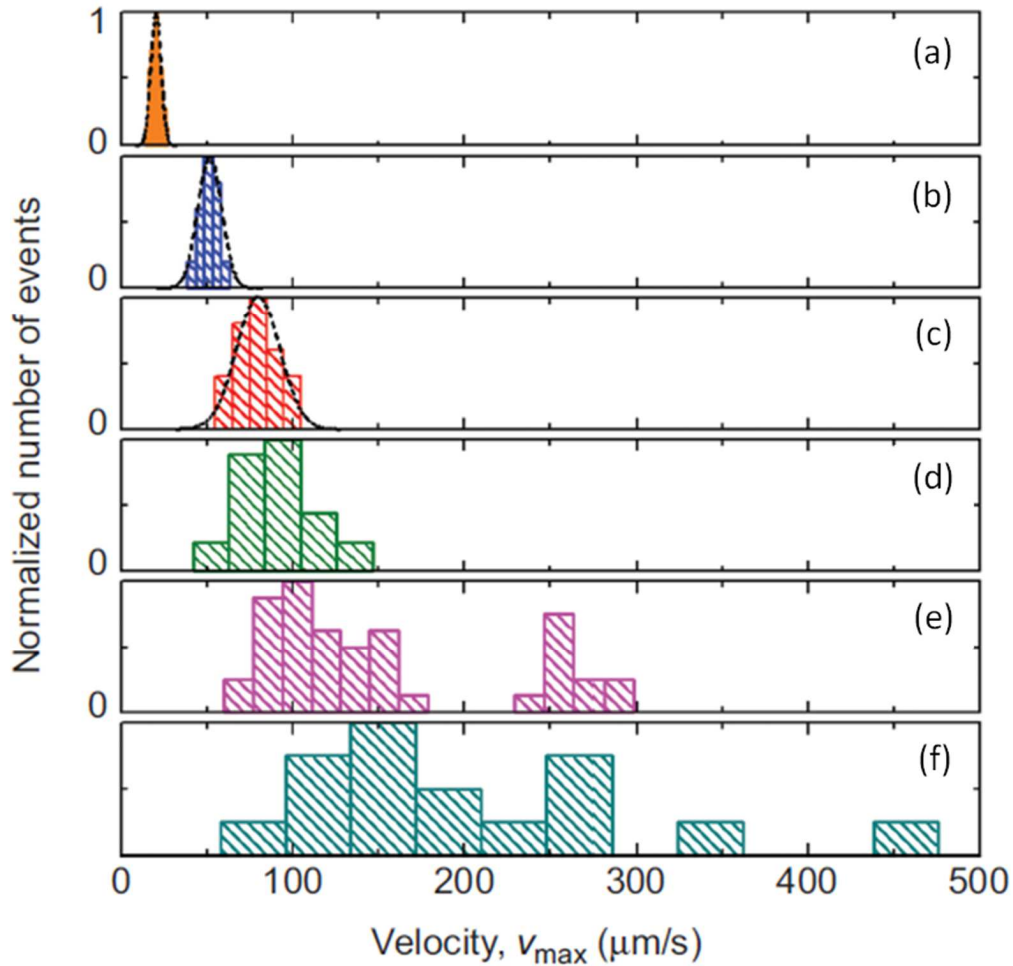

Figure 4.7: Probability distribution histograms for $v_{max}$ values measured for spheres with different mean diameters $D$: (a) 3 μm, (b) 5 μm, (c) 7 μm, (d) 10 μm, (e) 15 μm, and (f) 20 μm. For each mean diameter up to 20-30 measurements were performed using spheres with random ~1% diameter variation. The fitting curves in (a-c) are normalized Gaussian probability density distributions, $f \sim \exp[(v_{max} - v_{max0})^2/2\sigma^2]$, where $v_{max0}$ is the average $v_{max}$ and $\sigma$ is the standard deviation. It can be seen that for small sphere with $3 \leq D \leq 7$ μm the distributions have Gaussian shape with narrow standard deviation $\sigma/v_{max0} \sim 15\%$. For large spheres with $15 \leq D \leq 20$ μm the distributions are extremely wide. The case of 10 μm spheres can be considered as a transition between these two situations. [5]



Such a significant increase of $v_{max}$ in resonant cases can be used for developing novel devices capable of sorting microspheres with WGMs positions overlapped with the laser wavelength $\lambda_0$ with ultra-high precision, $\Delta\lambda/\lambda_0 \delta 1/Q$. Taking into account the ~1% diameter variation in the initial suspension, the WGMs resonances in such selected spheres may have different angular $l$ numbers. However, such WGMs can still be efficiently coupled [101] in structures and devices formed by multiple spheres in a contact position.

## 4.7. Summary

We developed a microfluidic platform integrated with a tapered microfiber for evanescent coupling and optical propulsion investigations. We have studied the taper-to-microsphere coupling properties of polystyrene sphere with diameters from 3 to 20 µm in water immersion. Polystyrene microsphere is ideal for the study of optical propulsion due its similar mass density to water, providing the condition of easy floating and moving. Our results show that despite the small refractive index contrast between polystyrene and water, high-$Q$ (~$10^3$) WGMs can be efficiently excited in large ($15 \leq D \leq 20$ µm) spheres. Up to ~55% of the guided optical power has been shown that coupled into the sphere.

With a modest guided power of 43 mW at the tapered region we observed optical trapping towards the tapered fiber and optical propulsion along the taper for polystyrene spheres of all sizes from 3 to 20 µm. We found that the optical propulsion with constant velocity was typical for sphere diameters in the range $3 \leq D \leq 7$ µm. The propulsion velocity increases linearly with the size of the sphere in this range. Experimentally we observed giant optical propulsion velocities for some of the 15-20 µm polystyrene



spheres. The normalized propulsion velocities measured in our work ~10 mm/(s×W) exceed previous observations by more than an order of magnitude. The magnitude of the corresponding forces reaches 60% of the maximal force estimated at the total absorption limit. We interpret these observations by the resonant enhancement of optical force due to evanescent coupling to WGMs in microspheres. This interpretation is consistent with our numerical calculations of the peak forces in a simplified 2D model of surface electromagnetic waves evanescently coupled to circular cavities. It is also supported by the statistical analysis of the propulsion velocities measured for multiple spheres of different mean diameters with ~1% size variation.

The effects of resonant enhancement of optical forces can be used for sorting microcavities with WGMs positions that are resonant with the wavelength of the laser source with ~$1/Q$ relative precision. By using a tunable laser the spheres with the desired positions of WGM peaks can be selected. Depending on the application, the method of sorting cavites by using resonant light pressure can be a much more accurate and flexible technique compared to standard in-plane fabrication of coupled microrings and microdisks [100, 103, 186]. Microspheres with resonant WGMs can be used as building blocks of photonic devices such as delay lines [102], ultra-narrow spectral filters, laser-resonator arrays [188], waveguides [102, 158, 236, 237], lasers [245], focusing devices [246, 247], microspectrometers [248], and sensors [249]. Such coherent resonant spheres are also required in biomedical applications [250] where they can be used as markers, fluorescent labels, and spectral fingerprints.

CHAPTER 5: SPECTRALLY RESOLVED OPTICAL MANIPULATION WITH
RESONANTLY ENHANCED LIGHT FORCES

## 5.1. Introduction

Resonant enhancement of optical forces exerted on microdroplets has been observed

by Arthur Ashkin and Joseph Dziedzic more than three decades ago [1]. However, there

have not been many following research on this effect because the resonant force peaks

are usually weak and difficult to observe [239, 251]. In recent years, the situation has

changed dramatically since the resonant light forces became the essentials of many cavity

optomechanics effects including dynamic back action [2], carousel trapping [58], and

resonant propulsion [5, 6, 189, 190, 252-255] of microparticles. Theoretically, stronger

resonant force peaks have been predicted for various evanescent coupled systems

[119-121, 147-149].

The knowledge of the shape of resonant features in the spectral response of optical

forces is critical for understanding the mechanisms of various cavity optomechanics

effects. For fixed planar structures such as coupled microring-strip waveguides the force

spectrum displays an asymmetric Fano line shape [256, 257]. For free-moving particles in

the vicinity of evanescent couplers, such as surface waveguides, tapered fibers, or prisms,

the resonant optical forces can be used to spatially separate particles according to desired

resonant properties [7, 8]. The principle of particle sorting is based on extreme sensitivity

of the magnitude of the optical force to the laser wavelength detuning ($\Delta\lambda$) from the

resonant wavelengths of whispering gallery modes (WGMs) in microspheres. The sorting



precision is determined by an inverse of the WGM's quality factor ($1/Q$). Taking into account that $Q \sim 10^3$-$10^5$ is common for spheres of only a few microns sizes, such precision far exceeds the current fabrication capability of ~1% standard deviation for sphere diameters. This method can also over perform the in-plane processing of coupled microdisks [100], microrings [258-260], and photonic crystal defects [261] where the deviation in cavity dimensions typically cannot be made less than ~0.1%. The approach of using resonant light pressure can be applied to sorting microspheres made from various materials, both in air and in liquid environment [7, 8]. Microspheres having identical resonant wavelengths can be used as building blocks for coupled-cavity devices where the disorder and localization effects are strongly suppressed [17]. Examples of such devices are coupled resonator optical waveguides (CROWs) [102, 106, 111, 158], high-order spectral filters [21], delay lines [262], and sensors [159, 250].

The motivation of this work is to understand the basic physics of the resonant propulsion effects, and for this reason we focus on the polystyrene microspheres whose mass density is very similar to water, which facilitates optical manipulation of spheres in a water environment. In Chapter 4 we have demonstrated efficient evanescent coupling to WGMs in large polystyrene spheres with diameter D>15 μm which allows us to study the resonant properties of optical forces. It should be noted that the forces exerted on particles in a liquid can be rather complicated involving: i) optical propelling forces along the evanescent coupler, ii) liquid drag forces, iii) transverse forces balanced for stable particle separation from the coupler surface, iv) possible rotational forces. An additional factor is photophoretic force caused by non-uniform heating of the particles. For absorbing metallic particles, the photophoretic forces can be much stronger than the



optical forces [232]. For dielectric particles, however, the negligible absorption significantly diminishes the role of photophoretic forces.

Optical propulsion of dielectric microspheres has been studied in various evanescent field couplers [128-145, 231]. However, the effect of resonant enhancement of the optical forces was too weak in these experiments. The ultra-high peaks of resonant forces observed in propulsion experiments with tapered fiber coupler has been discussed in Chapter 4 [5]. However, the detuning between the laser line and WGMs in different spheres was varied randomly in that study due to the 1-2% size deviation of microspheres.

In this work, we directly measured the spectral shape of the peaks of the resonant optical forces exerted on microspheres. It was achieved by a precise control of the wavelength detuning between the laser line and WGMs in individual spheres. For small detuning ($|\Delta\lambda|<\lambda_0/Q$, where $\lambda_0$ is the WGM resonance wavelength), we demonstrated the existence of a stable radial trapping of microspheres near the tapered fiber in the course of millimeter-scale propulsion. In contrast, for larger detuning we observed the lack of radial trapping. We demonstrate a very broad and weakly pronounced spectral peak in the force for 10 µm polystyrene spheres and strong and sharp force peak with $Q>10^3$ for 20 µm spheres. We show that the shape of the peak of the spectral response of propulsion forces replicates, with the opposite sign, the shape of the dip in fiber transmission spectra. We observed significantly higher magnitude of the peak forces compared to previous measurements [5] due to controllable resonant conditions and showed that the peak force reaches the total absorption limit ($F\sim P_0/c$) for the light momentum flux.



## 5.2. 2-D Theoretical Modelling and Discussion of 3-D Case

We first consider a two-dimensional (2-D) model to explain the origin of longitudinal ($F_x$) and transverse ($F_y$) forces exerted on microspheres in evanescent couplers. The scheme of the model is shown in the inset of Fig. 5.1 [6]. Our approach and results bear similarity with previous force calculations performed for different evanescent couplers [119-121, 147-149]. We consider a thin dielectric slab coupled to a cylindrical resonator [263].

This model is the closest 2-D analog of our experiments with tapered fiber couplers. It should be noted, however, that the splitting of WGMs well-known for 3-D real physical spheres cannot be accounted for by this 2-D model. WGMs in microspheres are characterized by three modal numbers, $q$, $l$ and $m$ [17]. The radial number $q$ represents the number of the intensity maxima in the radial direction. The angular number $l$ represents the number of wavelengths fitted in the equatorial plane. The azimuthal number $m$ takes ($2l+1$) values from $-l$ to $+l$ and represents the effective spread of the mode away from the equatorial plane towards the poles of the microsphere [20, 75, 76, 112, 264, 265]. In a perfect free-standing sphere, the azimuthal modes are degenerate. However, in real experimental situations the degeneracy of azimuthal modes is often lifted. It can take place due to elliptical deformation of microspheres. Spectral overlap between several simultaneously excited azimuthal modes usually leads to dramatically decrease of measured $Q$-factors compared to that from single-mode calculations. Another factor of WGMs broadening is related to the coupling with the waveguide [189]. Despite these broadening mechanisms, the experimental linewidth of WGM resonances in the individual spheres can be two orders of magnitude narrower (for $Q\sim10^4$) than the typical



variations of WGM positions in suspensions of microspheres with ~1% size dispersion. This opens a possibility to sort spheres resonant with the laser source with $\Delta\lambda/\lambda_0 \sim 10^{-4}$ accuracy. In this work we considered the case of perfectly circular cavity as a simplest example illustrating the origin of the resonant optical forces. The closest 3-D analogy of our 2-D calculations would be perfect free-standing spheres with degenerate azimuthal modes. For such spheres the $q$ and $l$ numbers are analogous to corresponding numbers in the 2-D geometry of circular resonators.

As illustrated in the inset of Fig. 5.1, the incident mode at frequency $\omega$ is guided by the slab with thickness $L$ and refractive index $n_g$. The scattering cylinder has radius $R$ and refractive index $n_s$ and is separated from the slab surface by distance $d$. We choose $n_g = n_s = 1.30$, $kd = 1.5$ ($k=\omega/c$) and assume vacuum environment ($n_b =1$). The chosen refractive indices provide an index contrast of $n_s/n_b=1.3$, which is higher than that in our experimental cases (~1.2). By numerical modeling, we checked that the results obtained for index contrasts of 1.3 and 1.2 are qualitatively similar. But lower contrast requires using larger $R$ to achieve similar high $Q$ which resulted in a significantly increase of computational time.

The Maxwell equations in 2-D can be split into two independent sets: transverse electric (TE) and transverse magnetic (TM). We consider only the TM case. The number of modes supported by the slab increases with the size parameter $kL$. For $n_g = 1.30$, the slab supports only one TM mode if $kL<3.78$. We take $kL = 2$ that gives the phase index of this mode $n_{ph} = c/v_{ph} \approx 1.074$, where $v_{ph}$ is the phase velocity of the mode.



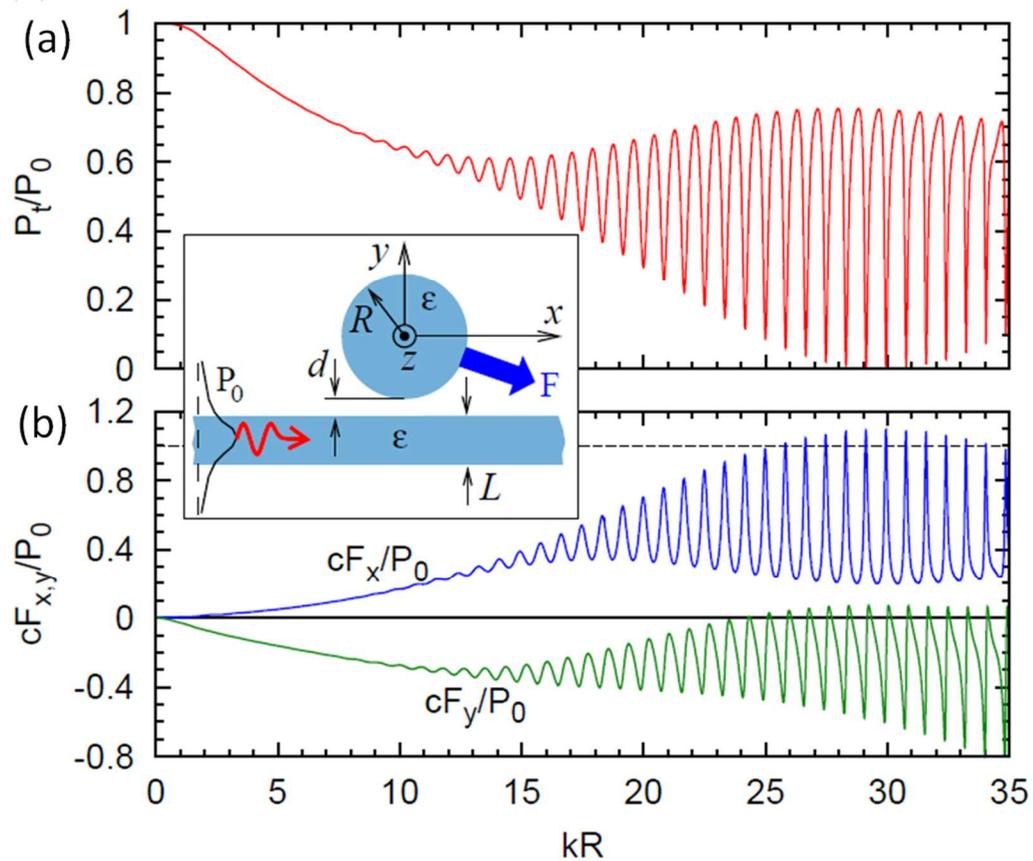

Figure 5.1: (a) Transmitted guided power $P_t/P_0$, (b) longitudinal force, $cF_x/P_0$, and transverse force, $cF_y/P_0$, as function of the particle size parameter $kR$. Inset illustrates the geometry and parameters of the 2-D model. [6] The calculations were performed in collaboration with Dr. Alexey Maslov.

Under resonant conditions with WGMs, the guided mode of the slab is coupled to the resonator. The coupling to WGMs results in the reduction of the transmitted power, creation of the reflected guided mode as well as of bulk radiation that escapes to the upper and lower half-spaces. The model accounts rigorously for the excitation of the WGMs as well as the subsequent re-radiation of the energy into guided and bulk waves. The power contained in the reflected guided mode is also enhanced at frequencies of the WGM resonances but still remains several orders of magnitude smaller than the incident



power for the parameters ($kR$~20-30) used in this work. To find the distribution of the bulk radiation and the optical forces, we use the analytical theory that was previously developed for a cylinder interacting with the surface wave of a plasma half-space [189] or with a slab waveguide [263]. The theory is based on representing the scattered fields that expand outside the resonator by effective surface currents and by utilizing a Green's function for a source located near the slab. The fields inside the cylinder are expanded in terms of cylindrical functions. Matching the fields at the boundary of the cylinder gives the effective currents and the fields inside the cylinder. To adopt the formulas in Ref. [189] to present model, we replaced the corresponding Green's function with the one that describes the emission near the slab. The calculated fields can then be used to find the propelling and trapping forces using either the Lorentz formula or the Maxwell stress tensor.

The dependence of the transmitted power and forces on the parameter $kR$ is shown in Fig. 5.1. For small $kR$, the transmitted power decreases monotonically with $kR$ due to increased scattering of light by the cylinder. In this limit, the cylinder is too small to support sufficiently high-$Q$ WGMs and can be considered as a wavelength-scale scattering particle. From $kR$>15 the transmission starts to exhibit narrow resonances dips. The dips are due to the excitation of the first order ($q$=1) WGMs in the resonator. At 27<$kR$<32, the transmission loss at the dips can reach almost 100%. This situation corresponds to a critical coupling condition when the power is coupled to WGMs in microspheres almost completely and subsequently scattered in the surrounding space. The fraction of the power reflected in the backward direction is too small to contribute markedly to the balance of photon fluxes. Despite the large total power of the bulk



radiation, its momentum flow along the incidence direction is rather small due to the lack of directionality [189, 263]. Therefore, when on resonance with WGMs the momentum flow, $n_{ph}P_0/c$, of initial waveguide mode is transferred almost entirely to the cylinder and the normalized propelling force peaks reach the values $cF_x/P_0 \approx 1.1$ as presented in the calculated results in Fig. 5.1(b) for $27 < kR < 32$. The peaks positions of the propelling force, $F_x$, also correlate well with the dips in the transmission spectrum.

Unlike the propelling force that exhibits well defined resonant peaks, the transverse (trapping) force $cF_y/P_0$ has a more complicated behavior. Its spectral features are clearly resonant in nature, but highly asymmetric. A phenomenological theory shows that the transverse force on the resonator can be decomposed into two terms: symmetric and anti-symmetric [266]. Depending on specific parameters, their combination gives an asymmetric line shape. Similar line shapes have also been reported in Refs. [119-121, 148, 149, 256, 257]. The results presented in Fig. 5.1(b) show that the transverse force can be attractive ($F_y < 0$) and has a resonant nature. However, the repulsive peaks of the resonant optical forces have also been reported in previous studies [121]. Generally, the spectral shape, magnitude and direction of the transverse forces are sensitive to the specific structural parameters ($k$, $R$, $n_s$, $n_g$ and $d$), and require calculations in specific cases.

Most of these considerations should also be applicable in a 3-D geometry of tapered microfiber coupled to microspheres realized in our experimental work. We should note, however, that in a 3-D geometry of tapered fibers the non-resonant optical forces should be additionally suppressed relative to the resonant peaks. Away from the resonance, the microsphere experiences only a fraction of the total power carried by the evanescent field



with radial symmetry. On the other hand, close to critical coupling, the optical power can still be passed from the taper to the dielectric microsphere almost entirely. This means that in the 3-D geometry of microsphere-to-microfiber couplers we expect well pronounced negative (attractive) force peaks with weaker forces in between the peaks compared to calculations in 2-D geometry. The splitting of azimuthal WGMs discussed earlier can also lead to broadening of the peak forces in 3-D geometry due to overlap of the peaks associated with different azimuthal modes.

### 5.3. Experimental Apparatus and Procedure

The experimental setup is illustrated in Fig. 5.2 [6]. We select the tapered microfiber for propulsion experiments with microspheres for several reasons. It allows fabricating extremely thin tapers (micron-scale diameter) with a large fraction of power guided as evanescent waves. The tapers obtained by wet etching are only a few millimeters long and when fixed in a platform it avoids problems such as accidental breaking. They also allow readily connecting to fiber-integrated devices including white light source, tunable laser source, spectrometer, and photodiode.

The propelling force can be determined due to the fact that the terminal velocity ($v$) of the particle is reached when the propelling force ($F_x$) is equal to the drag force, $3\pi\mu Dv$, where $\mu$ is the dynamic viscosity. Thus, the goal is to measure $v$ and calculate $F_x$. The propulsion of small ($D<10$ μm) polystyrene particles is a steady process, and the measurements of $v$ are straightforward [5]. For larger spheres ($15<D<25$ μm), however, the WGM resonances become more pronounced ($Q>10^3$) resulting in significant increase of $v$ for some of the on-resonance spheres [5]. The propulsion also becomes less steady for larger spheres. This behavior may have different explanations including possible



additional friction caused by touching of the taper by more massive spheres or by variations in the coupling efficiency. In practice, it means that only the maximal propulsion velocity should be considered as a measure of the optical forces for unrestricted motion of spheres.

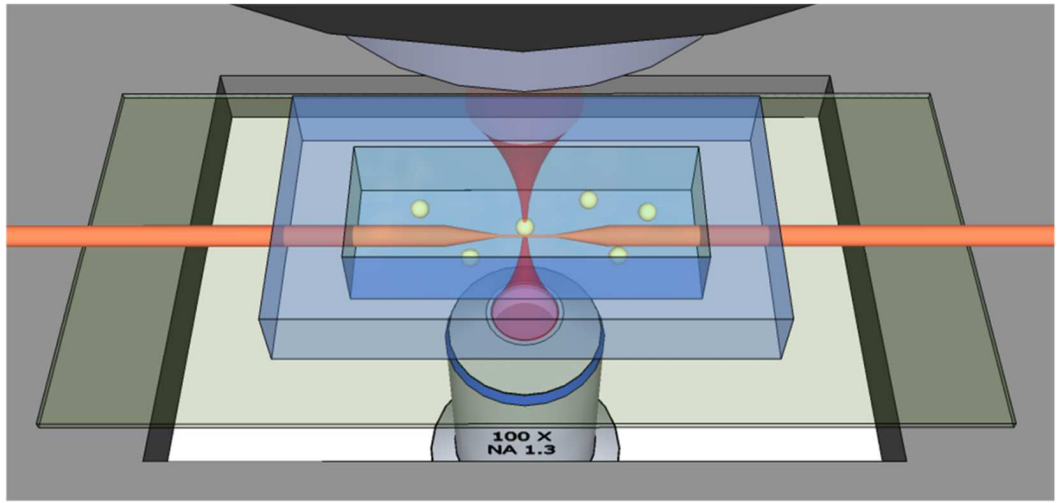

Figure 5.2: Overview of the experimental platform with tapered fiber coupler illustrating the manipulation of microspheres by the optical tweezers. [6]

In our previous studies, we analyzed recorded movies frame-by-frame to find the maximal propulsion velocities of different spheres [5]. We relied on slight variations of the diameter from sphere to sphere which translated into random variation of the detuning $\Delta\lambda$ between the laser and WGMs in the spheres. So that some of the spheres turned out to be on resonance with the laser operating at a fixed wavelength.

In this work, we developed a direct control of parameter $\Delta\lambda$ for each sphere. It required two modifications of our experimental apparatus. First, the addition of optical



tweezers to our tapered fiber platform, as shown in Fig. 5.2, enhanced our manipulation capability. Optical tweezers can precisely trap and move microspheres in an aqueous environment. In particular, an individual sphere can be trapped and brought to the vicinity of the tapered fiber. Second, spectroscopic characterization of WGMs in individual spheres was utilized. For transmission measurements the fiber was connected to a broadband white light source (AQ4305; Yokogawa Corp. of America, Newnan, GA, USA) and an optical spectral analyzer (AQ6370C-10; Yokogawa Corp. of America).

The microfluidic platform sketched in Fig. 5.2 was fabricated using a plexiglas frame with a single mode fiber SMF-28e fixed in a sidewall of this frame [5, 9]. The center of the fiber was etched in a droplet of hydrofluoric acid to a waist diameter of ~1.5 μm with several millimeters in length. The frame was sealed with a 100 μm thickness glass plate at bottom and filled with distilled water. The spheres used in our experiments are polystyrene microspheres (Duke Standards 4000 Series, Thermo Fisher Scientific, Fremont, CA, USA) with refractive index of 1.57 at $1.2<\lambda<1.3$ μm and mass density of 1.05 g/cm$^3$. Such density is very close to that of water, which is an important condition for observation of sustained optical propulsion along the taper. The platform was placed on an optical tweezers set-up built by an Nd:YAG laser and an oil-immersed objective lens with 1.3 numerical aperture (NA). White light illumination was provided from the top and imaging was realized with the same objective and a CMOS camera.

The sequence of our propulsion experiments is illustrated step-by-step in Figs. 5.3(a-c) [6, 252, 253]. First, the WGM resonances were characterized for a specific trapped sphere, as shown in Fig. 5.3(a). While the sphere was held by the optical tweezers in close vicinity to the taper, white light was coupled into the fiber and



transmission spectrum was measured. The spectrum in Fig. 5.3(a) shows two sets of first-order ($q$=1) WGMs with different polarization, $\text{TE}_q^l$ and $\text{TM}_q^l$.

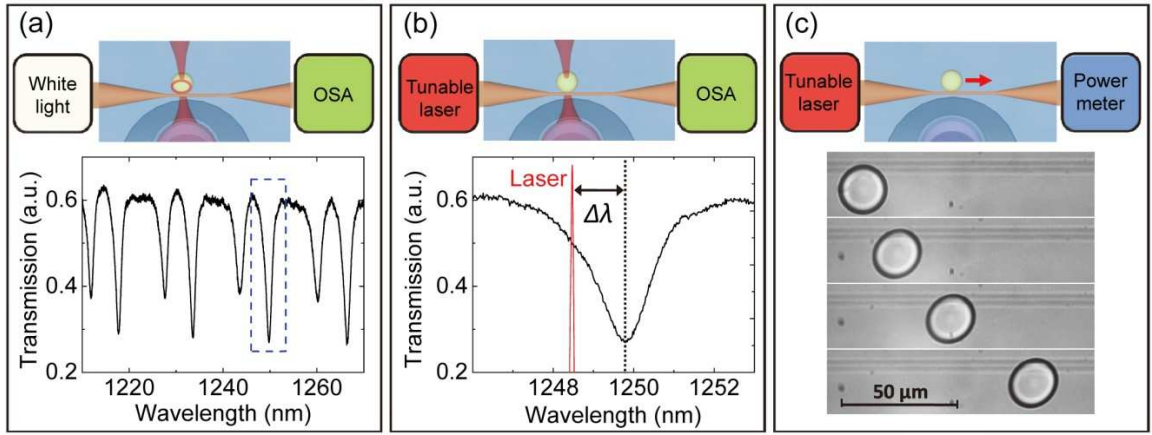

Figure 5.3: The sequence of experimental procedures: (a) Characterization of WGMs in a given sphere by fiber-transmission spectroscopy; (b) Setting the detuning of the laser emission line (narrow peak), $\Delta\lambda$, from the center of the dip in transmission spectrum; (c) Propelling of the 20 µm sphere along the tapered fiber represented by the snapshots taken with 100 ms time intervals. [6]

Second, the wavelength detuning $\Delta\lambda$ was set, as shown in Fig. 5.3(b). To one end the fiber was coupled to a single mode tunable semiconductor laser (TOPTICA Photonics AG, Grafelfing, Germany) operating in 1180-1260 nm wavelength range. The optical power in the tapered region was limited to ~18 mW. Fig. 5.3(b) illustrates a blue shift $\Delta\lambda$ of the laser emission line relative to the resonance dip. Since the polystyrene microspheres used in our experiments have a size deviation of 1-2%, the WGMs resonant wavelengths are shifted from sphere to sphere. Therefore, we selected for each sphere a well-pronounced resonance dip in the transmission spectrum and measured $\Delta\lambda$ from its center.



Third, we switched off the optical tweezers beam to release the sphere and recorded its motion of possible propulsion along the taper. We found, however, that the stable propulsion was observable only for small detuning ($|\Delta\lambda|<\lambda_0/Q$, where $\lambda_0$ is the average wavelength) for 20 μm diameter spheres, as an example illustrated in Fig. 5.3(c).

### 5.4. Stable Radial Trapping of Microspheres

This Section is devoted to study transverse forces which are extremely important for achieving stable propulsion along the taper. In the course of propulsion the microspheres need to be separated from the taper by a small liquid gap. We first summarize main results obtained previously in a plane-parallel geometry for spheres moving in water in the vicinity of a totally internally reflecting prism [119-121, 147-149]. Stable transverse trapping (perpendicular to the surface of the prism) requires a combination of long-range attraction with a short-range repulsion. The long-range attraction has been attributed to the gradient optical forces which have ~100 nm spatial extents determined by the exponential decay of the evanescent field [119-121, 147-149]. In addition, gravity provided downward force bringing the spheres closer to the surface of the prism. The short-range repulsion was ascribed either to electrostatic double-layer effect with the Debye screening length of about 40 nm or to radiation pressure effects caused by the imperfection-induced scattering of light from the dielectric interface [148].

As discussed earlier in Section 5.2, the transverse force calculated using a simplified 2-D model is not fully applicable to 3-D microfiber geometry. However, qualitatively we expect a similar interplay of forces, with the peaks of attractive force on resonance with WGMs in spheres combining with short-range double-layer electrostatic repulsion to create radial trapping of microspheres in the course of their long-range propulsion [5].



This scenario, however, only takes place for resonant or near-resonant conditions, $|\Delta\lambda| < \lambda_0/Q$.

The situation away from the resonance is more complicated since the transverse optical forces are weakened in the 3-D microfiber geometry case, as discussed in Section 5.2. For the same optical power in the taper, the depth of the radial optical trap can be reduced in non-resonant situations. It should be noted that a distinction between the resonant and non-resonant case is possible only for sufficiently large spheres with high $Q$-factors of WGMs.

Experimentally, we realized resonant and non-resonant situations using polystyrene spheres with $D$=20 μm with $Q$~$10^3$ [6, 267]. Two typical scenarios of behavior of spheres corresponding to WGMs off-resonance or on-resonance with laser line are presented in Figs. 5.4(c) and 5.4(d), respectively. The comparison of these two cases demonstrates completely different response of spheres after release. Under off-resonance condition the propulsion effect was not able to observe. As seen in Fig. 5.4(c), the sphere slowly drifted away from the taper likely due to fluctuations of the background flux. When the sphere separated from the taper by a few microns, it could no longer experience the gradient trapping force that only exists in the near field vicinity. After separating from the fiber, the sphere slowly descended to the bottom due to gravity. For large detuning, $|\Delta\lambda| > \lambda_0/Q$, the optical gradient force appears to be not sufficient to hold the sphere around the fiber. In contrast, at the same guided power limited at 18 mW under conditions of small detuning ($|\Delta\lambda| < \lambda_0/Q$) we observed stable propulsion effect. As illustrated in Fig. 5.4(d), the sphere was radially trapped around the taper when WGM in sphere on resonance with the laser line. Stably trapped microspheres became subjects of resonantly enhanced



propelling force and would move along the taper with high velocity.

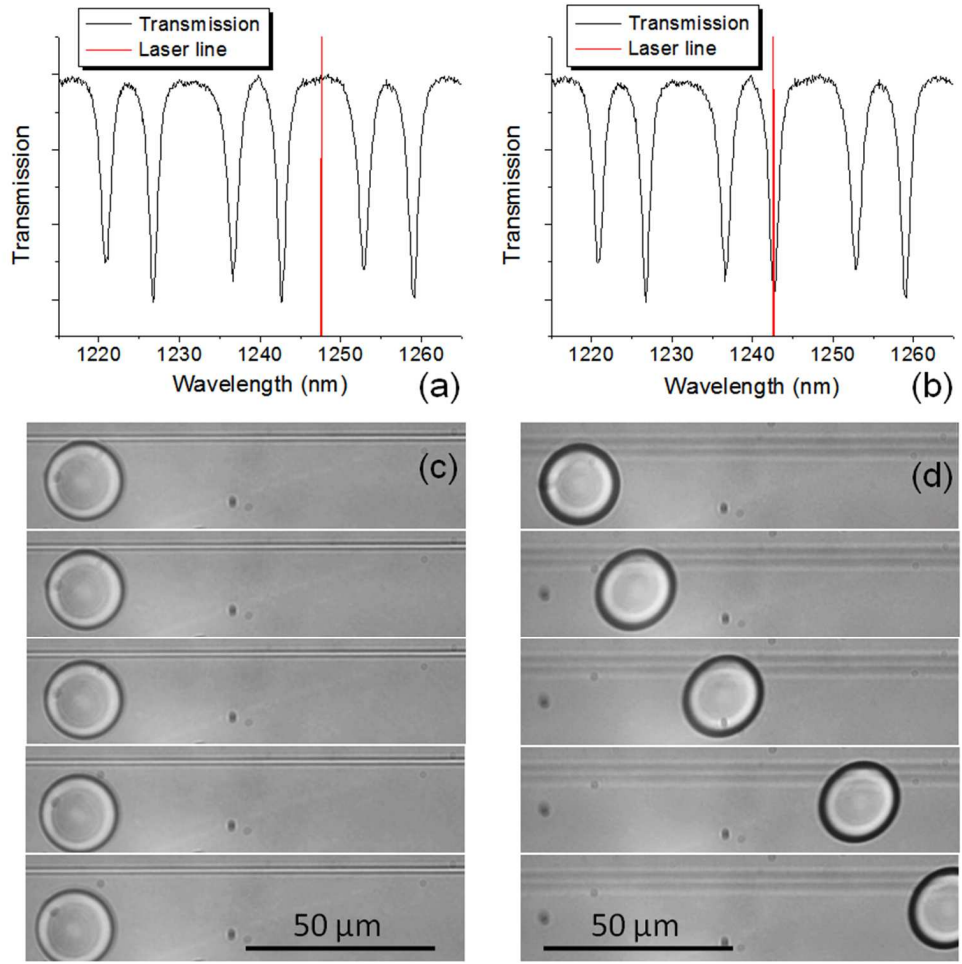

Figure 5.4: Transmission spectra of the fiber-microsphere coupler and the positions of laser line in (a) off-resonance and (b) on-resonance situations. Sequence of snapshots separated by 100 ms time intervals illustrating motion of microspheres after release from optical tweezers for (c) off-resonance and (d) on-resonance cases. The sphere appears as an ellipse in (d) because of the image distortion induced by the CMOS camera for rapidly moving objects. [267]

To study the strength and stability of the radial trap in resonant cases we set the detuning to zero ($\Delta\lambda$=0) and performed synchronous measurements of the propulsion



velocity and the total transmitted power through the fiber in the course of the sphere propulsion. The idea of this experiment is that the transmitted power can represent as a measure of the gap separating the sphere from the taper. In the weak-to-critical coupling regime, smaller transmitted power implies a smaller gap between the sphere and the fiber surface [5, 9].

For these studies, we slightly modified the sequence of experimental procedures to synchronize the measurements of the instantaneous propulsion velocity and the transmitted optical power. The power meter in Fig. 5.3 (c) was replaced by a photodiode connected to an oscilloscope. Before releasing the sphere near the taper, the tunable laser beam was first blocked. As a result, the sphere tended to slowly move away from the taper after release. The tunable laser beam was immediately opened, however, usually came with ~0.1-0.3 s delay which could translate into gaps of <0.5 μm between the sphere and taper as an initial condition for the propulsion process.

Motion of spheres were recorded by the CMOS camera with average 50 ms time intervals between the frames and exposure time per frame smaller than 10 ms. By reviewing these movies frame by frame, we can calculate the instantaneous velocity of sphere moving along the taper between each frame. The recording of photocurrent was triggered by the tunable laser signal with negligible delay. In this method, we are able to synchronize photocurrent and instantaneous propulsion velocity and plot them as a function of time in the same graph.

The results of simultaneous measurements of the propulsion velocity and the transmitted power are presented in Fig. 5.5 for two representative cases [6]. They were selected from 40 different propulsion cases. A trigger signal threshold was set so that the



oscilloscope would start recording at the same time when the laser for propulsion was turned on. An initial abrupt increase of the transmission was seen in all case, which represents the time of switching on the tunable laser beam. The high transmitted power at the beginning indicates weak coupling due to a gap between the sphere and the taper, which is the initial position of the sphere. The following gradual decrease of the transmission on the period of 100-200 ms can be explained by reducing gap between the sphere and taper resulting in stronger coupling. It is seen that the propulsion velocity was gradually increasing at this stage showing amplified propelling force due to increased coupling strength.

As shown in Fig. 5.5(a), after that period a relatively stable small level of photocurrent in combination with a constant level of high propulsion velocity were measured, which indicates that the sphere settled in a stable radial trap. It is seen that the sphere stays in the stable radial trap for ~0.3 s (in some cases up to 2-3 s) covering sub-millimeter distances (in some cases up to 1 mm).

A different scenario is illustrated in Fig. 5.5(b). After the initial period of fast approaching the taper in the first 100 ms, the sphere did not settle in a stable radial trap. On the contrary, it continued to approach the taper at a slower pace during the time interval from 100 ms to 500 ms. As expected, the propulsion velocity was slowly increasing during this time. At about 180 ms the sphere suddenly moved away from the taper as a result of perturbation. Taper surface roughness, flux turmoil, and non-uniform charging condition may contribute to such variation in experiments. However, the depth of the transverse trap was sufficient to bring the sphere back to the taper. It indicates the balance of multiple transverse forces that results in a self-restoring radial trap. This case



shows that only at the end of this evolution the sphere approached the stable radial

trapping condition.

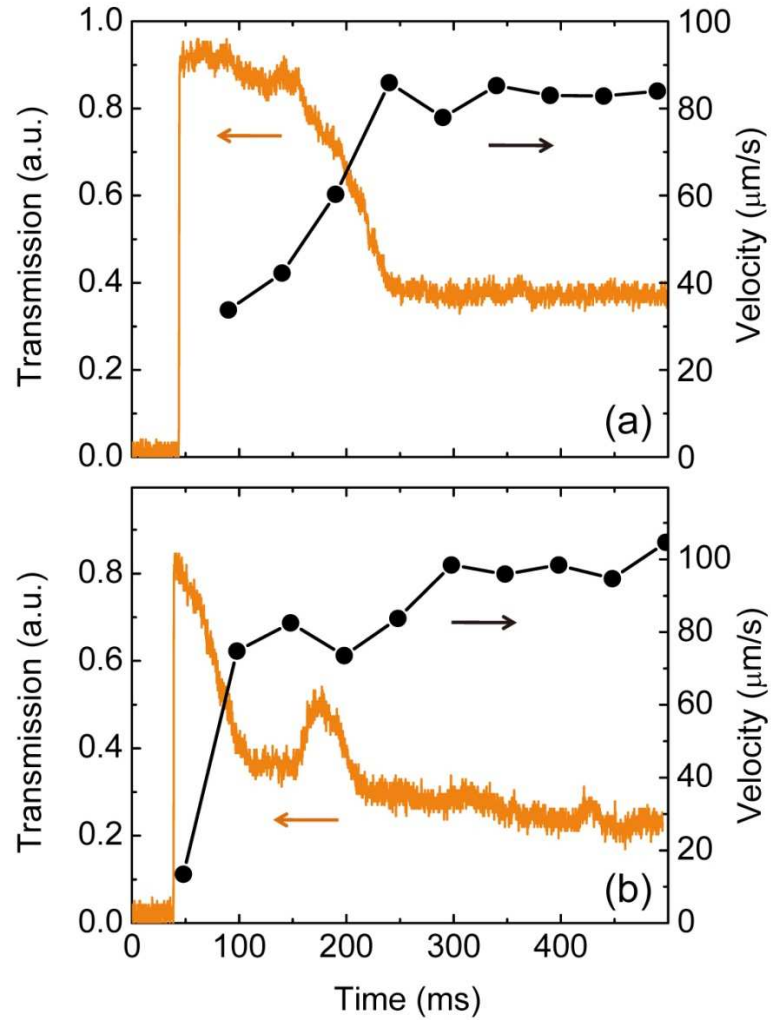

Figure 5.5: Fiber power transmission (photocurrent) and the propulsion velocity measured for $\Delta\lambda=0$ as a function of time for 20 μm spheres. (a) Stable radial trap achieved after 200 ms. (b) Stable radial trap is slowly approached after 500 ms. [6]



## 5.5. Linear Dependence of Propulsion Velocity on Optical Power

Propulsion velocity depends on many parameters of the system such as the sphere size, index and associated $Q$-factors of WGMs, the laser detuning from the WGM peaks, and the guided power in the tapered region of the fiber. It could be assumed that in the limit of small optical power the latter dependence should be linear for the same detuning since the terminal velocity ($v$) is proportional to the total momentum flux. In this case, the power dependence can be excluded from the analysis by normalizing the velocity by the laser power. It also enables comparison of particle propulsion results obtained by using different optical powers.

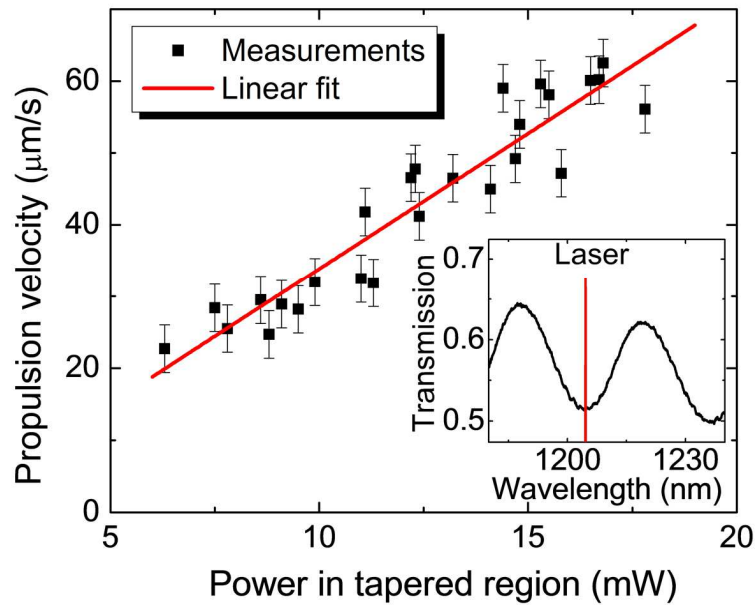

Figure 5.6: Optical power dependence of propulsion velocity for 10 μm spheres with a linear fit. Insert: typical transmission spectrum for 10 μm sphere in water and laser emission line tuned at resonance dip (zero wavelength detuning). [6]



We checked the power linearity of the propulsion effects for various conditions. An example of such study for polystyrene spheres with 10 μm mean diameter and zero detuning is presented in Fig. 5.6 [6].

The WGM-defined dips in the fiber transmission spectra are very broad and shallow, as shown in the inset of Fig. 5.6. For each sphere the laser emission line was tuned to the middle of one of the spectral dips. The propulsion velocity was found to be linearly dependent on the laser powers within 6-18 mW range. For smaller powers the spheres were found difficult to be retained near the taper due to insufficient depth of the radial trap and thus could not be propelled.

### 5.6. Optical Propulsion with Resonant Wavelength Detuning

We have seen in Section 5.4 dramatic difference in comparison of optical propulsion of 20 μm spheres under on-resonance and off-resonance conditions. In this Section, we present results of measurements of propulsion velocity as a function of the laser wavelength detuning $(\Delta\lambda)$ relative to the center of a given resonance dip in fiber transmission spectra. The velocity measurements were performed after the sphere settled in a stable radial trap, as illustrated in Fig. 5.5. The measured velocities were normalized by the total optical power guided in the tapered region, presented win the unit of $\mu m \cdot s^{-1} \cdot mW^{-1}$.

The propulsion velocity measurements for polystyrene spheres of 10 μm and 20 μm mean diameters are summarized in Figs. 5.7(a) and 5.7(b) respectively [6]. Multiple points for the same detuning represent multiple measurements with different spheres.

The coupling between a tapered fiber and a 10 μm sphere in water is weak, indicated by the broad resonance with a shallow depth seen in Fig. 5.7(a). For the entire range of



detuning from -15 nm to +15 nm, the radial trapping followed by the optical propulsion could be observed. The measured velocities were found to be within 2-5 $\mu m \cdot s^{-1} \cdot mW^{-1}$ range. The spectral shape of the velocity peak inversely replicates the shape of the dip in the fiber transmission spectrum, as illustrated in Fig. 5.7(a).

The case of water-immersed 10 $\mu m$ polystyrene microspheres represents a transition from non-resonant optical propulsion studied previously for smaller spheres [129, 131, 133, 139, 231] to resonant optical propulsion observed for larger spheres with 15-20 $\mu m$ diameters [5]. This can be understood by relating our experimental observations to the modeling results presented in Fig. 5.1, when taking into account the smaller index contrast for water-immersed polystyrene spheres ($\sim$1.2) compared to that (1.30) used in the modeling. For $\lambda_0$=1.25 $\mu m$, one can estimate $kR$~25 in our experimental case. Additional calculations performed with the 1.2 index contrast (not presented in Fig. 5.1) show extremely weak oscillations of both forces, $F_x$ and $F_y$, for $kR\leq$25, which is in agreement with previous experiments performed on spheres with $D$<10 $\mu m$ [129, 131, 133, 139, 231]. The force oscillations tend to increase with $kR$, however they remain weak at $kR$=25 in qualitative agreement with the results of propelling $D$=10 $\mu m$ spheres shown in Fig. 5.7(a).

In contrast, the case of water-immersed 20 $\mu m$ polystyrene microspheres represents the value of $kR$~50 where the resonances dips in fiber transmission are extremely sharp and distinct demonstrating $Q$~$10^3$. Calculations for the 1.2 index contrast predict strong peak forces for $kR$~50. The magnitude of the force peaks in this case is comparable to that calculated for the 1.3 index contrast in the 27$\leq kR \leq$32 range presented in Fig. 5.1(b).



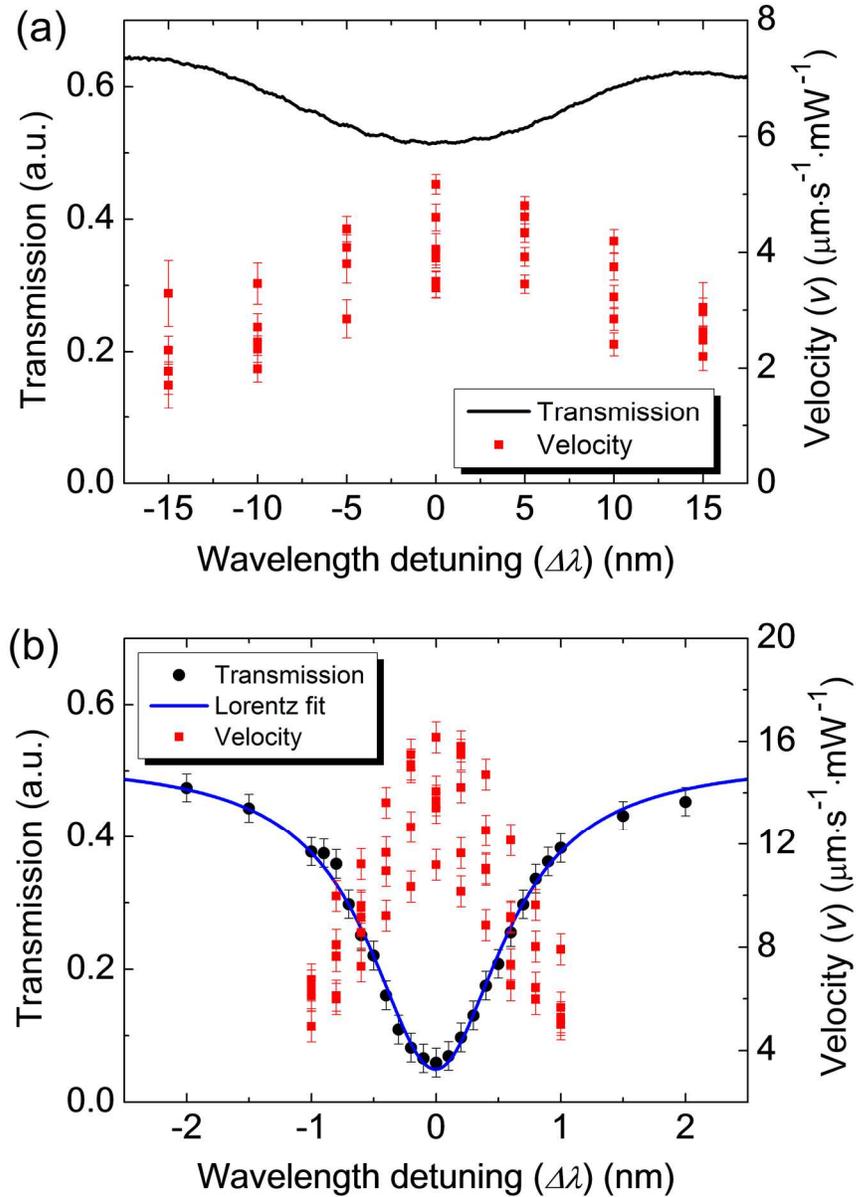

Figure 5.7: Transmission of the fiber with the water-immersed polystyrene microsphere held at the waist of the taper by optical tweezers (solid lines) and maximal propulsion velocities ($v$) normalized by the power in the tapered region (squares) measured as a function of the laser wavelength detuning ($\Delta\lambda$) from the center of the WGM resonance for spheres with different mean diameters: (a) 10 μm and (b) 20 μm. For each detuning the velocity measurements were repeated with different spheres leading to multiple data points represented by squares. The curve in (b) represents a Lorentz fit of the experimental transmission measurements. [6]



Experimentally, the coupling to 20 μm spheres was found to be rather close to a critical coupling regime, as illustrated in Fig. 5.7(b). We utilized the same tunable laser for the spectral transmission and propulsion experiments. The transmission of the fiber coupled to the microsphere was normalized by the fiber transmission without the microsphere, as illustrated by black circles in Fig. 5.7(b). The Lorentz fit of the transmission dip with 90% (10 dB) depth and the 1.1 nm width is also shown in Fig. 5(b).

The propulsion velocity measurements performed with multiple spheres of 20 μm mean diameter as a function of the wavelength detuning are presented in Fig. 5.7(b). An extraordinary high velocity of 16 $\mu m \cdot s^{-1} \cdot mW^{-1}$ was observed when the laser was tuned exactly at the center of the resonance dip in the transmission spectrum. This peak value exceeds previous measurements of non-resonant propulsion velocities with various evanescent couplers [129, 131, 133, 139, 231] by more than an order of magnitude. Although the resonant enhancement of the propulsion velocities was observed in our previous studies [5], the peak velocity in Fig. 5.7(b) exceeds our previous measurements for 20 μm spheres by a factor of 1.6. It can be explained by the fact that in present work we were able to precisely set the detuning to zero for each sphere, whereas in our previous studies we relied on the random detuning between the fixed laser wavelength and WGMs in different spheres. The distribution of the velocity over the limited detuning range, $|\Delta\lambda| < \lambda_0/Q$, was found to be inversely proportional to the shape of the resonance dip in the transmission spectrum. As discussed previously, the propulsion could not be observed for spheres detuned by more than 1 nm away from the center of their WGMs resonances because of the loss of sufficient radial trapping for a given power.



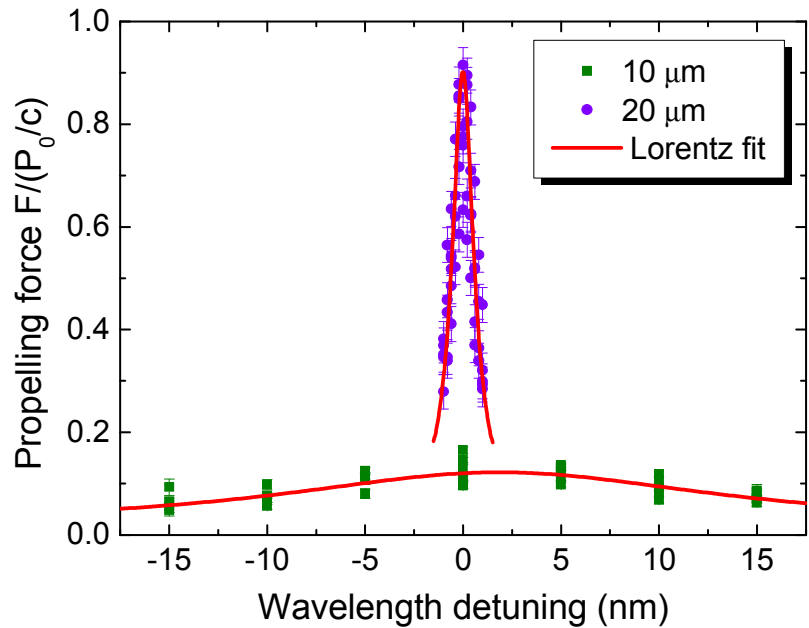

Figure 5.8: Comparison of optical propelling force for 10 μm and 20 μm spheres as a function of the laser wavelength detuning from the WGM resonance. [6]

A simple method of estimating the propelling forces ($F_x$) is based on the assumption that the propelling force is balanced by the liquid drag force, $3\pi\mu Dv$, when the terminal velocity is reached. It is convenient to plot the propelling force in the unit of $P_0/c$ which is equivalent to the momentum flux of the incident radiation in vacuum. The momentum flux for a guided mode is larger than $P_0/c$ by an additional factor which is equal to the value of the phase index of the mode $n_{ph} = c/v_{ph}$. However, the consideration of not including the phase index into the normalization constant when plotting the experimental data comes from the Abraham-Minkowski controversy in which the index of the medium appears either in numerator or denominator of the photon momentum definition [268, 269]. For the modes guided by the tapered fiber, the phase index lies between the



refractive index of the taper material (silica glass, $n$=1.45) and that of the surrounding medium (water, $n$=1.33). In the limit of a very thin taper, the phase index should be slightly larger than that of water and, therefore, the value $cF_x/P_0$=1 should correspond to the conversion efficiency of $1/1.33\approx75\%$ from the incident photon flux to the propelling force. The resultant propelling forces (in unit of $P_0/c$) are presented in Fig. 5.8 as a comparison of 10 and 20 μm spheres [6].

For 10 μm spheres, the force peak is broad and relatively flat. The magnitude of non-resonant optical force with large detuning is less than 10% of $cF_x/P_0$. With partial resonant enhancement when the laser tuned to the center of resonance the force peak is still limited at $cF_x/P_0\sim0.15$ due to weak coupling. On the other hand, for 20 μm spheres the optical forces show a sharp peak with narrow width. The peak force value corresponds to $cF_x/P_0\sim0.95$ due to strong resonant enhancement. As was discussed in Section 5.2, this is a result of very efficient coupling with WGMs in microspheres. It should be noted that observation of such extremely high efficiencies provides strong support for our model of the observed effects. The results showed that stronger coupling leads to larger propulsion force, and thus directly confirmed the resonant enhancement of optical forces. While the propulsion did not occur when the laser detuned 1 nm away from the center of resonance, an extraordinary high efficiency of 95% of $cF_x/P_0$ was realized when the laser line matched the wavelengths of WGMs in spheres. Such extreme sensitivity of optical forces for wavelength detuning can be utilized to develop sorting techniques for obtaining microspheres with ultra-high uniform WGMs resonances.



## 5.7. Summary

This work provides a direct experimental proof of existence of ultra-high resonant propulsion forces in microspherical photonics. In contrast to our previous work [5], we took full control of the propulsion experiments. It was achieved by manipulation of individual spheres using optical tweezers and by setting the amount of detuning between the laser and WGMs in each sphere. As a result, we measured the shape of the spectral response of the propulsion forces and established that it correlates precisely with the coupling properties of the microspheres. We also proved that the amplitude of the peak forces is in agreement with the model calculations.

The sorting technology developed in this work allows selecting building blocks of various photonic structures with extraordinary uniform resonant properties. Standard sorting technologies are based on sphere diameters and cannot provide the accuracy required for efficient resonant tunneling between WGMs with sufficiently high $Q$-factors [17, 101]. Such building blocks can be also viewed as "photonic atoms" [93] based on analogy between the quantum mechanics and the classical electromagnetics of the WGMs in spherical particles [176]. This work shows that the level of the uniformity of such photonic atoms sorted by light allow their use in photonic applications similar to indistinguishable quantum mechanical atoms. The photonic dispersions can be engineered in such structures based on tight-binding approximation leading to applications in coupled resonator optical waveguides [102, 106, 111, 158], high-order spectral filters [21], delay lines [262], sensors [159, 250], and laser-resonator arrays [188].



Sorting of microspheres resonant with the laser wavelength can be achieved for WGMs with different $l$ numbers. However, it does not limit the applications of microspherical photonics since coupling between WGMs with different $l$ numbers is rather efficient [101]. If initial size dispersion of microspheres is sufficiently small, sorting of spheres with the same $l$ number might become possible. This means that the spheres sorted by light would have diameters identical within $\sim 1/Q$ relative accuracy. Such spheres might become the most accurate calibration standards in metrology applications.

From the point of view of fundamental studies, particularly interesting structures are represented by large-scale 2-D and 3-D arrays of spheres where WGMs are coupled on a massive scale. The optical transport through such coupled-cavity networks is very complicated and poorly studied which involves multiple paths for photons leading to interference, localization [270] and percolation [101] of light. The optical gain in such structures can be realized by doping the spheres with dye molecules or active ions. The combination of gain, scattering loss and optical nonlinearity in such coupled-cavity networks can result in unidirectional propagation properties [187].

By using tunable laser source it should also be possible to select spheres resonating at different wavelengths. They can be assembled in structures where both position of spheres and their individual resonant properties are controlled according to a certain design. This opens a way to explore quantum-optics analogies in photonics [96] and to create novel structures such as parity-time synthetic lattices [271], coupled resonator ladders [272] or waveguides based on an effective gauge field for photons [273].

It is likely that the applications of such structures and devices will be developed



using high-index ($n$>1.9) microspheres which can possess $Q$~$10^4$ in sufficiently small particcles with $D$~3 µm suitable for compact coupled-cavity devices. Although the specific details of the sorting set-up may vary depending on the properties of microspheres [7, 8], the present work shows the basic principles of resonant light forces useful for developing such applications.

CHAPTER 6: CONCLUSIONS AND OUTLOOK

In Chapter 2, we developed a robust tapered microfiber coupler platform, which, with its subsequent modifications, served as the generic part for all experimental setups used in this dissertation, including WGMs characterization of single microspheres and coupled microcavity arrays, and optical trapping and propulsion of dielectric microparticles. All the resonant effects being investigated are based on WGMs in spheres. Therefore, the realization of efficient taper-to-sphere evanescent coupling and the understanding of the resonant properties of individual spheres are essential building blocks of our work. We characterized WGMs in dielectric microspheres made from a variety of materials and with different diameters and found that the high index barium titanate glass ($n$~1.9-2.1) spheres are preferable for developing high-$Q$ compact photonic devices. Even in water medium, which usually required in biomedical and microfluidic applications, the BTG spheres demonstrate $Q$~$3 \times 10^3$ for 5 μm diameter. We also believe that our characterization of $Q$-factors of WGMs for spheres made from different materials and immersed in different media is very useful for the research groups working on sensor applications of microspherical photonics. Some spheres have been characterized prior to our work [83, 86, 161, 166]. To the best of our knowledge, however, the systematic studies have not been performed for sufficiently wide range of index contrasts with sufficiently wide range of spheres' diameters. Actually, our results support an exponential dependence of the experimentally measured (loaded) values of the WGM



*Q*-factors for spheres smaller than 20 μm. Because of the limited spectral resolution of our system, our characterization was generally limited by $Q<10^5$, however, taking into account wide range of studied spheres, our measurements have good predictive power regarding the regime of coupling (undercoupled, critical, or overcoupled) and values of *Q*-factors available in different environments. We found that gradually other groups started using our results [9] as a reference for characterization of microspheres with tapered fiber coupler.

In Chapter 3, we studied strongly coupled WGMs of photonic molecules formed by different configurations of almost identical photonic atoms. These studies were performed to illustrate the future prospects of microspherical photonics. To guide our fabrication and characterization effort, we used 2-D FDTD modeling performed with identical circular cavities with very similar dimensions and refractive index contrasts to the experimental situations. We experimentally assembled photonic molecules which can be considered as photonic analogs of the electronic molecules formed by identical atoms. The simplest example of such photonic molecules is given by a bisphere which can be considered as "photonic hydrogen" molecule with a hybridization of WGMs which have symmetric and antisymmentric modes similar to its electronic counterpart. More complicated configurations of coupled WGMs were built and characterized using sorted microspheres with almost identical (within 0.05%) wavelength positions of WGMs. These molecules were tested using fiber transmission spectroscopy. The results are found to be in very good agreement with the 2-D FDTD modeling that shows good prospect of micropherical photonics for developing structures and devices with engineered photonic dispersions. Such structures can be used as multi-wavelength sensors, slow-light



waveguide, and laser-resonator coupled arrays.

In Chapter 4 and Chapter 5, we have observed and studied the optical trapping and propulsion effects with polystyrene microspheres. Initially, as described in Chapter 4, we relied on random deviations of sphere diameters in our optical propulsion experiments. The basic idea was to allow different spheres with different resonant properties to pass near the tapered fiber, and only those spheres which have WGMs resonant with the laser emission wavelength could be propelled at higher velocities [5]. Eventually, we took full control of the propulsion experiments so that both spatial location of spheres and spectral position of their WGMs could be precisely controlled in our experiments [6], which has been described in Chapter 5. The resonant enhancement of optical forces has been demonstrated in experiment with large polystyrene spheres and explained by theoretical calculations. We propose that due to its giant magnitude, extreme sensitivity, and large peak-to-background ratio, the resonant propulsion force can be utilized for sorting microspheres with extraordinary uniform resonant properties unachievable with current fabrication techniques. We believe that our work stimulated a great interest in studies of basic physics and applications of this resonant propulsion effect. In particular, this interest is explained by potential applications of this effect for developing methods of massive-scale sorting of microspheres with the positions of WGM peaks resonant with the wavelength of the laser source. This work provides a direct experimental proof of such principle of particle sorting. However, due to the low refractive index of polystyrene and the degradation of $Q$-factor by water medium, only large spheres ($D$>15 μm) can be sorted by the resonant propulsion in a liquid environment studied in our experiment.



From the point of view of developing microspherical photonics, it is desirable to be able to sort high index spheres ($n>1.9$) which can possess $Q>10^4$ in sufficiently small $D\sim3$ μm particles, suitable for the construction of compact coupled-cavity devices. These spheres made from high index materials, however, have much larger mass densities than water or other liquids. They quickly sink in the liquid thus it will be difficult to manipulate with the tapered microfiber coupler. For this reason, it is likely that sorting of such high index spheres will be achieved by modified experimental techniques [7].

Some of these designs of sphere sorting devices have been suggested in Ref. [8]. For example, a slab waveguide fabricated on the surface of a substrate can be used for sphere sorting. The substrate surface needs to be treated to eliminate electrostatic charge that would otherwise trap microparticles. The substrate is positioned with an inclined angle on top of a glass plate, as illustrated in Fig. 6.1. Laser light is coupled into the waveguide and microspheres are released from the top of the substrate.

The spheres will roll down due to gravity, however, when in contact with the surface waveguide they will experience optical forces. For spheres whose WGMs are not resonant with the laser the propulsion forces are small thus they will continue to roll down without much change of the direction due to small duration of interaction. On the other hand, those on-resonance spheres will experience significantly enhanced propulsion force and thus accelerate in the light propagation direction. The horizontal velocity will result in the change of falling direction as indicated by dashed lines in Fig. 6.1. Therefore, the spheres will be separated on the glass plate according to the detuning between their WGMs and the laser where a smaller wavelength detuning leads to a further forward position.



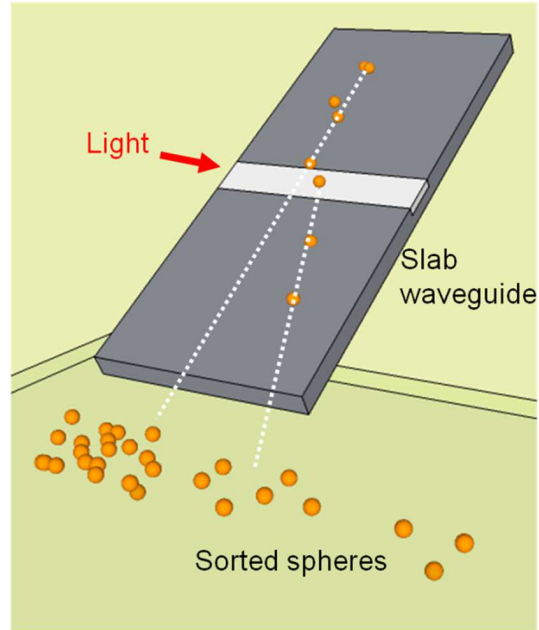

Figure 6.1: Schematics of experimental set-up for sphere sorting with slab waveguide. Position of the sphere is determined by the magnitude of the propulsion force which is extreme sensitive to its WGM detuning from the laser.

It has been recently demonstrated in theoretical calculation that a focused free-space laser beam can also resonantly propel microspheres efficiently [7]. It was shown that the produced forces strongly depend on the detuning between WGMs and laser wavelengths. When the spheres traverse the focused beam the difference of obtained horizontal velocity between the on- and off-resonance cases will lead to a measurable separation of the particles equal to the product of the velocity difference and the time of free fall under gravity. In order to enhance the sorting efficiency we can use a one-dimensional focused beam created by a cylindrical lens, as illustrated in Fig. 6.2. Such design will expand the region of light-sphere interaction to a horizontal plane instead of a focus point, which reduces the possibility of spheres falling through without experiencing the light forces.



The mechanism of sphere sorting by resonant light forces is very similar in different experimental implementation. However, with a variety of modified designs we can perform the sorting in vacuum, air, or liquid environments [8]. It is particularly important that this method can also be applied to the sorting of heavier high index spheres.

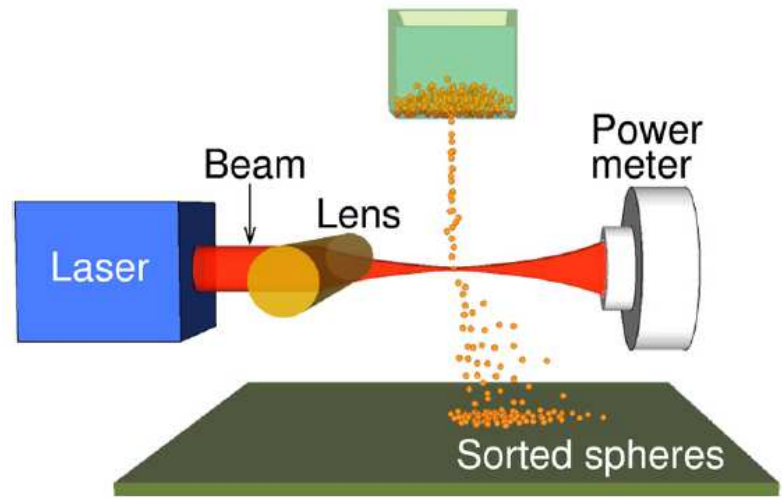

Figure 6.2: Schematics of sphere sorting by one-dimensional focused free-space laser beam. The difference of obtained horizontal velocities when traversing the focused beam between the on- and off-resonance spheres will lead to a measurable separation. [7]

In Chapter 3, we were able to select a few polystyrene microspheres of 25 μm mean diameter with 0.05% uniformity of WGM resonances positions by spectroscopic comparison, and successfully assembled several 1-D and 2-D photonic molecules for the characterization of coupled cavity modes. However, this sorting method is time consuming and not practical to obtain a large quantity of resonant spheres.

By applying the sorting mechanism developed in this dissertation, an automatic parallel spheres sorting can be realized and a large number of spheres with desired



resonant properties can be acquired. Such spheres can be used as building blocks for the construction of complex 2-D and even 3-D arrays where WGMs are coupled on a massive scale.

The optical properties of such large-scale coupled cavity networks are complicated and poorly studied. We have demonstrated in simulation with a few 1-D and 2-D configurations that the resonant properties, photonic dispersion, optical transport, and spatial distribution can all be engineered in such coupled photonic molecules. The capability of obtaining desirable resonant spheres in a large quantity opens up great opportunities for the experimental studies of the fundamentals in quantum optics [96] as well as for the experimental explorations of novel photonic structure designs such as coupled resonator ladders [272] and parity-time synthetic lattices [271]. Optical gain and nonlinearity can also be introduced in such structures by choosing spheres made from active materials or by doping the spheres with dye molecules or active ions. The combination of gain, scattering loss and nonlinearity can result in novel optical properties such as unidirectional propagation in coupled bi-cavities system [187]. There is more to explore with the gain and nonlinearity in such coupled cavity networks.

Besides coupling through WGMs, microsphere arrays support another non-resonant type of optical modes referred to as photonic "nanojet-induced" [203, 236, 237] or "periodically-focused" [247] modes. Due to these modes, the chains of spheres can operate as polarization filters [274], low-loss waveguides with focusing capability [203, 236, 237] and laser surgical devices [246, 275-277]. Arrays of microspheres can also be used in super-resolution imaging applications [278-284]. All these non-resonant microspherical applications may potentially benefit from using sorted high-$Q$ resonant



spheres that can reduce optical losses, ensure optical quality and surface smoothness, and facilitate self-ordering of spheres in such arrays.

## APPENDIX: LIST OF PUBLICATIONS

Journal Articles:

1.  Oleksiy V. Svitelskiy, **Yangcheng Li**, Arash Darafsheh, Misha Sumetsky, David Carnegie, Edik Rafailov, and Vasily N. Astratov, "Fiber coupling to $BaTiO_3$ glass microspheres in an aqueous environment," Optics Letters, 36(15), 2862-2864 (2011).

2.  **Yangcheng Li**, Oleksiy V. Svitelskiy, Alexey V. Maslov, David Carnegie, Edik Rafailov, and Vasily N. Astratov, "Giant resonant light forces in microspherical photonics," Light: Science & Applications, 2, e64 (2013).

3.  **Yangcheng Li**, Alexey V. Maslov, Nicholaos I. Limberopoulos, Augustine M. Urbas, and Vasily N. Astratov, "Spectrally resolved resonant propulsion of dielectric microspheres," Laser & Photonics Reviews, 9(2), 263-273 (2015).

4.  **Yangcheng Li**, Farzaneh Abolmaali, Kenneth W. Allen, and Vasily N. Astratov, "Spectral signatures of photonic molecules," to be submitted.

5.  Kenneth W. Allen, Navid Farahi, **Yangcheng Li**, Nicholaos I. Limberopoulos, Dennis E. Walker Jr., Augustine M. Urbas, Vladimir Liberman, and Vasily N. Astratov, "Movable thin films with embedded high-index microspheres for super-resolution microscopy," submitted.

6.  Kenneth W. Allen*, Navid Farahi*, **Yangcheng Li**, Nicholaos I. Limberopoulos, Dennis E. Walker Jr., Augustine M. Urbas, Alexey V. Maslov, and Vasily N. Astratov, "Overcoming diffraction limit of imaging nanoplasmonic arrays by microspheres and microfibers," submitted.

Conference Proceedings:

1.  Oleksiy V. Svitelskiy, **Yangcheng Li**, Misha Sumetsky, David Carnegie, Edik Rafailov, and Vasily N. Astratov, "A microfluidic platform integrated with tapered optical fiber for studying resonant properties of compact high index microspheres," IEEE Proceedings of 13th International Conference on Transparent Optical Networks (ICTON), paper We.P.1 (2011).

2.  Oleksiy V. Svitelskiy, **Yangcheng Li**, Misha Sumetsky, David Carnegie, Edik Rafailov, and Vasily N. Astratov, "Resonant coupling to microspheres and light pressure effects in microfluidic fiber-integrated platforms," Proceedings of IEEE Photonics Conference, 185-186 (2011).